\documentclass[12pt,a4paper]{article}
\usepackage[T1]{fontenc}
\usepackage{amssymb}
\usepackage{amsmath}
\usepackage{amsthm} % used for definition/theorem/corollary/... numbering

\usepackage{stmaryrd}
\usepackage{graphicx}
\usepackage{here}
\usepackage[margin=1in]{geometry}
\usepackage[colorlinks,allcolors=blue]{hyperref}
\usepackage{cite}
\numberwithin{equation}{section}
\newcommand{\be}{\begin{equation}}
\newcommand{\ee}{\end{equation}}
\newcommand{\ben}{\begin{enumerate}}
\newcommand{\een}{\end{enumerate}}
\newcommand{\bi}{\begin{itemize}}
\newcommand{\ei}{\end{itemize}}
\newcommand{\bea}{\begin{eqnarray}}
\newcommand{\eea}{\end{eqnarray}}
\newcommand{\nn}{\nonumber}
\newcommand{\bm}{\begin{pmatrix}}
\newcommand{\emm}{\end{pmatrix}}

\newcommand{\demi}{\frac{1}{2}}
\newcommand{\phii}{\varphi}

\newcommand{\der}{\partial}

\newcommand{\calM}{{\cal M}}
\newcommand{\calO}{{\cal O}}
\newcommand{\calP}{{\cal P}}
\newcommand{\calQ}{{\cal Q}}

\newcommand{\mM}{\mathrm{M}}
\newcommand{\mO}{\mathrm{O}}
\newcommand{\mU}{\mathrm{U}}
\newcommand{\SL}{\mathrm{SL}}
\newcommand{\SU}{\mathrm{SU}}
\newcommand{\SO}{\mathrm{SO}}

\newcommand{\su}{\mathfrak{su}}
\newcommand{\CC}{\mathbb{C}}
\newcommand{\HH}{\mathbb{H}}
\newcommand{\OO}{\mathbb{O}}
\newcommand{\RR}{\mathbb{R}}
\newcommand{\VV}{\mathbb{V}}
\newcommand{\ZZ}{\mathbb{Z}}

\newcommand{\Ker}{\mathrm{Ker}}
\newcommand{\Ima}{\mathrm{Im}}

\begin{document}

\hrule
\begin{center}
\Large{\bfseries{\scshape{From the Lorentz Group to the Celestial Sphere}}}
\end{center}
\hrule
~
\begin{center}
\large{\bfseries{\scshape{Blagoje Oblak$^*$}}}
\end{center}
~\\
~\\
~
\begin{centering}
\begin{minipage}{\textwidth}\small \it  \begin{center}
   Physique Th\'eorique et Math\'ematique\\
   Universit\'e Libre de Bruxelles and International Solvay Institutes\\
   Campus Plaine C.P. 231, B-1050 Bruxelles, Belgium
 \end{center}
\end{minipage}
\end{centering}
~\\
~\\
~\\
~
\begin{center}
\begin{minipage}{.9\textwidth}
\begin{center}{\bfseries{Abstract}}\end{center}
In these lecture notes we review the isomorphism between the 
(connected) Lorentz group and the set of conformal transformations of the sphere. More precisely, 
after establishing the main properties of the Lorentz group, we show that it is isomorphic to the group 
$\SL(2,\CC)$ of complex $2\times 2$ matrices with unit 
determinant. We then classify conformal transformations of the sphere, define the notion of 
null infinity in Minkowski space-time, and show that the action of Lorentz transformations on the celestial 
spheres at null infinity is precisely that of conformal transformations. In particular, we discuss the 
optical phenomena observed by the pilots of the {\it Millenium Falcon} during the jump to lightspeed.\\
~\\
~\\

This text, aimed at undergraduate 
students, was written for the seventh Brussels Summer School of 
Mathematics$^{\dagger}$ that took place at Universit\'e Libre de Bruxelles in August 2014. An abridged 
version has been 
published in the proceedings of the school, ``Notes de la septi\`eme BSSM'', printed in July 2015.
\end{minipage}
\end{center}

\vfill
\noindent
\mbox{}
\raisebox{-3\baselineskip}{%
  \parbox{\textwidth}{\mbox{}\hrulefill\\[-4pt]}}
{\scriptsize $^*$ Research Fellow of the Fund for Scientific Research-FNRS
Belgium. E-mail: boblak@ulb.ac.be\\
\scriptsize $^{\dagger}$ Website: http://bssm.ulb.ac.be/}

\newpage
\tableofcontents

\newpage
\section*{Introduction}
\addcontentsline{toc}{section}{Introduction}

The Lorentz group is essentially the symmetry group of special relativity. It is commonly defined as a set of 
(linear) transformations acting on a four-dimensional vector space $\RR^4$, representing changes of 
inertial frames in Minkowski space-time. But as 
we will see 
below, one can exhibit an isomorphism between the Lorentz group and the group of 
conformal transformations of the sphere $S^2$; the latter is of course two-dimensional. 
This 
isomorphism thus relates the action of a group on a {\it four}-dimensional space to its action on 
a {\it two}-dimensional manifold. At first sight, such a relation seems surprising: loosely speaking, one 
expects to have lost some information in going from four to two dimensions. In particular, the isomorphism 
looks like a coincidence of the group structure: there is no obvious geometric 
relation between the original four-dimensional space on the one hand, and the sphere on the other hand.\\

The purpose of these notes is to show that such a relation actually exists, and is even quite natural. 
Indeed, by defining a notion of ``celestial 
spheres'', one can derive a direct link between four-dimensional Minkowski space and the 
two-dimensional sphere. In short, the celestial sphere of an inertial observer in Minkowski space is the 
sphere of all directions towards which the observer can look, and coordinate transformations between 
inertial observers ({\it i.e.}~Lorentz transformations) correspond to conformal transformations of this 
sphere \cite{Terrell,PenroseSphere,Penrose,HeldNewman}. In this work we will review this construction 
in a self-contained way.\\

Keeping this motivation in mind, the text is organized as follows. In section \ref{secSpec}, we review 
the basic principles of special relativity and define the natural symmetry groups that follow, namely the 
Poincar\'e group and its homogeneous subgroup, the Lorentz group \cite{HenneauxGroupe,HenneauxRG,Hladik}. 
This will also be an excuse to discuss 
certain 
elegant properties of the Lorentz group that are seldom exposed in elementary courses on special relativity, 
in particular regarding the physical meaning of the notion of ``rapidity'' \cite{Levy,LevyRap,LevySpeed}. In 
section \ref{secIsom}, we 
then establish the isomorphism between the connected Lorentz group and the group $\SL(2,\CC)$ of complex, two 
by 
two matrices of 
unit determinant, quotiented by its center $\ZZ_2$. We also derive the analogue of this 
result in three space-time dimensions. Section \ref{secConf} is devoted to the construction of conformal 
transformations of the sphere; it is shown, in particular, that such transformations span a group 
isomorphic to $\SL(2,\CC)/\ZZ_2$ $-$ a key result in the realm of two-dimensional conformal field 
theories \cite{DiFran,Blum}. 
At that point, the stage will be set for the final link between the Lorentz group and the sphere, which 
is established in section \ref{secLorentzSphere}. The conclusion, section \ref{secCon}, relates these 
observations to some recent developments in quantum gravity $-$ in particular BMS 
symmetry 
\cite{BvM,Sachs,BarnichTroessaert01,BarnichTroessaert02,Banks,Strominger01,He,Kapec} and 
holography \cite{tHooft,Susskind,Maldacena,Witten,Gubser}.\\

The presentation voluntarily starts with fairly elementary considerations, in order to be accessible 
(hopefully) to undergraduate 
students. Though some basic knowledge of group theory and special relativity should come in handy, no prior 
knowledge of differential geometry, general relativity or conformal field theory is assumed. In particular, 
sections \ref{secSpec} and \ref{secIsom} are mostly based on the undergraduate-level lecture notes 
\cite{HenneauxGroupe}.

\section{Special relativity and the Lorentz group}
\label{secSpec}

In this section, after reviewing the basic principles of special relativity (subsection 
\ref{subsecSpec}), we define the associated 
symmetry groups (subsection \ref{subsecPoin}) and introduce in particular the Lorentz group. In subsection 
\ref{subsecSub}, we then 
define certain natural subgroups of the latter. Subsection \ref{ssBoosts} is devoted to the notion 
of Lorentz boosts and to the associated additive parameter, which turns out to have the physical meaning of 
``rapidity''. Finally, in subsection \ref{ssStd} we show that any Lorentz transformation preserving the 
orientation of space and the direction of time flow can be written as 
the product of two rotations and a boost, and then use this result in subsection \ref{ssCon} to classify the 
connected components of the Lorentz group. All these results are well known; the acquainted reader may 
safely jump directly to section \ref{secIsom}. The presentation of this section is mainly inspired from the 
lecture notes \cite{HenneauxGroupe} and 
\cite{HenneauxRG}; more specialized references will be cited in due time.

\subsection{The principles of special relativity}
\label{subsecSpec}

\subsubsection{Events and reference frames}

In special relativity, natural phenomena take place in the arena of space-time. The latter consists of 
points, called {\it events}, which occur at some position in space, at some moment in time. 
Events are seen by observers who use coordinate systems, also called {\it reference frames}, to 
specify the location of an event in space-time. In the realm of special relativity, reference frames 
typically consist of three orthonormal spatial coordinates\footnote{The presentation here is confined to 
four-dimensional space-times, but the generalization to $d$-dimensional space-times is straightforward: 
simply take spatial coordinates $(x^1,...,x^{d-1})$.} $(x^1,x^2,x^3)$ and one time coordinate $t$, measured 
by a clock carried by the observer. For practical purposes, the speed of light in the vacuum,
\be
c=299\,792\,458\;\text{m/s},
\label{c}
\ee
is used as a conversion factor to express time as a quantity with dimensions of distance. This is done by 
defining a new time coordinate $x^0\equiv ct$.\\

Thus, in a given reference frame, an event occurring in space-time is labelled by its four coordinates 
$(x^0,x^1,x^2,x^3)$, 
collectively denoted as $(x^{\mu})$. (From now one, greek indices run over the values $0$, $1$, $2$, $3$.) Of 
course, the event's existence is independent of the observers who see it, but its coordinates are not: if 
Alice and Bob are two observers looking at the same event, Alice may use a set of four 
numbers $(x^{\mu})$ to describe its location, but Bob will in general use different coordinates 
$(x'^{\mu})$ to locate the same event. Besides, if we do not specify further the relation between 
Alice and 
Bob, there is no link whatsoever between the coordinates they use. What we need are restrictions on 
the possible reference frames used by Alice and Bob; the principles of special relativity will then 
apply only to those observers whose reference frames satisfy the given restrictions.

\begin{figure}[H]
\begin{center}
\includegraphics[width=0.60\textwidth]{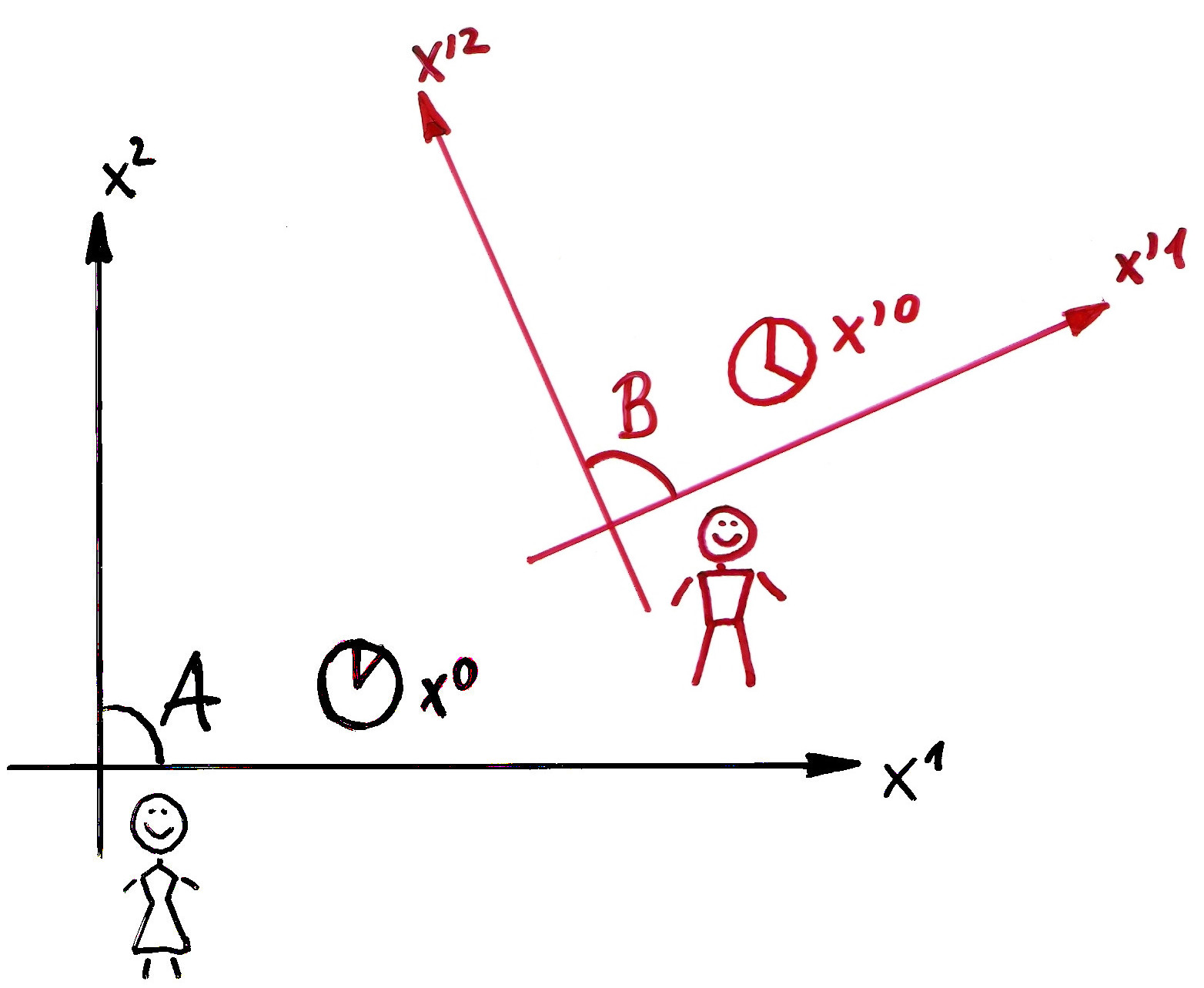}
\caption{Alice (drawn in black) and Bob (drawn in red) in space-time, with their respective reference frames 
(including clocks); for 
simplicity, only two spatial dimensions are represented. Looking at the same event, Alice and Bob will 
typically associate different coordinates with it. In general, there is no way to 
relate the coordinates of Alice's frame to those of Bob's frame.}
\end{center}
\end{figure}

\subsubsection{The principles of special relativity}

We now state the defining assumptions of special relativity. The three first basic assumptions are 
homogeneity of space-time, isotropy of space and causality \cite{Levy,Hladik}. The remaining 
principles, discussed below, lead to constraints on the relation between reference frames. To expose these 
principles, we first need to define the notion of inertial frames.\\

According to the principle of {\it inertia}, a body left to itself, without forces acting 
on it, should move in space in a constant direction, with a constant velocity. Obviously, this principle 
cannot hold in all reference frames. For example, suppose Alice observes that the principle of inertia is 
true in her reference frame (for example by throwing tennis balls in space and observing that they move in 
straight 
lines at constant velocity). Then, if Bob is accelerated with respect to Alice, he will naturally use a 
comoving frame and the straight motions seen by Alice will become curved motions in his reference frame. 
Therefore, if two reference frames are accelerated with respect to each other, the principle of 
inertia cannot hold in both frames. More generally, we call {\it inertial frame} a reference frame in 
which 
the principle of inertia holds \cite{Hladik}; the results of special relativity apply only to such frames. 
Accordingly, 
an observer using an inertial frame is called an inertial observer; in physical terms, it is an observer 
falling freely in empty space. We will see in 
subsection \ref{subsecPoin} what restrictions are imposed on the relation between coordinates of inertial 
frames; at present, we already know that, if two such frames move with respect to each other, then this 
motion must take place along a straight line, at constant velocity.\\

Given this definition, we are in position to state the two crucial defining principles of special relativity. 
The 
first, giving its name to the theory, is the principle of {\it relativity} (in the restricted sense 
\cite{Einstein}), which 
states that the laws of Nature must take the same form in all inertial frames. In other words, according to 
this principle, there exists no privileged inertial frame in the Universe: there is no experiment that would 
allow an experimenter to distinguish a given inertial frame from the others. This is a principle of {\it 
special} relativity in that it only applies to inertial frames; a principle of {\it general} relativity would 
apply to all possible reference frames, inertial or not. The latter principle leads to the theory of general 
relativity, which we will not discuss further here.\\

The second principle is Einstein's 
historical ``second postulate'', which states that the speed of light in the vacuum takes the same value 
$c$, written in (\ref{c}), in all inertial frames. In fact, if one assumes that Maxwell's theory of 
electromagnetism holds, then the second postulate is a consequence of the principle of relativity. Indeed, 
saying that the speed of light (in the vacuum) is the same in all inertial frames is really saying that the 
laws of electromagnetism are identical in all inertial frames \cite{Hladik}.

\subsection{The Poincar\'e group and the Lorentz group}
\label{subsecPoin}

We now work out the relation between coordinates of inertial frames; the set of all such relations will form 
a group, called the Poincar\'e group. We will see that the second postulate is crucial in determining the 
form of this group, through the notion of ``space-time interval''.

\subsubsection{Linear structure}

Suppose $A$ and $B$ are two inertial frames, {\it i.e.}~the principle of inertia holds in both of them. Then, 
a particle moving along a straight line at constant velocity, as seen from $A$, must also move at constant 
velocity along a straight line when seen from $B$. Thus, calling $(x^{\mu})$ the 
space-time coordinates of $A$ and $(x'^{\mu})$ those of $B$, the relation between these coordinates must be 
such that any straight line in the coordinates $x^{\mu}$ is mapped on a straight line in the coordinates 
$x'^{\mu}$. The most general transformation satisfying this property is a projective map \cite{HenneauxRG}, 
for which
\be
x'^{\mu}=\frac{a^{\mu}+{\Lambda^{\mu}}_{\nu}x^{\nu}}{b+c_{\mu}x^{\mu}}
\quad\forall\,\mu=0,1,2,3.
\label{proj}
\ee
(From now on, summation over repeated indices will always be understood.) Here $a^{\mu}$, 
${\Lambda^{\mu}}_{\nu}$, $b$ and $c_{\mu}$ 
are constant coefficients. If we insist that points having finite values of coordinates in $A$ remain 
with 
finite coordinates in $B$, we must set $c_{\mu}=0$. Then, absorbing the constant $b$ in the parameters 
$a^{\mu}$ and ${\Lambda^{\mu}}_{\nu}$, the transformation (\ref{proj}) reduces to
\be
x'^{\mu}={\Lambda^{\mu}}_{\nu}x^{\nu}+a^{\mu}.
\label{inhom}
\ee
Thus, the principle of inertia endows 
space-time with a linear structure. In order for the 
transformation 
(\ref{inhom}) to be invertible, we must also demand that the matrix $\Lambda=({\Lambda^{\mu}}_{\nu})$ be 
invertible. 
Apart from that, using only the principle of inertia, we cannot go further at this point. The principle of 
relativity will set additional restrictions on $\Lambda$.

\subsubsection{Invariance of the interval}

Let again $A$ be an inertial frame and let $\calP$ and $\calQ$ be two events in space-time. Call 
$\Delta x^{\mu}$ the components of the vector going from $\calP$ to $\calQ$ in the frame $A$. Then, we call 
the number
\be
\Delta s^2
\equiv
-(\Delta x^0)^2+(\Delta x^1)^2+(\Delta x^2)^2+(\Delta x^3)^2
\equiv
\eta_{\mu\nu}\Delta x^{\mu}\Delta x^{\nu}
\label{ds2}
\ee
the {\it square of the interval} between $\calP$ and $\calQ$. The matrix
\be
\eta=(\eta_{\mu\nu})=\bm -1 & 0 & 0 & 0 \\ 0 & 1 & 0 & 0 \\ 0 & 0 & 1 & 0 \\ 0 & 0 & 0 & 1 \emm
\label{Flush}
\ee
appearing in this definition is called the {\it Minkowski metric} matrix. The terminology 
associated with definition (\ref{ds2}) may seem inconsistent, in that we call ``square of the 
interval between two events'' a quantity that seems to depend 
not only on the events, but also on the coordinates chosen to locate them (in the present case, the 
separation $\Delta x^{\mu}$). This is not the case, however, thanks to the following important result:

\paragraph{Proposition.} Let $A$ and $B$ be two inertial frames, $\calP$ and $\calQ$ two events, the 
square 
of the interval between them being $\Delta s^2$ in the coordinates of $A$ and $\Delta s'^{\,2}$ in those of 
$B$. 
Then,
\be
\Delta s^2=\Delta s'^{\,2}\quad\text{(``invariance of the interval'').}
\label{invInt}
\ee
In other words, the number (\ref{ds2}) does not depend on the inertial coordinates used to define it.

\begin{proof}
First suppose that $\calP$ and $\calQ$ are light-like separated, {\it i.e.}~that there exists a light ray 
going 
from $\calP$ to $\calQ$ (or from $\calQ$ to $\calP$). Then, $\Delta s^2=0$ by construction. But, by 
Einstein's second postulate, the speed of light is the same in both reference frames $A$ and $B$, so $\Delta 
s'^{\,2}=0$ as well. Thus,
\be
\Delta s^2=0\quad\text{iff}\quad\Delta s'^{\,2}=0.
\nn
\ee
Now, since $A$ and $B$ are inertial frames, the relation between their coordinates must be of the linear form 
(\ref{inhom}); in particular, $\Delta x^{\mu}={\Lambda^{\mu}}_{\nu}\Delta x^{\nu}$. Therefore 
$\Delta s'^{\,2}$ 
is a polynomial of second order in the components $\Delta x^{\mu}$. But we have just seen that $\Delta s^2$ 
and $\Delta s'^{\,2}$ have identical roots; since polynomials having identical roots are necessarily 
proportional 
to each other, we know that there exists some number $K$ such that
\be
\Delta s'^{\,2}=K\Delta s^2.
\label{ds2K}
\ee
This number $K$ depends on the matrix $\Lambda$ appearing in $\Delta x^{\mu}={\Lambda^{\mu}}_{\nu}\Delta 
x^{\nu}$, which itself depends on the velocity $\vec v$ of the frame $B$ with respect to the frame $A$. (This 
velocity is constant, since accelerated frames cannot be inertial.) But space is isotropic by assumption, 
so $K$ actually depends only on the modulus $\|\vec v\|$ of $\vec v$, and not on its direction. In 
particular, $K(\vec v)=K(-\vec v)$. Since the velocity of $A$ with respect to $B$ is $-\vec v$, we know that
\be
\Delta s^2=K\Delta s'^{\,2}\stackrel{\text{(\ref{ds2K})}}{=}K^2\Delta s^2,
\nn
\ee
implyiing that $K^2=1$. Since real transformations cannot change the signature of a quadratic form, $K$ 
cannot be negative, so $K=1$.
\end{proof}

\subsubsection{Lorentz transformations}

We now know that coordinate transformations between inertial frames must preserve the square of the interval; 
let us work out the consequences of this statement for the matrix $\Lambda$ in (\ref{inhom}). To simplify 
notations, we will see $(x^{\mu})$ and $(a^{\mu})$ as four-component column vectors, so that (\ref{inhom}) 
can be written as
\be
x'=\Lambda\cdot x+a.
\label{inhomBis}
\ee
Similarly, seeing $(\Delta x^{\mu})$ as a vector $\Delta x$, the square of the interval (\ref{ds2}) 
becomes $\Delta s^2=\Delta x^t\cdot\eta\cdot\Delta x$. (The superscript ``$t$'' denotes transposition.) Then, 
since $\Delta x'=\Lambda\cdot\Delta x$, 
demanding invariance of the square of the interval under (\ref{inhomBis}) amounts to 
the equality
\be
\Delta s'^{\,2}
=
\Delta x^t\cdot\left(\Lambda^t\eta\Lambda\right)\cdot\Delta x
\stackrel{!}{=}
\Delta x^t\cdot\eta\cdot\Delta x=\Delta s^2,
\nn
\ee
to be satisfied for any $\Delta x$. Because $\eta$ is non-degenerate, this 
implies that the matrix $\Lambda$ satisfies
\be
\Lambda^t\eta\Lambda=\eta.
\label{LeL}
\ee

\paragraph{Definition.} The {\it Lorentz group} (in four dimensions) is
\be
\boxed{\mO(3,1)\equiv L\equiv\left\{\Lambda\in\mM(4,\RR)|\Lambda^t\eta\Lambda=\eta\right\},}
\label{Lorentz}
\ee
where $\mM(4,\RR)$ denotes the set of real $4\times 4$ matrices. More generally, the Lorentz 
group in $d$ space-time dimensions, $\mO(d-1,1)$, is the group of real $d\times d$ matrices $\Lambda$ 
satisfying property (\ref{LeL}) for the $d$-dimensional Minkowski metric matrix 
$\eta=\text{diag}(-1,\underbrace{1,1,...,1}_{d-1\text{ times}})$.

\paragraph{Remark.} This definition is equivalent to saying that the rows and columns of a Lorentz 
matrix form a 
Lorentz basis of $\RR^d$, that is, a basis $\{e_0,e_1,e_2,...,e_{d-1}\}$ of $d$-vectors $e_{\alpha}$ such 
that 
$e_{\alpha}^{\mu}\eta_{\mu\nu}e_{\beta}^{\nu}=\eta_{\alpha\beta}$. The Lorentz group in $d$ space-time 
dimensions is a Lie group of real dimension $d(d-1)/2$. This is analogous to the 
orthogonal group $\mO(d)$, defined as the set of $d\times d$ matrices $\calO$ satisfying 
$\calO^t\calO=\mathbb{I}$, where $\mathbb{I}$ is the identity matrix. In particular, the rows and columns of 
an orthogonal matrix form an orthonormal basis of $\RR^d$.

\paragraph{Definition.} The group consisting of inhomogeneous transformations (\ref{inhomBis}), where 
$\Lambda$ belongs to the 
Lorentz group, is called the {\it Poincar\'e group} or the inhomogeneous Lorentz group. Its abstract 
structure is that of a semi-direct product $\mO(3,1)\ltimes\RR^4$, where $\RR^4$ is the group of 
translations, the group operation being given by
\be
(\Lambda,a)\cdot(\Lambda',a')=\left(\Lambda\cdot\Lambda',a+\Lambda\cdot a'\right).
\nn
\ee
Of course, this definition is readily generalized to $d$-dimensional space-times upon replacing 
$\mO(3,1)$ by $\mO(d-1,1)$ and $\RR^4$ by $\RR^d$. We will revisit the definition of the Lorentz and 
Poincar\'e groups at the end of subsection \ref{confG}, with the tools of pseudo-Riemannian geometry. Apart 
from that, in the rest of these notes, we will mostly need only 
the homogeneous Lorentz group (\ref{Lorentz}) and we will not really use the Poincar\'e group. We stress, 
though, that the latter is crucial for particle physics and quantum field theory \cite{Wigner,Weinberg}.

\subsection{Subgroups of the Lorentz group}
\label{subsecSub}

The defining property (\ref{LeL}) implies that each matrix $\Lambda$ in the Lorentz group satisfies 
$\det(\Lambda)=\pm1$. This splits the Lorentz group in two disconnected subsets, corresponding to matrices 
with determinant 
$+1$ or $-1$. In particular, Lorentz matrices with determinant $+1$ 
span a subgroup of the Lorentz group $\mO(3,1)=L$, called the {\it proper} Lorentz group and denoted 
$\SO(3,1)$ or $L_+$. It is the set of Lorentz transformations that preserve 
the orientation of space.\\

Another natural subgroup of $L$ can be isolated using (\ref{LeL}), though in a somewhat less 
obvious way. Namely, consider the $0-0$ component of eq. (\ref{LeL}),
\be
{\Lambda^{\mu}}_0\eta_{\mu\nu}{\Lambda^{\nu}}_0
=
-\left({\Lambda^0}_0\right)^2+{\Lambda^i}_0{\Lambda^i}_0
\stackrel{!}{=}\eta_{00}=-1.
\nn
\ee
This implies the property
\be
\left({\Lambda^0}_0\right)^2=1+{\Lambda^i}_0{\Lambda^i}_0\geq 1,
\label{ineq}
\ee
valid for any matrix $\Lambda$ in the Lorentz group. The inequality is saturated only if 
${\Lambda^i}_0=0$ for all $i=1,2,3$. Since the inverse of relation (\ref{LeL}) implies 
$\Lambda\eta\Lambda^t=\eta$ for any Lorentz matrix $\Lambda$, we also find
\be
\left({\Lambda^0}_0\right)^2=1+{\Lambda^0}_i{\Lambda^0}_i\geq 1,
\label{ineqBis}
\ee
with equality iff ${\Lambda^0}_i=0$ for all $i=1,2,3$. Thus, in particular, $|{\Lambda^0}_0|=1$ iff 
${\Lambda^0}_i={\Lambda^i}_0=0$ for all $i$, in which case the spatial components ${\Lambda^i}_j$ of 
$\Lambda$ form a matrix in $\mO(3)$. Just as the determinant property $\det(\Lambda)=\pm1$, the 
inequality in (\ref{ineq}) splits the Lorentz group in two disconnected components, corresponding to matrices 
with positive or negative ${\Lambda^0}_0$. Note that the product 
of two 
matrices $\Lambda$, $\Lambda'$, with positive ${\Lambda^0}_0$ and ${\Lambda'^{\,0}}_0$, is itself a 
matrix with 
positive $0-0$ component:
\bea
{\left(\Lambda\cdot\Lambda'\right)^0}_0
& = &
{\Lambda^0}_{\mu}{\Lambda'^{\,\mu}}_0=
{\Lambda^0}_{0}{\Lambda'^{\,0}}_0+{\Lambda^0}_{i}{\Lambda'^{\,i}}_0\nn\\
& \geq &
{\Lambda^0}_{0}{\Lambda'^{\,0}}_0-\left|{\Lambda^0}_{i}{\Lambda'^{\,i}}_0\right|\nn\\
& \stackrel{\text{(\ref{ineq}),(\ref{ineqBis})}}{>} &
\sqrt{{\Lambda^0}_{i}{\Lambda^0}_{i}}\sqrt{{\Lambda'^{\,i}}_0{\Lambda'^{\,i}}_0}
-\left|{\Lambda^0}_{i}{\Lambda'^{\,i}}_0\right|\geq0.
\eea
(In the very last inequality we applied the Cauchy-Schwarz lemma to the spatial vectors whose components 
are ${\Lambda^0}_{i}$ and 
${\Lambda'^{\,i}}_0$.) Therefore, the set of Lorentz matrices $\Lambda$ with positive ${\Lambda^0}_0$ forms a 
subgroup of the Lorentz group, called the {\it orthochronous} Lorentz group and denoted 
$\mO(3,1)^{\uparrow}$ or $L^{\uparrow}$. As the name indicates, elements of $L^{\uparrow}$ are Lorentz 
transformations that preserve the direction of the arrow of time.\\

Given these subgroups, one defines the proper, orthochronous Lorentz group
\be
\SO(3,1)^{\uparrow}=L_+^{\uparrow}\equiv L_+\cap L^{\uparrow},
\nn
\ee
which is of course a subgroup of $L$. In fact, we will see at the end of this subsection that this is the 
maximal connected subgroup of the Lorentz group. The rows and columns of Lorentz matrices belonging 
to $L_+^{\uparrow}$ form Lorentz bases with a future-directed time-like unit vector, and with positive 
orientation. The group of orientation-preserving rotations of 
space, $\SO(3)$, is a natural subgroup of $L_+^{\uparrow}$, consisting of matrices of the form
\be
\bm
1 & 0 \\ 0 & R
\emm,
\quad\text{with }R\in\SO(3).
\label{rot}
\ee
Note that $L_+$ can be generated by adding to $L_+^{\uparrow}$ the time-reversal 
matrix
\be
T=\bm -1 & 0 & 0 & 0 \\ 0 & 1 & 0 & 0 \\ 0 & 0 & 1 & 0 \\ 0 & 0 & 0 & 1 \emm.
\label{T}
\ee
Similarly, $L^{\uparrow}$ can be obtained by adding to $L_+^{\uparrow}$ the parity matrix
\be
P=\bm 1 & 0 & 0 & 0 \\ 0 & -1 & 0 & 0 \\ 0 & 0 & -1 & 0 \\ 0 & 0 & 0 & -1 \emm.
\label{P}
\ee
More generally, the whole Lorentz group $L$ can be obtained by adding $T$ and $P$ to $L_+^{\uparrow}$. (Note 
that $T$ and $P$ do not commute with all matrices in $L_+^{\uparrow}$. This should be contrasted with the 
case of $\mO(3)$ and $\SO(3)$, where the three-dimensional parity operator belongs to the center of $\mO(3)$.)

\subsection{Boosts and rapidity}
\label{ssBoosts}

We have just seen that any rotation, acting only on the space coordinates, is a Lorentz transformation. In 
the language of inertial frames, this is obvious: if the spatial axes of a frame $B$ are rotated 
with respect to those of an inertial frame $A$ (and provided the time coordinates in $A$ and $B$ coincide), 
then $B$ is certainly an inertial frame. The same would be true even in Galilean relativity \cite{Hladik}. In 
order to see effects specific to Einsteinian special relativity, we need to consider 
Lorentz transformations involving inertial frames in relative motion.

\begin{figure}[H]
\begin{center}
\includegraphics[width=0.50\textwidth]{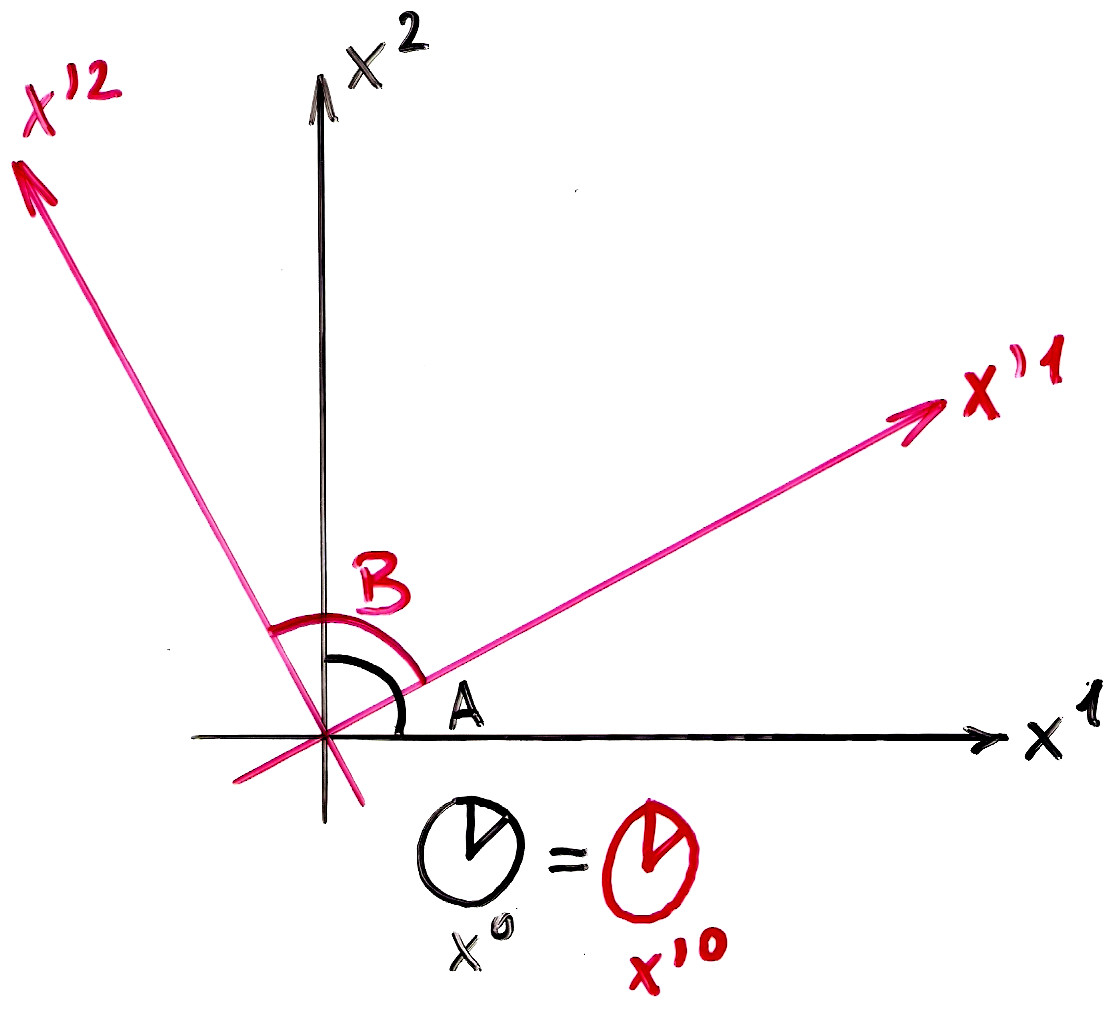}
\caption{Two inertial frames $A$ and $B$ related by a rotation of their spatial axes (the third space 
direction is omitted). The clocks of $A$ and $B$ are synchronized.}
\end{center}
\end{figure}

\subsubsection{Boosts}

Call $A$ and $B$ the inertial frames used by Alice and Bob, with respective coordinates $(x^{\mu})$ 
and $(x'^{\,\mu})$. Suppose Bob moves with respect to Alice 
in a 
straight line, at constant velocity $v$. Without loss of generality, we may assume that the origins of the 
frames $A$ and $B$ coincide. (If they don't, just apply a suitable space-time translation to bring them 
together.) By rotating the spatial 
axes of $A$ and $B$, we can also choose their coordinates to satisfy $x^2=x'^2$ and $x^3=x'^3$. Then, the 
only coordinates of $A$ and $B$ that are related by a 
non-trivial transformation are $(x^0,x^1)$ and $(x'^0,x'^1)$. Finally, using parity and time-reversal if 
necessary, 
we may choose the same orientation for the spatial frames of $A$ and $B$, and the same orientation for their 
time arrows.

\begin{figure}[H]
\begin{center}
\includegraphics[width=0.40\textwidth]{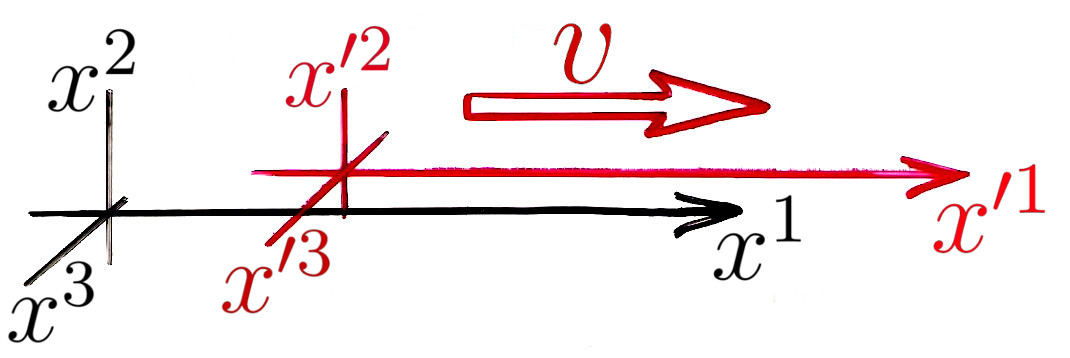}
\caption{The frame $B$ (in red) is boosted with respect to $A$ (in black) with velocity $v$ along the $x^1$ 
axis. The coordinates $x^2$ and $x^3$ coincide with $x'^2$ and $x'^3$. In principle, the clocks of $A$ and 
$B$ (not represented in this figure) need not tick at the same rate.}
\end{center}
\end{figure}

Under these assumptions the relation between 
the coordinates of $A$ and those of $B$ takes the form
\be
x'^{\mu}={\Lambda^{\mu}}_{\nu}x^{\nu}
\quad\text{with}\quad
\left({\Lambda^{\mu}}_{\nu}\right)
=
\bm
p & q & 0 & 0 \\ r & s & 0 & 0 \\ 0 & 0 & 1 & 0 \\ 0 & 0 & 0 & 1
\emm,
\nn
\ee
where $p$, $q$, $r$ and $s$ are some real coefficients such that $\Lambda$ belongs to $L_+^{\uparrow}$ $-$ in 
particular, $p>0$. Since $B$ moves with respect to $A$ at constant velocity $v$ (along 
the $x^1$ direction), the coordinate $x'^1$ of $B$ must vanish when $x^1=vt=vx^0/c$. By virtue of linearity, 
we may write
\be
x'^1=\gamma(v)\cdot\left(x^1-\frac{v}{c}x^0\right),
\nn
\ee
where $\gamma(v)$ is some $v$-dependent, positive coefficient (on account of the fact that the directions 
$x^1$ 
and $x'^1$ coincide). Demanding that $\Lambda$ satisfies relation (\ref{LeL}), with the restrictions $p>0$ 
and $\gamma(v)>0$, then yields
\be
\Lambda
=
\bm
\gamma(v) & -\gamma(v)\cdot v/c & 0 & 0 \\ -\gamma(v)\cdot v/c & \gamma(v) & 0 & 0 \\ 0 & 0 & 1 & 0 \\ 0 & 0 
& 0 & 1
\emm,
\quad\text{and}\quad
\gamma(v)=\frac{1}{\sqrt{1-v^2/c^2}}.
\label{boost}
\ee
A Lorentz transformations of this form is called a {\it boost} (with velocity $v$ in the direction $x^1$). In 
particular, reality of $\Lambda$ requires $|v|$ to be smaller than $c$: boosts faster than light 
are forbidden.\\

Boosts give rise to the 
counterintuitive phenomena of {\it time dilation} and {\it length contraction}. Let us briefly describe the 
former. Suppose Bob, moving at velocity $v$ with respect to Alice, carries a clock and measures a time 
interval $\Delta t'$ in his reference frame; for definiteness, suppose he measures the time elapsed between 
two consecutive ``ticks'' of his clock, and let the clock be located at the origin of his reference frame. 
Call $\calP$ the event ``Bob's clock ticks for the first time at his location at that moment'', and 
call $\calQ$ the event ``Bob's clock ticks for the second time (at his location at that time)''. Then, 
in Alice's coordinates, the time interval $\Delta t$ separating the events $\calP$ and $\calQ$ is not equal 
to 
$\Delta t'$; rather, according to (\ref{boost}), one has $\Delta t=\gamma(v)\Delta t'$. Since $\gamma(v)$ is 
always larger than one, this means that Alice measures a longer duration than Bob: Bob's time is 
``dilated'' compared to Alice's time, and $\gamma(v)$ is precisely the dilation factor. This phenomenon is 
responsible, for instance, 
for the fact that cosmic muons falling into Earth's atmosphere can be detected at the level of the oceans 
even though their time of flight (as measured by an observer standing still on Earth's surface) is about a 
hundred times longer than their proper lifetime. In subsection \ref{boostSphere}, we will see that boosts 
also lead to surprising optical effects on the celestial sphere.

\subsubsection{Notion of rapidity}

Although the notion of velocity used above is the most intuitive one, it is not the most practical one from a 
mathematical viewpoint. In particular, composing two boosts with velocities $v$ and $w$ (in the same 
direction) does {\it not} yield a boost with velocity $v+w$. It would be convenient to find an alternative 
parameter to specify boosts, one that would be additive when two boosts are combined. This leads to the 
notion of {\it rapidity} \cite{LevyRap,HenneauxGroupe},
\be
\chi(v)\equiv\text{argtanh}(v/c),
\label{rap}
\ee
in terms of which the boost matrix (\ref{boost}) becomes
\be
\Lambda
=
\bm
\cosh\chi & -\sinh\chi & 0 & 0 \\ -\sinh\chi & \cosh\chi & 0 & 0 \\ 0 & 0 & 1 & 0 \\ 0 & 0 & 0 & 1
\emm
\equiv L(\chi).
\label{boostBis}
\ee
One verifies that the composition of two such boosts with rapidities $\chi_1$ and $\chi_2$ is a 
boost of the same form, with rapidity $\chi_1+\chi_2$.\\

Rapidity is thus the additive parameter specifying Lorentz boosts. It exhibits the fact that boosts 
along a given axis form a non-compact, one-parameter 
subgroup of $L_+^{\uparrow}$. It also readily provides a formula for the addition of velocities: the 
composition of two boosts with velocities $v$ and $w$ is a boost with rapidity $\chi(v)+\chi(w)$; 
equivalently, according to (\ref{rap}), the velocity $V$ of the resulting boost is
\be
V=c\cdot\tanh(\chi(v)+\chi(w))
=
c\cdot\tanh\left(\text{argtanh}(v/c)+\text{argtanh}(w/c)\right)
=
\frac{v+w}{1+vw/c^2}.
\nn
\ee
This is the usual formula for the addition of velocities (in the same direction) in special relativity 
\cite{Hladik}.\\

As practical as rapidity is, its physical meaning is a bit obscure: the definition 
(\ref{rap}) does not seem related to any measurable quantity whatsoever. But in fact, there exist at least 
three different natural definitions of the notion of ``speed'', and rapidity is one of them 
\cite{LevySpeed}. To illustrate these definitions, consider an observer, Bob, who travels by train from 
Brussels to Paris \cite{FerrariRG}, and 
measures his speed during the journey. For simplicity, we will assume that the motion takes place along the 
$x$ axis of Alice, an inertial observer standing still on the ground.\\

A first notion of speed he might want 
to define is an ``extrinsic'' one: he lets Alice measure the distance between Brussels and Paris, and the two 
clocks of the Brussels and Paris train stations are synchronized. Then, looking at the clocks upon 
departure and upon arrival, he defines his {\it velocity} as the ratio of the distance measured by Alice to 
the duration of his trip, measured by the clocks in Brussels and Paris. The infinitesimal version of velocity 
is the usual expression $v=dx/dt$, where $x$ and $t$ are the space and time coordinates of an inertial 
frame which, in general, is not 
related to Bob. (In the present case, these are the coordinates that Alice, or any inertial observer 
standing still on the ground, would likely use.)\\

A second natural definition of speed is given by {\it proper velocity}. To define this notion, Bob still 
lets Alice 
measure the distance between Brussels and Paris, but now he divides this distance by the duration that he 
himself has measured using his wristwatch. The infinitesimal version of (the $x$ component of) proper 
velocity is $u=dx/d\tau$, 
where $\tau$ denotes Bob's proper time, defined by
\be
d\tau^2
=
dt^2-\frac{1}{c^2}dx^2
\label{starr}
\ee
along Bob's trajectory. (In (\ref{starr}), it is understood that Bob's trajectory is written as $(t,x(t))$ 
in the coordinates $(t,x)$ of Alice, but the value of $d\tau$ would be the same in any inertial frame with 
the same direction for the arrow of time, by virtue of Lorentz-invariance of the interval, 
eq.~(\ref{invInt}).) If Bob's motion occurs at constant speed, the relation 
between the $x$ component of
proper velocity and standard velocity is $u=\gamma(v)\cdot v$, as follows from time dilation.\\

Finally, Bob may decide not to believe Alice's measurement of distance, and that he wants to 
measure everything by himself. Of course, sitting in the train, he cannot measure the 
distance between Brussels and Paris using a measuring tape. He therefore carries an accelerometer and 
measures his proper acceleration at each moment during the journey. Starting from rest in Brussels (at 
proper time $\tau=0$ say), he can then integrate this acceleration from $\tau=0$ to $\tau=s$ to obtain a 
measure of his speed at proper time $s$. This notion of speed is precisely the {\it rapidity} introduced 
above \cite{LevySpeed}, up to a conversion factor given by the speed of light. Indeed, assuming that Bob 
accelerates in the direction of positive $x$, his proper 
acceleration at proper time $\tau$ is the Lorentz-invariant quantity \cite{HenneauxRG,MTW}
\be
a=\sqrt{\left(\frac{d^2x}{d\tau^2}\right)^2-c^2\left(\frac{d^2t}{d\tau^2}\right)^2}.
\nn
\ee
It is the value of acceleration that would be measured by a ``locally inertial observer'', that is, an 
observer whose velocity coincides with Bob's velocity at proper time $\tau$, but who is falling freely 
instead of following an accelerated trajectory. The integral of this quantity along proper time, 
provided Bob accelerates in the direction of positive $x$, is thus
\bea
\label{Nabla}
I(s)
& = &
\int_0^sd\tau\sqrt{\left(\frac{d^2x}{d\tau^2}\right)^2-c^2\left(\frac{d^2t}{d\tau^2}\right)^2}
=
\int_0^sd\tau\frac{d^2x}{d\tau^2}\left[1-c^2\left(\frac{d^2t/d\tau^2}{d^2x/d\tau^2}\right)\right]^{1/2}\\
\label{iss}
& = &
\int_0^sd\tau\frac{d^2x}{d\tau^2}\left[1-\frac{1}{c^2}\left(\frac{dx/d\tau}{dt/d\tau}\right)\right]^{1/2},
\eea
where we used the definition (\ref{starr}) of proper time, which implies
\be
\left(\frac{dx}{d\tau}\right)^2-c^2\left(\frac{dt}{d\tau}\right)^2=c^2
\quad
\Rightarrow
\quad
\frac{dx}{d\tau}\frac{d^2x}{d\tau^2}
-c^2\frac{dt}{d\tau}\frac{d^2t}{d\tau^2}=0.
\nn
\ee
But $\frac{dx/d\tau}{dt/d\tau}=\frac{dx}{dt}=v$; since Bob's proper velocity along $x$ vanishes at 
proper time $\tau=0$, the integral (\ref{iss}) can be written as
\bea
I(s)
& = &
\int_0^{u(s)}du\sqrt{1-v^2/c^2}
=
\int_0^{u(s)}d\left(\gamma(v)\cdot v\right)\sqrt{1-v^2/c^2}\nn\\
\label{BigStar}
& = &
\int_0^{v(s)}\frac{dv}{1-v^2/c^2}
=
c\cdot\text{argtanh}(v(s)/c)
\stackrel{\text{(\ref{rap})}}{=}c\cdot\chi(v(s)),
\eea
where $u(s)$ and $v(s)$ are Bob's proper velocity and velocity at proper time $s$. This is precisely the 
relation we wanted to prove: up to the factor $c$, the integral (\ref{Nabla}) of proper acceleration 
coincides with rapidity. In subsection 
\ref{boostSphere}, we will also see that rapidity is the simplest parameter describing the effect of Lorentz 
boosts on the celestial sphere.

\subsection{Standard decomposition theorem}
\label{ssStd}

We now derive the following important result, which essentially states that any proper, orthochronous Lorentz 
transformation can be written as a combination of rotations together with a standard 
boost (\ref{boost}). We closely follow \cite{HenneauxGroupe}.

\paragraph{Theorem.} Any $\Lambda\in L_+^{\uparrow}$ can be written as a product
\be
\Lambda=R_1L(\chi)R_2,
\label{dec}
\ee
where $R_1$ and $R_2$ are rotations of the form (\ref{rot}) and $L(\chi)$ is a Lorentz boost of the 
form (\ref{boostBis}). The decomposition (\ref{dec}) is called {\it standard decomposition} of a proper, 
orthochronous Lorentz transformation. There are many such decompositions for a given $\Lambda$.

\begin{proof}
Let $\Lambda\in L_+^{\uparrow}$. Let $\vec a$ denote the vector in $\RR^3$ whose components are the 
coefficients $({\Lambda^k}_0)$. If $\vec 
a=0$, then ${\Lambda^0}_k=0$ and $\Lambda$ is of the form (\ref{rot}). In that case the decomposition 
(\ref{dec}) is trivially satisfied with $R_1=L(\chi)=I$ and $R_2=\Lambda$. If $\vec a\neq0$, let ${\vec e}_1$ 
denote one of the two unit vectors proportional to $\vec a$ (${\vec 
e}_1=\lambda\vec a$, $({\vec e}_1)^2=1$); write its components as $(\alpha_1,\alpha_2,\alpha_3)$. Let also 
${\vec e}_2$ and ${\vec e}_3$ be two vectors in $\RR^3$ such that the set $\{{\vec e}_1,{\vec e}_2,{\vec 
e}_3\}$ be an orthonormal basis of $\RR^3$ with positive orientation ({\it i.e.}~the $3\times 3$ matrix whose 
entries are the 
components of ${\vec e}_1$, ${\vec e}_2$ and ${\vec e}_3$ belongs to $\SO(3)$); denote their respective 
components as $(\beta_1,\beta_2,\beta_3)$ and $(\gamma_1,\gamma_2,\gamma_3)$. Then consider the rotation 
matrix
\be
{\bar R}_1
=
\bm
1 & 0 & 0 & 0 \\ 0 & \alpha_1 & \alpha_2 & \alpha_3 \\
0 & \beta_1 & \beta_2 & \beta_3 \\
0 & \gamma_1 & \gamma_2 & \gamma_3
\emm.
\nn
\ee
The product ${\bar R}_1\Lambda$ takes the form
\be
{\bar R}_1\Lambda
=
\bm
1 & 0 & 0 & 0 \\ 0 & \alpha_1 & \alpha_2 & \alpha_3 \\
0 & \beta_1 & \beta_2 & \beta_3 \\
0 & \gamma_1 & \gamma_2 & \gamma_3
\emm
\bm
{\Lambda^0}_0 & {\Lambda^0}_1 & {\Lambda^0}_2 & {\Lambda^0}_3 \\
{\Lambda^1}_0 & {\Lambda^1}_1 & {\Lambda^1}_2 & {\Lambda^1}_3 \\
{\Lambda^2}_0 & {\Lambda^2}_1 & {\Lambda^2}_2 & {\Lambda^2}_3 \\
{\Lambda^3}_0 & {\Lambda^3}_1 & {\Lambda^3}_2 & {\Lambda^3}_3
\emm
=
\bm
{\Lambda^0}_0 & {\Lambda^0}_1 & {\Lambda^0}_2 & {\Lambda^0}_3 \\
\times & \times & \times & \times \\
0 & \mu_1 & \mu_2 & \mu_3 \\
0 & \nu_1 & \nu_2 & \nu_3
\emm,
\label{productPlacement}
\ee
where the $\times$'s are unimportant numbers and where $(\mu_1,\mu_2,\mu_3)$ and $(\nu_1,\nu_2,\nu_3)$ are 
the components of two mutually orthogonal unit vectors in $\RR^3$ (because ${\bar R}_1\Lambda\in 
L_+^{\uparrow}$, so the rows and columns of this matrix form a Lorentz basis); let us denote these unit 
vectors by ${\vec f}_2$ and ${\vec f}_3$, respectively. Let also ${\vec f}_1$ be the (unique) vector such 
that $\{{\vec f}_1,{\vec f}_2,{\vec f}_3\}$ be an orthonormal basis of $\RR^3$ with positive orientation. 
Define the rotation ${\bar R}_2$ by
\be
{\bar R}_2
=
\bm
1 & 0 & 0 & 0 \\
0 & \lambda_1 & \mu_1 & \nu_1 \\
0 & \lambda_2 & \mu_2 & \nu_2 \\
0 & \lambda_3 & \mu_3 & \nu_3
\emm,
\nn
\ee
where $(\lambda_1,\lambda_2,\lambda_3)$ are the components of $\vec{f}_1$. Then, the product ${\bar 
R}_1\Lambda{\bar R}_2$ reads
\be
{\bar R}_1\Lambda{\bar R}_2
\stackrel{\text{(\ref{productPlacement})}}{=}
\bm
{\Lambda^0}_0 & {\Lambda^0}_1 & {\Lambda^0}_2 & {\Lambda^0}_3 \\
\times & \times & \times & \times \\
0 & \mu_1 & \mu_2 & \mu_3 \\
0 & \nu_1 & \nu_2 & \nu_3
\emm
\bm
1 & 0 & 0 & 0 \\
0 & \lambda_1 & \mu_1 & \nu_1 \\
0 & \lambda_2 & \mu_2 & \nu_2 \\
0 & \lambda_3 & \mu_3 & \nu_3
\emm
\nn
=
\bm
{\Lambda^0}_0 & \times & k & l \\
\times & \times & m & n \\
0 & 0 & 1 & 0 \\
0 & 0 & 0 & 1
\emm.
\label{matrixBB}
\ee
Since ${\bar R}_1\Lambda{\bar R}_2$ belongs to $L_+^{\uparrow}$, the two last rows of (\ref{matrixBB}) must 
be 
orthogonal to the two first ones, so the $2\times 2$ matrix
\be
\bm
k & l \\
m & n
\emm
\nn
\ee
must vanish. Thus ${\bar R}_1\Lambda{\bar R}_2$ is block-diagonal,
\be
{\bar R}_1\Lambda{\bar R}_2
=
\bm
B & 0 \\ 0 & 1
\emm,
\nn
\ee
with $B\in\SO(1,1)^{\uparrow}$. This implies that
\be
B=\bm \cosh\chi & -\sinh\chi \\ -\sinh\chi & \cosh\chi \emm
\nn
\ee
for some $\chi$, so that ${\bar R}_1\Lambda{\bar R}_2$ is a pure boost of the form (\ref{boostBis}). In other 
words, 
writing $R_1\equiv({\bar R}_1)^{-1}$ and $R_2\equiv ({\bar R}_2)^{-1}$, the decomposition (\ref{dec}) 
is satisfied.
\end{proof}

\paragraph{Remark.} The standard decomposition theorem holds in any space-time dimension $d\geq3$. The proof 
is a straightforward adaptation of the argument used in the four-dimensional case. (In $d=2$ space-time 
dimensions, there are no spatial rotations and the proper, orthochronous Lorentz group 
$\mathrm{SO}(1,1)^{\uparrow}$ only contains boosts in the spatial direction. In this sense, the standard 
decomposition theorem is trivially staisfied also in $d=2$.)

\subsection{Connected components of the Lorentz group}
\label{ssCon}

In a general topological group $G$ (and in particular in any Lie group), we call {\it connected component} of 
$g\in 
G$ the set of all elements in $G$ that can be reached by a continuous path starting at $g$. In particular, we 
denote by $G_e$ the connected component of the identity $e\in G$. A group is 
connected if it has only one connected component $-$ that of the identity $-$, in which case $G=G_e$.

\paragraph{Proposition.} $G_e$ is a normal subgroup of $G$. Furthermore, the set of connected components of 
$G$ coincides with the quotient group $G/G_e$.

\begin{proof}
We first prove that $G_e$ is a subgroup of $G$. Let $h_1$ and $h_2$ belong to $G_e$, and let
\be
\gamma_1:[0,1]\rightarrow G:t\mapsto\gamma_1(t)
\quad\text{and}\quad
\gamma_2:[0,1]\rightarrow G:t\mapsto\gamma_2(t)
\nn
\ee
be two continuous paths such that $\gamma_1(0)=\gamma_2(0)=e$ and $\gamma_1(1)=h_1$, $\gamma_2(1)=h_2$. Then 
the path
\be
\gamma_1\cdot\gamma_2^{-1}:[0,1]\rightarrow G:t\mapsto\gamma_1(t)\gamma_2(t)^{-1}
\nn
\ee
joins $e$ to $h_1h_2^{-1}$. Therefore $h_1h_2^{-1}$ belongs to $G_e$, and the latter is a subgroup of $G$.\\

Let us now show that $G_e$ is a normal subgroup. Let $h\in G_e$ and let $g\in G$. Consider the element 
$ghg^{-1}$ in $G$. Since $h$ belongs to the connected component of the identity, there exists a continuous 
path $\gamma:[0,1]\rightarrow G:t\mapsto\gamma(t)$ such that $\gamma(0)=e$ and $\gamma(1)=h$. But then the map
\be
g\gamma g^{-1}:[0,1]\rightarrow G:t\mapsto g\gamma(t)g^{-1}
\nn
\ee
is also a continuous path in $G$, joining $\gamma(0)=e$ to $\gamma(1)=ghg^{-1}$. Therefore $ghg^{-1}\in G_e$. 
Since this is true for any $h\in G_e$ and any $g\in G$, $G_e$ is a normal subgroup of $G$.\\

We now turn to the second part of the proposition. Suppose first that $g_1$ and $g_2$ belong to the same 
connected components in $G$; let 
$\gamma:[0,1]\rightarrow G:t\mapsto\gamma(t)$ be a continuous path such that $\gamma(0)=g_1$ and 
$\gamma(1)=g_2$. Then, $g_1^{-1}\gamma(t)$ is a continuous path joining the identity to $g_1^{-1}g_2$, so 
$g_1^{-1}g_2$ belongs to $G_e$. Therefore, the cosets $g_1G_e$ and $g_2G_e$, seen as elements of the quotient 
group $G/G_e$, coincide. (Since $G_e$ is a normal subgroup of $G$, $G/G_e$ is indeed a group.)\\

Conversely, suppose the cosets $g_1G_e$ and $g_2G_e$ coincide as elements of $G/G_e$. Then there exists an 
$h$ in $G_e$ such that $g_2=g_1h$, and a path $\gamma:[0,1]\rightarrow G_e$ such that $\gamma(0)=e$ and 
$\gamma(1)=h$. But then the path $g_1\gamma(t)$ joins $g_1$ to $g_2$, so $g_1$ and $g_2$ belong to the same 
connected component of $G$.
\end{proof}

Let us now apply this proposition to the Lorentz group. First observe that $L_+^{\uparrow}$ 
is connected, as follows 
from the standard decomposition theorem: in (\ref{dec}), each factor can be linked to the 
identity by a continuous path, and this remains true for the product of these factors\footnote{This 
argument relies in particular on the fact that the group $\SO(3)$ of orientation-preserving rotations is 
connected.}. Thus, $L_+^{\uparrow}$ 
is the connected subgroup of the Lorentz group, and the connected components of the latter coincide with the 
quotient $L/L_+^{\uparrow}$. As noted below expression (\ref{P}), each element of the Lorentz group $L$ can 
be reached by adding parity and/or time-reversal to the connected Lorentz group $L_+^{\uparrow}$. Thus, the 
quotient $L/L_+^{\uparrow}$ is the group $\ZZ_2\times\ZZ_2$ generated by $P$ and $T$ and the Lorentz group 
has exactly four connected components, denoted as follows:
\begin{center}
\begin{tabular}{rccc}
$L_+^{\uparrow}$: & $\det(\Lambda)=1$ & and & ${\Lambda^0}_0\geq1$,\\
$L_+^{\downarrow}$: & $\det(\Lambda)=1$ & and & ${\Lambda^0}_0\leq1$,\\
$L_-^{\uparrow}$: & $\det(\Lambda)=-1$ & and & ${\Lambda^0}_0\geq1$,\\
$L_-^{\downarrow}$: & $\det(\Lambda)=-1$ & and & ${\Lambda^0}_0\leq1$.
\end{tabular}
\end{center}
As already mentioned, $L_+=L_+^{\uparrow}\cup L_+^{\downarrow}$ and 
$L^{\uparrow}=L_+^{\uparrow}\cup L_-^{\uparrow}$ are subgroups of the Lorentz group. Note that 
$L_+^{\uparrow}\cup L_-^{\downarrow}$ is also a group.

\paragraph{Remark.} Analogous results hold for the Lorentz group $\mO(d-1,1)$ in any space-time dimension 
$d$. The properties 
$|\det(\Lambda)|=1$ and $|{\Lambda^0}_0|\geq1$ remain true and the definition of the proper Lorentz group 
$\SO(d-1,1)$ and the orthochronous Lorentz group $\mO(d-1,1)^{\uparrow}$ are straighforward 
generalizations of $L_+$ and $L^{\uparrow}$. Similarly, one defines 
$\SO(d-1,1)^{\uparrow}\equiv\SO(d-1,1)\cap\mO(d-1,1)^{\uparrow}$. The standard decomposition theorem 
(\ref{dec}) remains true, provided $\SO(3)$ is replaced by $\SO(d-1)$. In particular, $\SO(d-1,1)^{\uparrow}$ 
is the connected subgroup of the Lorentz group and the latter splits in four connected components. The 
transition between different components is realised by the generalization of the 
time-reversal and parity matrices (\ref{T}) and (\ref{P}). (In odd space-time dimensions, parity is not just 
$\text{diag}(1,-1,...,-1)$, since that matrix belongs to the proper Lorentz group. Rather, in odd dimensions, 
parity is $P=\text{diag}(1,1,-1,...,-1)$.)

\section{Lorentz groups and special linear groups}
\label{secIsom}

Having defined the Lorentz group, we now turn to its 
realization as the group $\SL(2,\CC)$ of volume-preserving linear transformations of $\CC^2$. Since the 
method used to derive this isomorphism has a wide range of applications, we will use it repeatedly 
in this section, proving three different isomorphisms along the way: first, in subsection \ref{SO3SU2}, we 
relate $\SO(3)$ to $\SU(2)$. Then, in subsection 
\ref{LSL3}, we turn to the isomorphism between $\SL(2,\RR)$ and the Lorentz group in three space-time 
dimensions. Finally, in subsection \ref{LSL}, we establish the announced link between 
$L_+^{\uparrow}$ and $\SL(2,\CC)$\footnote{These results have important implications for representation 
theory; we 
shall not 
discuss those 
implications here and refer to 
\cite{HenneauxGroupe,Weinberg,Corn} for more details.}. We end in subsection \ref{LorentzDiv} by mentioning 
(without proving them) higher-dimensional generalizations of these results and their relation to division 
algebras.\\

Before dealing with specific constructions, let us review a general group-theoretic result. 
Let $G$ and $H$ be groups, $f:G\rightarrow H$ a 
homomorphism. Then, the kernel $\Ker(f)$ of $f$ is a normal subgroup of $G$ and the quotient of $G$ by 
$\Ker(f)$ is isomorphic (as a group) to the image of $f$:
\be
G/\Ker(f)\cong\Ima(f).
\label{GKerf}
\ee
The proof is elementary, as it suffices to observe that the map
\be
G/\Ker(f)\rightarrow\Ima(f):g\Ker(f)\mapsto f(g)
\nn
\ee
is a bijective homomorphism, that is, the sought-for isomorphism. All isomorphisms exposed in this section 
will be obtained using that method: we will construct well-chosen homomorphisms that will lead us to the 
desired isomorphisms through relation (\ref{GKerf}).

\subsection{A compact analogue}
\label{SO3SU2}

Here we establish the isomorphism between $\SO(3)$ and the quotient of $\SU(2)$ by its center, following 
\cite{HenneauxGroupe}. This relation is important for our purposes both because of the simplicity of 
the example, and because of its role in the isomorphism between the Lorentz group in 
four dimensions and $\SL(2,\CC)$. We begin by 
reviewing briefly the main properties of the unitary group in two dimensions.

\subsubsection{Properties of $\SU(2)$}

The unitary group $\mU(2)$ in two dimensions is the group of linear transformations of $\CC^2$ that preserve 
the norm $\|(z,w)\|^2=|z|²+|w|^2$. It consists of $2\times 2$ complex matrices $U$ that are unitary in the 
sense that
\be
U^{\dagger}U=\mathbb{I}_2=2\times2\text{ unit matrix},
\label{unitary}
\ee
where $\dagger$ denotes hermitian conjugation ($U^{\dagger}=(U^{t})^*$). It follows that the lines and 
columns of each matrix $U\in\mU(2)$ define an orthonormal basis of $\CC^2$ for the scalar product 
$(z,w)\cdot(z',w')=z^*z'+w^*w'$. By virtue of the defining property (\ref{unitary}), each 
$U\in\mU(2)$ 
has $|\det(U)|=1$. In particular, we define the special unitary group $\SU(2)$ in two dimensions as the 
subgroup of $\mU(2)$ consisting of matrices $U$ with unit determinant, $\det(U)=1$.\\

For later purposes, we will need to know some topological properties of $\SU(2)$. Demanding that the matrix
\be
U=\bm \alpha & \beta \\ \gamma & \delta \emm
\nn
\ee
belong to $\SU(2)$ imposes the conditions 
$|\alpha|^2+|\beta|^2=|\gamma|^2+|\delta|^2=1$, 
$\alpha\gamma^*+\beta\delta^*=\alpha\beta^*+\gamma\delta^*=0$ and $\alpha\delta-\beta\gamma=1$. These 
requirements are solved by $\delta=\alpha^*$ and 
$\gamma=-\beta^*$, so each matrix in $\SU(2)$ can be written as
\be
U=\bm \alpha & \beta \\ -\beta^* & \alpha^* \emm,
\quad\text{with}\;\;|\alpha|^2+|\beta|^2=1.
\nn
\ee
Thus, each element of $\SU(2)$ is uniquely determined by four real numbers 
$\alpha_1, \alpha_2, \beta_1, \beta_2$ such that $\alpha_1^2+\alpha_2^2+\beta_1^2+\beta_2^2=1$. 
These numbers define a point on the unit $3$-sphere\footnote{Recall that the $n$-sphere $S^n$ is defined as 
the set of points in $\RR^{n+1}$ that are located at unit distance from the origin.}. Furthermore, this 
description is not redundant (two 
different quadruples lead to two different elements of $\SU(2)$), so $\SU(2)$ is homeomorphic\footnote{By 
definition, two topological spaces are {\it homeomorphic} if there exists a continuous bijection, mapping the 
first space on the second one, whose inverse is also continuous.} to $S^3$, 
as a 
topological space. In particular, $\SU(2)$ is connected and simply connected.\\

Finally, recall that the center of a group $G$ is the set of elements that commute with all elements of $G$. 
In particular, it is an Abelian normal subgroup of $G$. It is easy to show that the center of $\SU(2)$ 
consists of the two matrices
\be
\mathbb{I}_2=\bm 1 & 0 \\ 0 & 1 \emm,\quad -\mathbb{I}_2=\bm -1 & 0 \\ 0 & -1 \emm
\label{ZSU2}
\ee
and is thus isomorphic to $\ZZ_2$. This observation will be important in the next paragraph, and we will 
use it again once we turn to the Lorentz group in four dimensions.

\subsubsection{The isomorphism}

\paragraph{Theorem.} One has the following isomorphism:
\be
\SO(3)\cong\SU(2)/\ZZ_2,
\label{isoSU2}
\ee
where $\ZZ_2$ is the center of $\SU(2)$. In other words, $\SU(2)$ is the double cover of $\SO(3)$, and it is 
also its universal cover.

\begin{proof}
Consider the space $\VV$ of $2\times 2$ traceless Hermitian matrices. Each matrix $X\in\VV$ can be written 
as $X=x^i\sigma_i$ (with implicit summation over $i=1,2,3$), where the $x^i$'s are real coefficients, while 
the $\sigma_i$'s are Pauli matrices
\be
\sigma_1=\bm 0 & 1 \\ 1 & 0 \emm,\quad
\sigma_2=\bm 0 & -i \\ i & 0 \emm,\quad
\sigma_3=\bm 1 & 0 \\ 0 & -1\emm.
\label{Pauli}
\ee
The space $\VV$ is obviously a three-dimensional real vector space. Note that $\det(X)=-x^ix^i=-\|x\|^2$, 
where 
$\|.\|$ denotes the Euclidean norm in $\RR^3$. In addition, as a vector space, $\VV$ is isomorphic to the Lie 
algebra $\su(2)$ of $\SU(2)$. The group $\SU(2)$ naturally acts on its Lie algebra by the adjoint action, for 
which $U\in\SU(2)$ maps $X\in\VV$ on $UXU^{\dagger}\in\VV$. This action is a 
representation of $\SU(2)$, that is, a homomorphism from $\SU(2)$ into the linear group of $\RR^3$. 
Furthermore, it preserves 
the norm in $\VV\cong\RR^3$ in the sense that
\be
\det(UXU^{\dagger})=\det(X)=-\|x\|^2\quad\forall\, U\in\SU(2),\;\forall\, X\in H.
\nn
\ee
Thus, the adjoint action of $\SU(2)$ on $\RR^3$ consists of orthogonal transformations and we can define 
a homomorphism
\be
f:\SU(2)\rightarrow\mO(3):U\mapsto f[U],
\label{homSU}
\ee
where the $3\times 3$ matrix $f[U]$ is given by the condition
\be
Ux^i\sigma_iU^{\dagger}
=
{f[U]^i}_jx^j\sigma_i
\quad\forall\, x\in\RR^3,
\quad\text{that is,}
\quad U\sigma_jU^{\dagger}={f[U]^i}_j\sigma_i
\quad\forall j=1,2,3.
\label{fU}
\ee
It remains to compute the image and the kernel of the map $f$ so defined.\\

We begin with the image. By (\ref{fU}), the entries of the matrix $f[U]$ are quadratic combinations of the 
entries of $U$, so $f$ is continuous. Since $\SU(2)$ is connected, the image of $f$ must be contained 
in the connected subgroup of $\mO(3)$, that is, $\SO(3)$. To prove the isomorphism (\ref{isoSU2}), we need to 
show that the opposite inclusion holds as well, {\it i.e.}~that any matrix in $\SO(3)$ can be written as 
$f[U]$ for 
some $U\in\SU(2)$.\\

The latter statement actually follows from a geometric observation \cite{HenneauxGroupe}: any rotation 
$R(\vec n,\phii)$ of 
$\RR^3$ (around an axis $\vec n$, by an angle $\phii$) can be written as the product $R(\vec 
n,\phii)=ST$ of two reflexions $S$ and $T$ with respect to planes whose intersection is the rotation axis, 
the angle between the planes being half the angle of rotation. Explicitly, $S$ and $T$ can be written as
\be
S:\vec x\mapsto\vec x-2(\vec x\cdot\vec m)\vec m
\quad\text{and}\quad
T:\vec x\mapsto\vec x-2(\vec x\cdot\vec q)\vec q,
\label{rfx}
\ee
where $\vec m$ and $\vec q$ are unit vectors orthogonal to the planes corresponding to the reflexions $S$ and 
$T$, 
respectively. Then, define the matrices $M\equiv m^i\sigma_i$ and $Q\equiv q^i\sigma_i$. These matrices are 
Hermitian, traceless, have determinant $-1$ and square to unity. Defining similarly $X\equiv x^i\sigma_i$, 
the 
reflexions (\ref{rfx}) can be written as
\be
S:X\mapsto-MXM
\quad\text{and}\quad
T:X\mapsto-QXQ.
\nn
\ee
Therefore, the rotation $R(\vec n,\phii)=ST$ (with $T$ acting first) acts on $X$ according to
\be
R(\vec n,\phii):X\mapsto MQXQM.
\label{RX}
\ee
But now note that the product $U\equiv MQ$ is such that $U^{\dagger}=QM$ with 
$UU^{\dagger}=MQQM=\mathbb{I}_2$, and it 
has unit determinant. Hence $U=MQ$ belongs to $\SU(2)$ and the transformation (\ref{RX}) is of the form 
(\ref{fU}) defining the homomorphism $f$. Hence we can write the rotation $R(\vec n,\phii)$ as $f[U]$. This 
proves that $f$ is surjective on $\SO(3)$.\\

To conclude the proof of (\ref{isoSU2}) we turn to the kernel of $f$, that is, the inverse image of the 
unit element in $\SO(3)$. Saying that 
$f[U]$ is the identity in $\SO(3)$ is just saying that $UXU^{\dagger}=X$ for any $X$ in $\VV$, which in turn 
is 
equivalent to saying that $U$ commutes with all $X$'s. But, when this holds, $U$ also 
commutes with any $e^{iX}$. Since $X$ is Hermitian and traceless, this is the same as saying that $U$ 
commutes with all elements of $\SU(2)$, {\it i.e.}~that $U$ belongs to the center of $\SU(2)$. The latter 
consists of the two matrices (\ref{ZSU2}), which form a group $\ZZ_2$.
\end{proof}

\paragraph{Remark.} From the definition (\ref{fU}) of the homomorphism $f$ and the form of the Pauli 
matrices, one can easily read off the explicit expression of the orthogonal matrix $f[U]$, for $U$ an element 
of $\SU(2)$:
\be
f\left[\bm a & b \\ c & d \emm\right]
=
\bm
\mathrm{Re}(\bar ad+\bar bc) & \mathrm{Im}(a\bar d-b\bar c) & \mathrm{Re}(\bar ac-\bar bd) \\
\mathrm{Im}(\bar ad+\bar bc) & \mathrm{Re}(a\bar d-b\bar c) & \mathrm{Im}(\bar ac-\bar bd) \\
\mathrm{Re}(a\bar b-c\bar d) & \mathrm{Im}(a\bar b-c\bar d) & \demi(|a|^2-|b|^2-|c|^2+|d|^2)
\emm.
\label{Hoho}
\ee
This formula exhibits the fact that $f$ is insensitive to an overall change of sign 
in the entries of its argument, since the right-hand side only involves quadratic combinations of those 
entries. In particular, it implies that the kernel of $f$ must contain the matrices (\ref{ZSU2}).

\subsection{The Lorentz group in three dimensions}
\label{LSL3}

The Lorentz group in three space-time dimensions is the group $\mO(2,1)$, as defined in subsection 
\ref{subsecPoin}. Its connected subgroup is $\SO(2,1)^{\uparrow}$. We will show here that this connected 
group is isomorphic to the quotient of $\SL(2,\RR)$ by its center. Before doing that, we 
review a few topological properties of $\SL(2,\RR)$. For the record, the results of this subsection will play 
a minor role in the remainder of these notes, so they may be skipped in a first reading.

\subsubsection{Properties of $\SL(2,\RR)$}

The group $\SL(2,\RR)$ is the group of volume-preserving linear transformations of the plane $\RR^2$. It can 
be seen as the group of real $2\times2$ matrices with unit determinant:
\be
\SL(2,\RR)
=
\left\{
\left.
\bm a & b \\ c & d \emm
\in \text{M}(2,\RR)
\right|
ad-bc=1
\right\}.
\nn
\ee

\paragraph{Lemma.} The group $\SL(2,\RR)$ is connected, but not simply connected. It is homotopic to a 
circle; in particular, the fundamental group of $\SL(2,\RR)$ is isomorphic to $\ZZ$.

\begin{proof}
Let
\be
S=\bm a & b \\ c & d \emm
\in\SL(2,\RR).
\nn
\ee
Since $\det(S)\neq0$, the vectors $(a,b)$ and $(c,d)$ in $\RR^2$ are linearly 
independent. We can therefore find three real numbers $\bar\alpha$, $\bar\beta$ and $\bar\gamma$ such that 
the set
\be
\left\{
\bar\alpha\cdot (a,b),\bar\beta\cdot(a,b)+\bar\gamma\cdot(c,d)
\right\}
\label{orthBasisBis}
\ee
be an orthonormal basis of $\RR^2$. Equivalently, there exists a matrix
\be
\bar K=\bm \bar\alpha & 0 \\ \bar\beta & \bar\gamma \emm
\nn
\ee
such that the product
\be
\calO\equiv\bar KS=\bm \bar\alpha a & \bar\alpha b \\ \bar\beta a+\bar\gamma c & \bar\beta b+\bar\beta d\emm
\nn
\ee
be an orthogonal matrix (since the lines and columns of an orthogonal matrix form an orthonormal basis). We 
can choose $\bar\alpha^{-1}=\sqrt{a^2+b^2}$, making $\bar\alpha$ positive. Since 
$\det(\bar K)=\bar\alpha\bar\gamma=\text{det}(\calO)=\pm1$, we may choose the orientation of the basis 
(\ref{orthBasisBis}) so that $\bar\gamma=1/\bar\alpha$, {\it i.e.}~$\calO\in\SO(2)$. Thus, any matrix 
$S\in\SL(2,\RR)$ can be written as
\be
S=\bar K^{-1}\calO\equiv K\calO,
\quad\text{with}\;\calO\in\SO(2)\;\text{and}\;K=\bm k & 0 \\ m & 1/k \emm
\nn
\ee
for some $m\in\RR$ and $k\in\RR$ strictly positive. Now, the set of triangular matrices of the form
\be
\bm k & 0 \\ m & 1/k \emm
\quad\text{with }k>0\text{ and }m\in\RR
\nn
\ee
is homeomorphic to $\RR\times\RR^+$, which is connected and has the 
homotopy type of a point. On the other hand, $\SO(2)$ is homeomorphic to a circle. This shows that 
$\SL(2,\RR)$ is connected and homotopic to a circle. In particular, the fundamental group of $\SL(2,\RR)$ is 
$\ZZ$.
\end{proof}

Note that the center of $\SL(2,\RR)$ is the same as that of $\SU(2)$ (see eq. (\ref{ZSU2})): it 
consists of the identity matrix and minus the identity matrix, forming a group isomorphic to $\ZZ_2$.

\subsubsection{The isomorphism}

\paragraph{Theorem.} There is an isomorphism
\be
\SO(2,1)^{\uparrow}\cong\SL(2,\RR)/\ZZ_2,
\label{SOSL}
\ee
where $\ZZ_2$ is the center of $\SL(2,\RR)$. In other words, $\SL(2,\RR)$ is the double cover of the 
connected Lorentz group in three dimensions (but it is {\it not} its universal cover, since it is not simply 
connected).

\begin{proof}
Our goal is to build a well chosen homomorphism 
mapping $\SL(2,\RR)$ on $\SO(2,1)^{\uparrow}$. Consider, therefore, the space $\VV$ of real, traceless 
$2\times 2$ matrices, that is, the Lie algebra of $\SL(2,\RR)$. Each matrix $X$ in 
$\VV$ can be written as
\be
X=x^{\mu}t_{\mu}\quad\text{(implicit sum over $\mu=0,1,2$)},
\nn
\ee
where the $x^{\mu}$'s are real numbers, while the $t_{\mu}$'s are the following matrices:
\be
t_0\equiv\bm 0 & 1 \\ -1 & 0 \emm,
\quad
t_1\equiv\bm 0 & 1 \\ 1 & 0 \emm,
\quad
t_2\equiv\bm 1 & 0 \\ 0 & -1 \emm.
\label{GGen}
\ee
(These matrices are generators of the Lie algebra of $\SL(2,\RR)$.) Note that, with this convention, the 
determinant of $X$ is, up to a sign, the square of the Minkowskian norm of the corresponding vector 
$(x^{\mu})$:
\be
\det(X)=-\eta_{\mu\nu}x^{\mu}x^{\nu}\equiv-x^2.
\label{detX}
\ee
There is a natural action of $\SL(2,\RR)$ on the space $\VV$. Namely, with each $S\in\SL(2,\RR)$, associate 
the map
\be
\VV\rightarrow\VV:X\mapsto SXS^{-1}.
\label{SXS-1}
\ee
This is the adjoint action of $\SL(2,\RR)$. It is linear and it preserves the determinant, since 
$\det(SXS^{-1})=\det(X)$. In addition, thanks to (\ref{detX}), each map (\ref{SXS-1}) can be 
seen as a Lorentz transformation acting on the $3$-vector $(x^{\mu})$. We 
can thus define a map
\be
f:\SL(2,\RR)\rightarrow\mO(2,1):S\mapsto f[S],
\nn
\ee
where the $3\times 3$ matrix $f[S]$ is given by
\be
St_{\mu}x^{\mu}S^{-1}=t_{\mu}{f[S]^{\mu}}_{\nu}x^{\nu}\quad\forall\, (x^{\mu})\in\RR^3,
\nn
\ee
or equivalently,
\be
St_{\mu}S^{-1}=t_{\nu}{f[S]^{\nu}}_{\mu}\quad\forall\,\mu=0,1,2.
\label{fS}
\ee
Because $(ST)X(ST)^{-1}=S(TXT^{-1})S^{-1}$ for all matrics $S$, $T$ in $\SL(2,\RR)$, the map $f$ is obviously 
a homomorphism. Furthermore, 
by (\ref{fS}), the entries of $f[S]$ are quadratic combinations 
of the entries of $S$; therefore $f$ is continuous. In particular, since $\SL(2,\RR)$ is connected, 
the image of $f$ is certainly contained in the connected Lorentz group 
$\SO(2,1)^{\uparrow}$.\\

It remains to prove that $f$ is surjective on $\SO(2,1)^{\uparrow}$ and to compute its kernel. We begin with 
the former. Let therefore $\Lambda\in\SO(2,1)^{\uparrow}$. The standard decomposition theorem (\ref{dec}) 
adapted to $d=3$ states that $\Lambda$ can be written as $\Lambda=R_1L(\chi)R_2$, where $R_1$ and $R_2$ 
belong to the $\SO(2)$ subgroup of $\mO(2,1)$, while $L(\chi)$ is a standard boost of 
the form (\ref{boostBis}) with the last line and last column suppressed. To prove surjectivity of $f$ on 
$\SO(2,1)^{\uparrow}$, we need to show that there exist matrices $S_1$, $S_2$ and $S(\chi)$ in $\SL(2,\RR)$ 
such that
\be
f[S_1]=R_1,\quad f[S_2]=R_2\quad\text{and}\quad f[S(\chi)]=L(\chi).
\label{Ballo}
\ee
We begin with the rotations. If $S$ belongs to the $\SO(2)$ subgroup of $SL(2,\RR)$, {\it i.e.}
\be
S=\bm \cos\theta & \sin\theta \\ -\sin\theta & \cos\theta \emm
\label{Mario}
\ee
for some angle $\theta$, then formula (\ref{fS}) gives
\be
f[S]
=
\bm
1 & 0 & 0 \\
0 & \cos2\theta & -\sin2\theta \\
0 & \sin2\theta & \cos2\theta
\emm.
\label{MarioBis}
\ee
This implies that any rotation $R$ in $\SO(2,1)^{\uparrow}$ can be realised as $R=f[S]$ for some matrix $S$ 
of the form (\ref{Mario}) in $\SL(2,\RR)$. Thus, to prove surjectivity of $f$ as in (\ref{Ballo}), it only 
remains to find a matrix $S(\chi)$ such that $f[S(\chi)]=L(\chi)$ be a standard boost with rapidity $\chi$ in 
three space-time dimensions. Again, using (\ref{fS}), one verifies that the matrix
\be
S(\chi)
=
\bm
e^{-\chi/2} & 0 \\ 0 & e^{\chi/2}
\emm
\nn
\ee
is precisely such that
\be
f[S(\chi)]
=
\bm
\cosh\chi & -\sinh\chi & 0 \\
-\sinh\chi & \cosh\chi & 0 \\
0 & 0 & 1
\emm,
\label{MarioTris}
\ee
which was the desired relation. We conclude that $f$ is surjective on $\SO(2,1)^{\uparrow}$, as expected.\\

Finally, to establish (\ref{SOSL}), we need to show that the kernel of $f$ is isomorphic to $\ZZ_2$. The 
proof is essentially the same as for $\SU(2)$. Indeed, saying that $S\in\SL(2,\RR)$ belongs to the kernel of 
$f$ means that $S$ commutes with any linear combination of the generators (\ref{GGen}). But this 
implies that $S$ commutes with all elements of $\SL(2,\RR)$, {\it i.e.}~that $S$ belongs to the center of 
$\SL(2,\RR)$, which is just $\ZZ_2$.
\end{proof}

\paragraph{Remark.} Using (\ref{fS}), one can write down explicitly the homomorphism $f$ as
\be
f\left[\bm a & b \\ c & d \emm\right]
=
\bm
\demi(a^2+b^2+c^2+d^2) & \demi(a^2-b^2+c^2-d^2) & -ab-cd\\
\demi(a^2+b^2-c^2-d^2) & \demi(a^2-b^2-c^2+d^2) & -ab+cd\\
-ac-bd & bd-ac & ad+bc
\emm,
\nn
\ee
where the argument of $f$ on the left-hand side is a matrix in $\SL(2,\RR)$. We already displayed two 
special cases of this relation in equations (\ref{MarioBis}) and (\ref{MarioTris}). As in the analogous 
homomorphism 
(\ref{Hoho}) for $\SU(2)$, the fact that the right-hand side only involves quadratic combinations of the 
entries of the $\SL(2,\RR)$ matrix implies that $f$ is insensitive to overall signs, so that its 
kernel necessarily contains $\ZZ_2$.

\subsection{The Lorentz group in four dimensions}
\label{LSL}

We now turn to the analogue of the previous isomorphism for the Lorentz group in four dimensions, following 
\cite{HenneauxGroupe} once again. As usual, we 
will begin by reviewing certain topological properties of $\SL(2,\CC)$, turning to the isomorphism later.

\subsubsection{Properties of $\SL(2,\CC)$}

The group $\SL(2,\CC)$ is the set of volume-preserving linear transformations of the vector space $\CC^2$. 
It can be seen as the group of complex $2\times 2$ matrices with unit determinant:
\be
\SL(2,\CC)
=
\left\{
\left.
\bm a & b \\ c & d \emm
\in \text{M}(2,\CC)
\right|
ad-bc=1
\right\}.
\nn
\ee
\paragraph{Lemma.} The group $\SL(2,\CC)$ is connected and simply connected.

\begin{proof}
We use essentially the same technique as for the group $\SL(2,\RR)$ in subsection \ref{LSL3}. Let
\be
S=\bm a & b \\ c & d \emm
\in\SL(2,\CC).
\nn
\ee
Then $\text{det}(S)=1\neq0$, so the vectors $(a,b)$ and $(c,d)$ in $\CC^2$ are linearly 
independent. We can then find three 
complex numbers $\bar\alpha$, $\bar\beta$ and $\bar\gamma$ such that
\be
\left\{
\bar\alpha\cdot (a,b),\bar\beta\cdot(a,b)+\bar\gamma\cdot(c,d)
\right\}
\label{orthBasis}
\ee
be an orthonormal basis of $\CC^2$. In other words, there exists a matrix
\be
\bar K=\bm \bar\alpha & 0 \\ \bar\beta & \bar\gamma \emm
\nn
\ee
such that the product
\be
U\equiv\bar KS=\bm \bar\alpha a & \bar\alpha b \\ \bar\beta a+\bar\gamma c & \bar\beta b+\bar\beta d\emm
\nn
\ee
be unitary (since the lines and columns of a unitary matrix form an orthonormal basis). We may take 
$\bar\alpha^{-1}=\sqrt{|a|^2+|b|^2}$, so that $\bar\alpha$ is real and strictly positive. Since 
$\text{det}(\bar K)=\bar\alpha\bar\gamma=\text{det}(U)$ is a complex number with unit modulus, and since the 
second basis vector in the orthonormal basis (\ref{orthBasis}) is determined up to a phase, we may choose 
$\text{det}(\bar K)=1$, that is, $\bar\gamma=1/\bar\alpha$. Thus $U\in\SU(2)$ and $\bar\gamma$ is also a 
strictly positive real number. We conclude that any matrix in $\SL(2,\CC)$ can be written as
\be
S=\bar K^{-1}U\equiv KU,
\quad\text{with}\;U\in\SU(2)\;\text{ and }\;K=\bm k & 0 \\ m & 1/k \emm
\nn
\ee
for some $m\in\CC$ and $k\in\RR$ strictly positive. $\SU(2)$ is diffeomorphic to 
$S^3$, so it is connected and simply connected. 
Furthermore, the group of triangular matrices of the form
\be
\bm k & 0 \\ m & 1/k \emm\;\text{with }k>0\text{ and }m\in\CC
\nn
\ee
is homeomorphic to $\CC\times\RR^+$, which is also connected and simply connected. As a 
consequence, $\SL(2,\CC)$ itself is connected and simply connected.
\end{proof}

Note the topological difference between $\SL(2,\RR)$ and $\SL(2,\CC)$: both are connected, but only 
$\SL(2,\CC)$ is simply connected, while $\SL(2,\RR)$ is homotopic to a circle. This subtlety has important 
consequences for (projective) unitary representations of $\SL(2,\RR)$ and $\SL(2,\CC)$, and, accordingly, for 
those of the 
Lorentz groups in three and four dimensions \cite{Weinberg,Wigner,Binegar,Grigore}.\\

One can also verify that the center of $\SL(2,\CC)$ consists of the two matrices (\ref{ZSU2})
and is thus isomorphic to $\ZZ_2$, exactly as in the case of $\SU(2)$ and $\SL(2,\RR)$.

\subsubsection{The isomorphism}

\paragraph{Theorem.} There exists an isomorphism
\be
\boxed{L_+^{\uparrow}\cong\SL(2,\CC)/\ZZ_2,}
\label{isomL}
\ee
where $\ZZ_2$ is the center of $\SL(2,\CC)$. In other words, $\SL(2,\CC)$ is the double cover of the 
connected Lorentz group in four dimensions, and it is also its universal cover.

\begin{proof}
Proceeding as for the Lorentz group in three dimensions, we wish to build a 
homomorphism 
$f:\SL(2,\CC)\rightarrow \mO(3,1)$ and compute its image and its kernel. Consider, therefore, the vector 
space $\VV$ of $2\times 2$ Hermitian matrices. It is a real, four-dimensional vector space: any matrix 
$X\in\VV$ can be written as
\be
X=\begin{pmatrix} x^0+x^3 & -x^1-ix^2 \\ -x^1+ix^2 & x^0-x^3 \end{pmatrix},
\label{XXX}
\ee
where the $x^{\mu}$'s are real coefficients. This can also be written as $X=x^{\mu}\tau_{\mu}$, where 
$\tau_0$ denotes the $2\times 2$ identity matrix, while\footnote{The choice of signs is slightly 
unconventional here; it will eventually ensure that the action (\ref{ZconfBis}) of Lorentz transformations on 
celestial 
spheres coincides with the standard expression (\ref{Zconf}) of conformal transformations.} 
$\tau_1=-\sigma_1$, $\tau_2=\sigma_2$ and $\tau_3=\sigma_3$ in terms of the Pauli matrices (\ref{Pauli}). 
Then, just as in (\ref{detX}),
\be
\det(X)=\det(x^{\mu}\tau_{\mu})=(x^0)^2-(x^1)^2-(x^2)^2-(x^3)^2=-\eta_{\mu\nu}x^{\mu}x^{\nu}\equiv-x^2.
\label{detXmink}
\ee
Let us now define an action of $\SL(2,\CC)$ on 
$\VV$: for each matrix $S\in\SL(2,\CC)$, we consider the linear map
\be
\VV\rightarrow\VV:X\mapsto SXS^{\dagger}.
\label{SXSdag}
\ee
This action preserves the determinant 
because $\det(SXS^{\dagger})=\det(X)$. By (\ref{detXmink}), this amounts to preserving the 
square 
of the Minkoswkian norm of the four-vector $(x^{\mu})$, so the transformation (\ref{SXSdag}) can be seen as a 
Lorentz transformation acting on $(x^{\mu})$. We thus define a map
\be
f:\SL(2,\CC)\rightarrow\mO(3,1):S\mapsto f[S],
\label{hom}
\ee
where the $4\times 4$ matrix $f[S]$ is given by
\be
S\tau_{\mu}x^{\mu}S^{\dagger}
=
\tau_{\mu}{f[S]^{\mu}}_{\nu}x^{\nu}\quad\forall\, (x^{\mu})\in\RR^4,
\label{fSSS}
\ee
or equivalently,
\be
S\tau_{\mu}S^{\dagger}=\tau_{\nu}{f[S]^{\nu}}_{\mu}\quad\forall\,\mu=0,1,2,3.
\label{fSbis}
\ee
Since $(ST)X(ST)^{\dagger}=S(TXT^{\dagger})S^{\dagger}$, the map $f$ is obviously a homomorphism. 
Furthermore, by (\ref{fSbis}), the entries of $f[S]$ are quadratic combinations 
of the entries of $S$; so $f$ is continuous. In particular, since $\SL(2,\CC)$ is connected, 
the image of $f$ is contained in the connected Lorentz group 
$\SO(3,1)^{\uparrow}=L_+^{\uparrow}$.\\

It remains to prove that $f$ is surjective on $L_+^{\uparrow}$ and to compute its kernel. We have just seen 
that $\Ima(f)\subseteq L_+^{\uparrow}$ by continuity, so as far as surjectivity is concerned, we need only 
prove the opposite inclusion. Let therefore $\Lambda$ belong to $L_+^{\uparrow}$. By the standard 
decomposition 
theorem (\ref{dec}), we can write $\Lambda$ as a standard boost $L(\chi)$, for some value of the rapidity 
$\chi$, sandwiched between two (orientation-preserving) spatial rotations: $\Lambda=R_1L(\chi)R_2$. Thus, in 
order to prove surjectivity of $f$ on $L_+^{\uparrow}$, it suffices to find three matrices $S_1$, $S_2$ and 
$S(\chi)$ in $\SL(2,\CC)$ such that
\be
f[S_1]=R_1,\quad f[S_2]=R_2,\quad f[S(\chi)]=L(\chi)
\label{cond}
\ee
since in that case $f[S_1S(\chi)S_2]=R_1L(\chi)R_2=\Lambda$. Now, the restriction of the homomorphism 
(\ref{hom}) to the $\SU(2)$ subgroup of $\SL(2,\CC)$ is precisely the 
homomorphism (\ref{homSU}) that we used to prove the isomorphism $\SO(3)\cong\SU(2)/\ZZ_2$. We know, 
therefore, that there exist matrices $S_1$ and $S_2$ in $\SU(2)$ such that conditions (\ref{cond}) hold. As 
for the matrix $S(\chi)$, we make the educated guess
\be
S(\chi)
=
\bm \cosh(\chi/2) & \sinh(\chi/2) \\ \sinh(\chi/2) & \cosh(\chi/2) \emm
=
\cosh(\chi/2)\mathbb{I}_2-\sinh(\chi/2)\tau_1.
\nn
\ee
The image of $S(\chi)$ under $f$ can be read off from the property
\be
S(\chi)x^{\mu}\tau_{\mu}S(\chi)^{\dagger}
=
\left(x^0\cosh\chi-x^1\sinh\chi\right)\tau_0+
\left(-x^0\sinh\chi+x^1\cosh\chi\right)\tau_1+x^2\tau_2+x^3\tau_3.
\nn
\ee
Comparing with the definition (\ref{fSSS}) of $f$, we see that $f[S(\chi)]$ is precisely the standard 
Lorentz boost $L(\chi)$, as written in (\ref{boostBis}). In conclusion, the homomorphism $f$ is surjective on 
$L_+^{\uparrow}$.\\

The last missing piece of the proof is the computation of the kernel of $f$. By definition, the kernel 
consists of matrices $S$ such that $SXS^{\dagger}=X$ for any $X\in\VV$. Taking $X=\mathbb{I}_2$, we see that 
$S$ must belong to $\SU(2)$. Then, taking $X=\sigma_i$, we observe that $S$ must belong to the center of 
$\SU(2)$, that is, $\ZZ_2$.
\end{proof}

\paragraph{Remark.} As usual, the definition (\ref{fSbis}) can be used to compute explicitly the matrix 
$f[S]$, when $S$ belongs to $\SL(2,\CC)$. The result is
\bea
\label{THEhomomorphism}
& f\left[\bm a & b \\ c & d \emm\right]=  & \\
& = \bm
\demi\left(|a|^2+|b|^2+|c|^2+|d|^2\right) & -\text{Re}\left\{a\bar b+c\bar d\right\}
&
\text{Im}\left\{a\bar b+c\bar d\right\} & \demi\left(|a|^2-|b|^2+|c|^2-|d|^2\right)
\\
-\text{Re}\left\{\bar ac+\bar bd\right\} & \text{Re}\left\{\bar ad+\bar bc\right\}
&
-\text{Im}\left\{a\bar d-b\bar c\right\} & -\text{Re}\left\{\bar a c-\bar bd\right\}
\\
\text{Im}\left\{\bar ac+\bar bd\right\} & -\text{Im}\left\{\bar ad+\bar bc\right\}
&
\text{Re}\left\{a\bar d-b\bar c\right\} & \text{Im}\left\{\bar ac-\bar bd\right\}
\\
\demi\left(|a|^2+|b|^2-|c|^2-|d|^2\right) & -\text{Re}\left\{a\bar b-c\bar d\right\}
&
\text{Im}\left\{a\bar b-c\bar d\right\} & \demi\left(|a|^2-|b|^2-|c|^2+|d|^2\right)
\emm, &\nn
\eea
involving only quadratic combinations of the entries of the argument of $f$, which exhibits the fact that the 
kernel of $f$ must contain $\ZZ_2$.

\subsubsection{Examples}

For future reference, let us display two specific families of matrices in $\SL(2,\CC)$ corresponding to 
rotations around the $x^3$ axis and boosts along that axis, associated respectively with the Lorentz matrices
\be
\bm
1 & 0 & 0 & 0 \\
0 & \cos\theta & -\sin\theta & 0 \\
0 & -\sin\theta & \cos\theta & 0 \\
0 & 0 & 0 & 1
\emm
\quad\text{and}\quad
\bm
\cosh\chi & 0 & 0 & -\sinh\chi \\
0 & 1 & 0 & 0 \\
0 & 0 & 1 & 0 \\
-\sinh\chi & 0 & 0 & \cosh\chi
\emm.
\nn
\ee
Demanding that these matrices be of the form $f[S]$ for some $S\in\SL(2,\CC)$ determines $S$ uniquely, up to 
a sign, through formula (\ref{THEhomomorphism}). One thus finds that rotations by $\theta$ around $x^3$ are 
represented by
\be
S_{\text{rot}}
=
\pm
\bm
e^{-i\theta/2} & 0 \\
0 & e^{i\theta/2}
\emm
\label{ExRot}
\ee
while boosts with rapidity $\chi$ along $x^3$ are given by
\be
S_{\text{boost}}
=
\pm
\bm
e^{-\chi/2} & 0 \\ 0 & e^{\chi/2}
\emm
\label{ExBoost}
\ee
We will put these formulas to use in subsection \ref{boostSphere}, when describing the effect of Lorentz 
transformations on celestial spheres.

\subsection{Lorentz groups and division algebras}
\label{LorentzDiv}

In the two previous subsections, we proved the two very similar isomorphisms
\be
\SO(2,1)^{\uparrow}\cong\SL(2,\RR)/\ZZ_2
\quad\text{and}\quad
\SO(3,1)^{\uparrow}\cong\SL(2,\CC)/\ZZ_2.
\label{isomSeq}
\ee
From this viewpoint, going from three to four space-time dimensions amounts to changing $\RR$ into $\CC$. 
Now, from Hurwitz's theorem it is well known that $\RR$ and $\CC$ are only the two first entries of a 
list of four normed 
division algebras (see {\it e.g.}~\cite{LaFuente}): the two remaining algebras are the set $\mathbb{H}$ of 
quaternions and the 
set $\mathbb{O}$ of octonions. Given this classification and the apparent coincidence (\ref{isomSeq}), it is 
tempting to ask whether similar isomorphisms hold between certain higher-dimensional Lorentz groups and 
special linear groups of the form $\SL(2,\mathbb{H})$ or $\SL(2,\mathbb{O})$. This turns out to be the case 
indeed: one can prove that the connected Lorentz groups in six and ten 
space-time dimensions satisfy \cite{KugoTownsend,Sud,Baez,BaezHuerta01,BaezHuerta02}
\be
\SO(5,1)^{\uparrow}\cong\SL(2,\HH)/\ZZ_2
\quad\text{and}\quad
\SO(9,1)^{\uparrow}\cong\SL(2,\OO)/\ZZ_2.
\nn
\ee
We will not prove these isomorphisms here. We will not even attempt to explain the meaning of the last 
isomorphism in this list, given that octonions are not associative, so that what we call ``$\SL(2,\OO)$'' is 
not obvious. Let us simply mention, as a curiosity, that these isomorphisms are related to the fact 
that minimal supersymmetric gauge field theories (with minimally coupled massless spinors) can only be 
defined in space-time dimensions 3, 4, 6 and 10. More generally, the relation between spinors and division 
algebras spreads all the way up to superstring theory. We will not study these questions here 
and refer for instance to \cite{BaezHuerta01,BaezHuerta02} for many more details.

\section{Conformal transformations of the sphere}
\label{secConf}

This section is a differential-geometric interlude: setting the Lorentz group aside, we will show that the 
quotient $\SL(2,\CC)/\ZZ_2$ may be seen as the group of conformal transformations of the $2$-sphere. 
Accordingly, our battle plan will be the following. We will first define, in general terms, the notion of 
conformal transformations of a manifold 
(subsection \ref{confG}). We will then apply this definition to the plane and the sphere (subsections 
\ref{confP} 
and \ref{confS}) and classify the corresponding conformal transformations. These matters should 
be familiar to readers acquainted with conformal field theories in two dimensions, which we briefly mention 
in subsection \ref{CFT}. Although some basic knowledge of differential geometry may be useful at 
this point, it is not mandatory for understanding the text, as our presentation will not be 
cast in a mathematically rigorous language. We refer for instance to \cite{Lee,Bourgeois} for an introduction 
to differential geometry.

\subsection{Notion of conformal transformations}
\label{confG}

In short, a conformal transformation of some space is a transformation which preserves the angles. 
To define precisely what we mean by ``angles'' (and hence conformal transformations), we will now review at 
lightspeed the notions of manifolds and Riemannian metrics.

\subsubsection{Manifolds and metrics}

Roughly speaking, a (smooth) manifold is a topological space that looks locally like a Euclidean space 
$\RR^n$, the number $n$ being called the dimension of the manifold. Here, by ``locally'', we mean ``upon 
zooming in on the manifold'': any point on the manifold admits a neighbourhood that is homeomorphic to 
$\RR^n$. Two typical examples of $n$-dimensional manifolds are $\RR^n$ itself, and the 
sphere $S^n$. Thanks to the locally Euclidean structure, we can define, at each point $p$ of a 
manifold $\calM$, a vector space consisting of vectors tangent to $\calM$ at $p$; this vector space is called 
the {\it tangent space} of $\calM$ at $p$, denoted $T_p\calM$. If we think of a manifold $\calM$ as a smooth 
set 
of points embedded in some higher-dimensional Euclidean space $\RR^N$, then the tangent space $T_p\calM$ is 
literally the (affine) hyperplane in $\RR^N$ that is tangent to $\calM$ at $p$, endowed with the vector space 
structure inherited from $\RR^N$.

\begin{figure}[H]
\begin{center}
\includegraphics[width=0.60\textwidth]{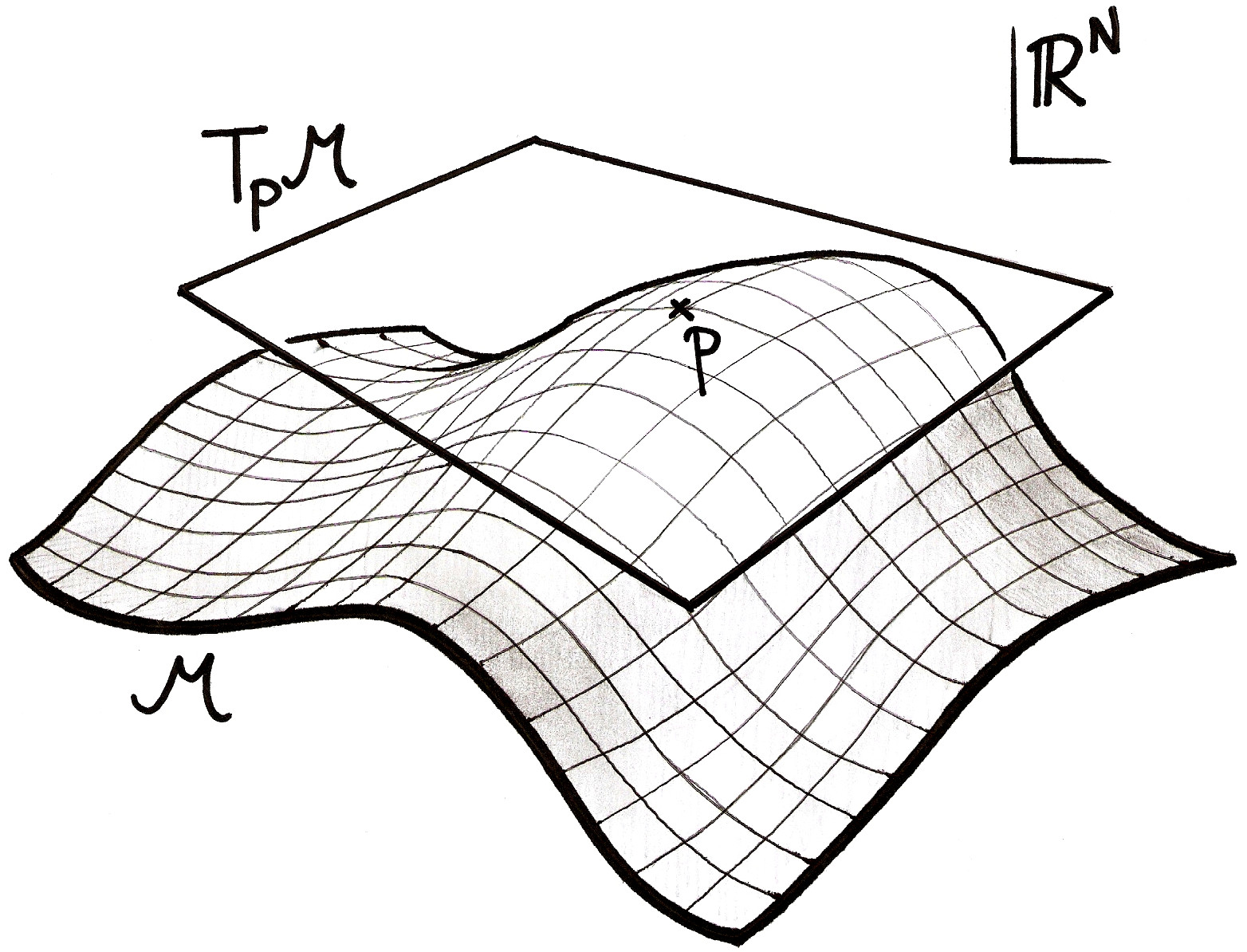}
\caption{A manifold $\calM$ embedded in $\RR^N$. The point $p$ belongs to the manifold, and the plane 
tangent to $\calM$ at $p$ is the tangent space $T_p\calM$. In this drawing, we take $N=3$ and the manifold is 
two-dimensional. The grid was added to emphasize the fact that the manifold looks, locally, like a plane 
$\RR^2$.}
\end{center}
\end{figure}

Given a vector space, it is natural to endow it with a scalar product, allowing one to compute norms of 
vectors and angles between vectors. Since a manifold $\calM$ has a tangent space at each point, one would 
like 
to define a scalar product in the tangent space at each point of $\calM$; a metric does precisely this job.

\paragraph{Definition.} A {\it (Riemannian) metric} $g$ on $\calM$ is the data of a scalar product 
in each tangent space of 
$\calM$, such that this scalar product varies smoothly on $\calM$ \cite{Bertelson}. More precisely, a metric 
is a symmetric, 
positive-definite, smooth tensor field
\be
g:\calM\rightarrow T^2(\calM):p\mapsto g_p,
\label{Crosss}
\ee
where $g_p$ is the aforementioned scalar product in $T_p\calM$:
\be
g_p:T_p\calM\times T_p\calM\rightarrow\RR:(v,w)\mapsto g_p(v,w).
\label{Satis}
\ee
The requirements of symmetry and positive-definiteness ensure that $g_p$ satisfies all the standard 
properties of a scalar product. This definition can be extended to {\it pseudo}-Riemannian metrics, that is, 
symmetric tensor fields such as (\ref{Crosss}) that are not necessarily positive-definite. In particular, we 
will see below that $d$-dimensional Minkowski space-time is the manifold $\RR^d$ endowed with the 
pseudo-Riemannian metric (\ref{Triangle}).

\subsubsection{Examples}

To illustrate concretely the above definition, let us consider a few simple examples of metrics on the 
manifold 
$\calM=\RR^2$. We can endow this manifold with global (Cartesian) coordinates $(x,y)$ such 
that any point $p\in\RR^2$ is identified with its pair of coordinates. Our first example is the 
Euclidean metric, whose expression in Cartesian coordinates is
\be
g=dx^2+dy^2.
\label{Eu}
\ee
To explain the meaning of this notation, let us pick a point $(x,y)$ in $\RR^2$ and two vectors $v$ and $w$ 
at that point, with respective components $(v_x,v_y)$ and $(w_x,w_y)$. Their scalar product is given by 
(\ref{Satis}), {\it i.e.}
\be
g_{(x,y)}(v,w)=\left(dx^2+dy^2\right)\left[(v_x,v_y),(w_x,w_y)\right].
\label{gxy}
\ee
By definition, upon acting on a vector, $dx$ gives the $x$-component of this vector. (In the standard 
language of differential geometry, $dx$ is the differential of $x$, that is, the one-form dual to the vector 
field $\der/\der x$ associated 
with the coordinate $x$ on $\RR^2$.) The notation $dx^2$ is 
then understood as the operation which, upon acting on two vectors, gives the product of their components 
along $x$. A similar 
definition holds for $dy$ and $dy^2$, except that they, of course, give $y$-components of vectors. Applying 
these rules to (\ref{gxy}), we find that the metric (\ref{Eu}) defines the standard Euclidean scalar product,
\be
g_{(x,y)}(v,w)=v_xw_x+v_yw_y.
\label{Euclideans}
\ee
Of course, one can define more generally the Euclidean metric on $\RR^d$ to be $g=(dx_1)^2+\cdots+(dx_d)^2$ 
in terms of Cartesian coordinates.

\begin{figure}[H]
\begin{center}
\includegraphics[width=0.60\textwidth]{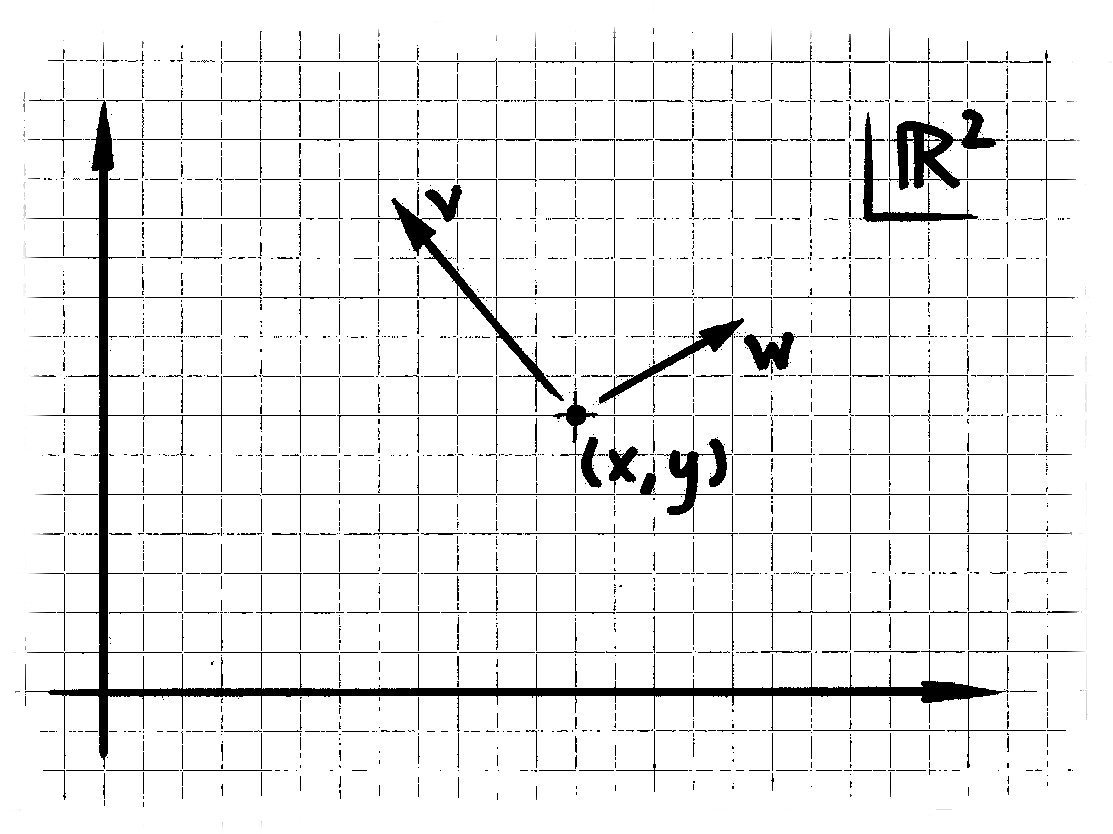}
\caption{The plane $\RR^2$ and the Cartesian coordinates used to label its points. A point with coordinates 
$(x,y)$ is also represented, and $v$, $w$ are two 
vectors at that point. Their scalar product with respect to the Euclidean metric 
(\ref{Eu}) is given by (\ref{Euclideans}).}
\end{center}
\end{figure}

A slightly less trivial example of metric on $\RR^2$ is given by
\be
g_{(x,y)}=\frac{dx^2+dy^2}{(1+x^2+y^2)^2},
\label{gxyy}
\ee
where $(x,y)$ is the point at which the metric is evaluated. If then $v$ and $w$ are two vectors at $(x,y)$, 
with the same components as before, their scalar product 
with respect to this new metric is
\be
g_{(x,y)}(v,w)=\frac{v_xw_x+v_yw_y}{(1+x^2+y^2)^2},
\nn
\ee
where we have used once more the rule saying that $dx$ (resp.~$dy$), upon acting on a vector, gives the 
$x$-component (resp.~$y$-component) of 
the vector. By contrast to the Euclidean scalar product (\ref{Euclideans}), this expression depends 
explicitly on the point $(x,y)$. In other words, if we take two families of vectors 
on $\RR^2$ with constant components $(v_x,v_y)$ and $(w_x,w_y)$ at each point of the plane, their scalar 
product will vary as we move on $\RR^2$.\\

Of course, the metric (\ref{gxyy}) that we picked was chosen for 
illustrative purposes only: any positive function on $\RR^2$ multiplying $dx^2+dy^2$ would give a 
(generally position-dependent) Riemannian metric on $\RR^2$. More generally, any position-dependent, real 
quadratic 
combination of $dx$'s and $dy$'s,
\be
A(x,y)dx^2+2B(x,y)dxdy+C(x,y)dy^2,
\label{ABC}
\ee
is a Riemannian metric on $\RR^2$ as long as $A$ and $AC-B^2$ are everywhere positive. If $v$ and 
$w$ are two vectors at $(x,y)$ with the same components as before, their scalar product with respect to the 
metric (\ref{ABC}) is $A(x,y)v_xw_x+2B(x,y)v_xw_y+C(x,y)v_yw_y$. Again, the generalization of these 
considerations to $\RR^d$ is straightforward: in terms of Cartesian coordinates $x_1$,...,$x_d$, the most 
general Riemannian metric on $\RR^d$ takes the form $g_{ij}(x_1,...,x_d)dx_idx_j$ (with implicit summation 
over $i,j=1,...,d$), where $(g_{ij})$ is a symmetric, positive-definite matrix at each point.

\subsubsection{Angles and conformal transformations}

Metrics can be used to define norms and angles on tangent spaces of a manifold. Indeed, suppose we are given 
a manifold $\calM$ endowed with a metric $g$. Let $p$ be a point in $\calM$ and let $v$ be a tangent vector 
of $\calM$ at $p$. Then, the norm of $v$ is naturally defined to be $\|v\|\equiv\left[g_p(v,v)\right]^{1/2}$. 
Furthermore, if $v$ and $w$ are two vectors at $p$, the angle $\theta$ between them is defined (up to a 
sign) by
\be
\cos\theta\equiv\frac{g_p(v,w)}{\|v\|\cdot\|w\|}.
\nn
\ee
Note that this definition is blind to the local normalization of the metric. Indeed, suppose we 
define two metrics $g$ and $g'$ on $\calM$, such that
\be
\left(g'\right)_p=\Omega(p)\cdot g_p\quad\forall\,p\in\calM,
\nn
\ee
where $\Omega$ is some smooth, positive real function on $\calM$. In other words, let us assume that $g$ 
and $g'$ are proportional, the proportionality factor being position-dependent. Then, these two metrics 
define the same angles. The proof is elementary: if $v$ and $w$ are two vectors at $p\in\calM$, then 
the cosine of the angle between these vectors is
\be
\frac{g_p(v,w)}{\left[g_p(v,v)\,g_p(w,w)\right]^{1/2}}
=
\frac{\Omega(p)g_p(v,w)}{\left[\Omega(p)g_p(v,v)\,\Omega(p)g_p(w,w)\right]^{1/2}}
=
\frac{g'_p(v,w)}{\left[g'_p(v,v)\,g'_p(w,w)\right]^{1/2}},
\nn
\ee
which is obviously independent of whether we choose to use the metric $g$ or the metric $g'$. This 
observation will be crucial in the following pages.\\

Given a manifold $\calM$, it is natural to wonder what modifications $\calM$ may undergo, such that these 
modifications ``preserve the structure'' of $\calM$. To answer this question, we must specify precisely what 
is the structure we wish to preserve. Clearly, a first feature we 
would like to preserve when deforming $\calM$ is its local Euclidean structure. This leads to 
the 
notion 
of {\it diffeomorphisms}: by definition, a diffeomorphism of a manifold $\calM$ is a smooth, invertible map 
$\phi:\calM\rightarrow\calM$ such that the inverse map $\phi^{-1}$ be smooth as well\footnote{A 
diffeomorphism is thus a smooth generalization of the notion of homeomorphism, the word ``smooth'' 
replacing the word ``continuous''.}. In this sentence, the 
word ``smooth'' means ``that preserves the local Euclidean structure in a continuous and differentiable 
way''. In heuristic terms, a diffeomorphism of $\calM$ is a smooth, invertible deformation 
of $\calM$ when the latter is seen as a rubber space.

\begin{figure}[H]
\begin{center}
\includegraphics[width=0.30\textwidth]{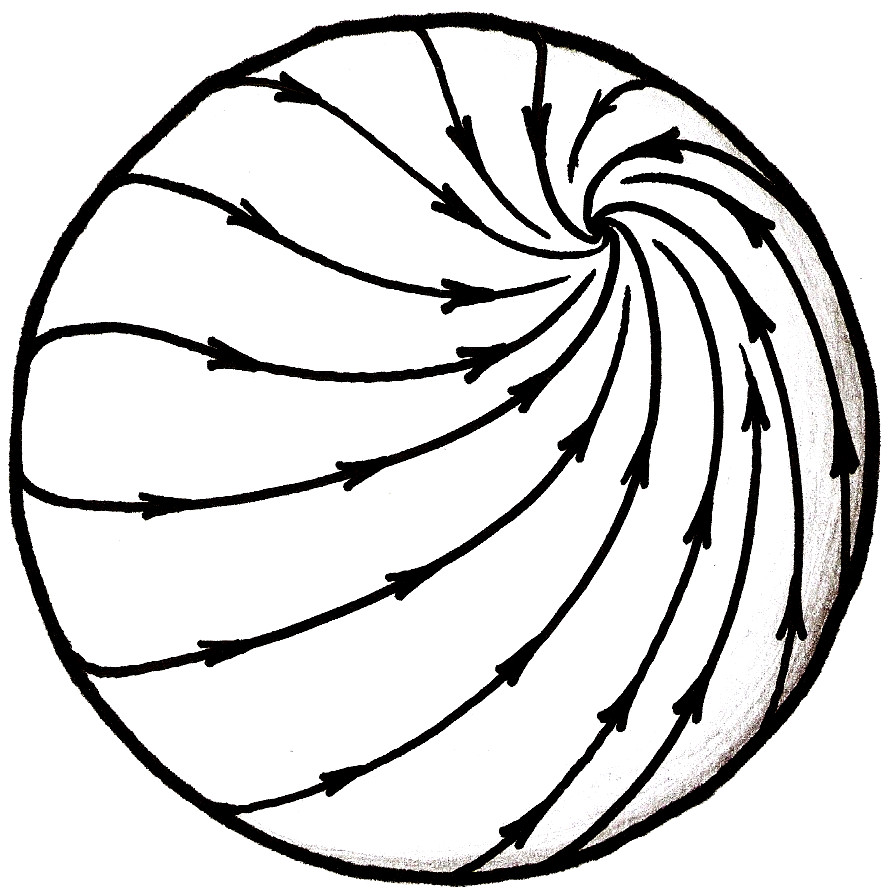}
\caption{A diffeomorphism of the sphere. The arrows represent how points of the sphere move as the 
diffeomorphism is applied. Under the action of the diffeomorphism, points are shuffled, shapes are 
distorted, but the motion is smooth, preserving the local Euclidean 
structure of the manifold.}
\end{center}
\end{figure}

Suppose now we pick a manifold $\calM$ endowed with a metric $g$, and consider a diffeomorphism 
$\phi:\calM\rightarrow\calM$ of 
that manifold. Since the diffeomorphism is a deformation of $\calM$, it will in general affect distances and 
angles on $\calM$; in other words, a general diffeomorphism does {\it not} preserve the metric on $\calM$ 
and maps the original metric $g$ on some new metric $g'$. (In precise terms, what we call the transformed 
metric is the pull-back of $g$ by $\phi$, that is, $g'\equiv\phi^*g$.) This gives a motivation for defining 
certain 
subclasses of diffeomorphisms that preserve some part (or the entirety) of the metric structure, {\it 
i.e.}~diffeomorphisms for which the new metric $g'$ has certain properties in common with the first metric 
$g$.

\paragraph{Definition.} A {\it conformal transformation} of $\calM$ is a diffeomorphism 
$\phi:\calM\rightarrow\calM$ such that the original metric $g$ and the transformed metric $g'$ define the 
same angles (possibly up to signs).\\

Given the property, shown above, that proportional metrics define identical angles (possibly up to 
signs), it is easy to write down an explicit formula for what we mean by a conformal transformation: it 
is a diffeomorphism for which the transformed metric $g'$ is related to the original metric $g$ as
\be
g'_p=\Omega(p)g_p\quad\forall\, p\in\calM,
\label{g'''}
\ee
where $\Omega$ is some smooth, positive function on $\calM$. When $\Omega(p)=1$ for all $p$ in $\calM$, we 
say that the diffeomorphism $\phi$ is an {\it isometry}: it preserves not only the angles, but also the norms 
defined by the metric $g$. Of course, conformal transformations and isometries can also be defined for 
pseudo-Riemannian metrics.

\paragraph{Remark.}

We are now equipped with the tools needed to restate in differential-geometric terms the definition of the 
Lorentz and Poincar\'e groups, originally described in subsection \ref{subsecPoin}. Namely, 
define $d$-dimensional Minkowski space to be the manifold $\RR^d$ endowed with a pseudo-Riemannian metric 
$g$ such that there exist global coordinates $(x^{\mu})$ on $\RR^d$ in which the metric takes the 
form
\be
g
=
\eta_{\mu\nu}dx^{\mu}dx^{\nu}
=
-(dx^0)^2+dx^idx^i,\quad i=1,...,d-1,
\label{Triangle}
\ee
with $(\eta_{\mu\nu})$ the Minkowski metric matrix written in (\ref{Flush}) for the case $d=4$. In the 
language of subsection 
\ref{subsecSpec}, the coordinates $(x^{\mu})$ are those of an inertial frame. Then the isometry group of this 
manifold is 
precisely the Poincar\'e group in $d$ dimensions, acting on $\RR^d$ according to (\ref{inhomBis}), and the 
stabilizer for this action is the Lorentz group $\mO(d-1,1)$. From this viewpoint, the property 
(\ref{invInt}) of 
invariance of the interval is simply the defining criterion for the transformation to be an isometry.

\subsection{Conformal transformations of the plane}
\label{confP}

To illustrate the definition of conformal transformations in the simplest possible case, let us consider the 
plane $\RR^2$ endowed with the Euclidean metric (\ref{Eu}). To make things technically simpler, we see 
$\RR^2$ as 
the complex plane $\CC$ and introduce a complex coordinate $z\equiv x+iy$, in terms of which the metric 
(\ref{Eu}) becomes $g=dzd\bar z$ (with $\bar z$ the complex conjugate of $z$). Then a generic diffeomorphism 
is a map
\be
\phi:\CC\rightarrow\CC:z\mapsto Z(z,\bar z),
\label{phiZ}
\ee
where the function $Z$ generally depends on both $z$ and $\bar z$. Demanding that $\phi$ be a conformal 
transformation imposes certain restrictions on this function, which we now work out.\\

Since $\phi$ maps $z$ on $Z$ and since the metric $g$ is just $dzd\bar z$, it is natural that the transformed 
metric be
\be
g'_z=\left.dZd\bar Z\right|_z,
\label{gprime}
\ee
where the subscript $z$ means that both sides are evaluated at the point $z$. (This is just the definition 
$g'=\phi^*g$ applied to (\ref{phiZ}).) Here the 
differential of $Z$ is
\be
\left.dZ\right|_z
=
\frac{\der Z}{\der z}dz+\frac{\der Z}{\der\bar z}d\bar z.
\nn
\ee
Plugging this expression (and its complex conjugate) in (\ref{gprime}), we find
\be
g'_z
=
\frac{\der Z}{\der z}\frac{\der\bar Z}{\der z}dz^2
+\frac{\der Z}{\der\bar z}\frac{\der\bar Z}{\der\bar z}d\bar z^2
+\left[\frac{\der Z}{\der z}\frac{\der\bar Z}{\der\bar z}+\frac{\der Z}{\der\bar z}
\frac{\der\bar Z}{\der z}\right]dzd\bar z.
\nn
\ee
According to the definition surrounding eq.~(\ref{g'''}), requiring that $\phi$ be a conformal transformation 
amounts to demanding that this expression be proportional 
to $g_z=dzd\bar z$. The terms involving $dz^2$ or $d\bar z^2$ must therefore vanish, which is the case 
if and only if
\be
\frac{\der Z}{\der\bar z}=0\quad\text{or}\quad\frac{\der Z}{\der z}=0.
\label{CauchyR}
\ee
In other words, the function $Z$ must depend either only on $z$, or only on $\bar z$. The latter possibility 
represents conformal transformations that change the orientation of $\RR^2$ (they map an angle $\theta$ on an 
angle $-\theta$), and we will discard them from now on. Thus, a diffeomorphism (\ref{phiZ}) is an 
orientation-preserving conformal transformation of $\RR^2$ provided $Z$ is a function of $z$ only, that is, 
a meromorphic function. 
Furthermore, locally, any such function is admissible\footnote{Upon writing $z=x+iy$ and $Z(z,\bar 
z)=X(x,y)+iY(x,y)$ where $X$ and $Y$ are real functions on the plane, the first equation in 
(\ref{CauchyR}) can be rewritten as the two Cauchy-Riemann equations for $X$ and $Y$.}.\\

Of course, this is not the end of the story since (\ref{phiZ}) must be a smooth bijection. This restricts the 
form of $Z(z)$ even further. To begin with, $Z(z)$ must be regular, so $Z(z)$ must be an analytic function
\be
Z(z)=A+Bz+Cz^2+Dz^3+\cdots.
\label{Zz}
\ee
The zeros of $Z(z)$ are the points that are mapped on the origin $Z=0$. Since $Z(z)$ must be an injective 
map, there can be only one such zero, say $z^*$. If this zero is degenerate, then the map $Z(z)$ will not be 
injective in a neighbourhood of that zero. (If $z$ is sufficiently close to $z^*$ and if $z^*$ is a zero of 
$Z(z)$ with 
order $n>1$, then $z$ is mapped by $Z$ on $n$ different points, and $Z(z)$ cannot be injective.) Thus, in 
(\ref{Zz}), the coefficients of all powers of $z$ higher than one must vanish, {\it i.e.}~$C=D=0$, {\it etc.} 
In other words, the function $Z(z)$ 
must be linear in $z$. Finally, requiring $Z(z)$ to be surjective imposes that the coefficient of 
the $z$-linear term be non-zero. We conclude that all conformal transformations of the plane are 
of the form
\be
Z(z)=az+b,\quad\text{with }a,\,b\in\CC\quad\text{and }a\neq 0.
\label{confPlane}
\ee
These transformations naturally split in three classes:
\begin{center}
\begin{tabular}{lll}
Translations & $z\mapsto z+b$, & $b\in\CC$;\\
Rotations & $z\mapsto e^{i\theta}z$, & $\theta\in\RR$;\\
Dilations & $z\mapsto e^{-\chi}z$, & $\chi\in\RR$.
\end{tabular}
\end{center}
We will see in the next subsection that these transformations may also be seen as (a subclass of) conformal 
transformations of the 
sphere.\\

Before going further, let us note one important detail: in deriving the set of conformal transformations 
(\ref{confPlane}), the fact that the metric $g$ on $\CC$ was the Euclidean metric (\ref{Eu}) played a minor 
role. Indeed, we would have obtained the exact same set of transformations for {\it any} metric of the form 
$\Omega(z,\bar z)dzd\bar z$ on the plane, since conformal transformations are blind to the multiplication of 
metrics by (positive) functions. The only crucial point was that the metric be proportional to $dzd\bar z$, 
since it is this property that led to the condition (\ref{CauchyR}). The further restrictions leading to 
(\ref{confPlane}), on the other hand, originated from topological (hence metric-independent) considerations. 
These observations will be 
essential in the following subsection.

\subsection{Conformal transformations of the sphere}
\label{confS}

We now turn to the main goal of this section: the classification of conformal transformations of the sphere 
$S^2$. By definition, the latter is a two-dimensional manifold consisting of all points 
with Cartesian coordinates $(x_1,x_2,x_3)$ in $\RR^3$ such that $x_1^2+x_2^2+x_3^2=r^2$, where $r$ is some 
fixed (positive) radius. (The notation $S^2$ is usually reserved for the unit sphere, with radius $r=1$, but 
here we will denote any sphere by $S^2$, regardless of its radius.)

\subsubsection{Stereographic coordinates}

The standard way to locate points on a sphere of radius $r$ relies on polar coordinates 
$\theta\in[0,\pi]$ and $\phii\in[0,2\pi)$ defined by
\be
\left\{
\begin{array}{ccl}
x_1 & = & r\sin\theta\cos\phii,\\
x_2 & = & r\sin\theta\sin\phii,\\
x_3 & = & r\cos\theta
\end{array}
\right.
\nn
\ee
for any point $(x_1,x_2,x_3)$ belonging to the sphere.

\begin{figure}[H]
\begin{center}
\includegraphics[width=0.40\textwidth]{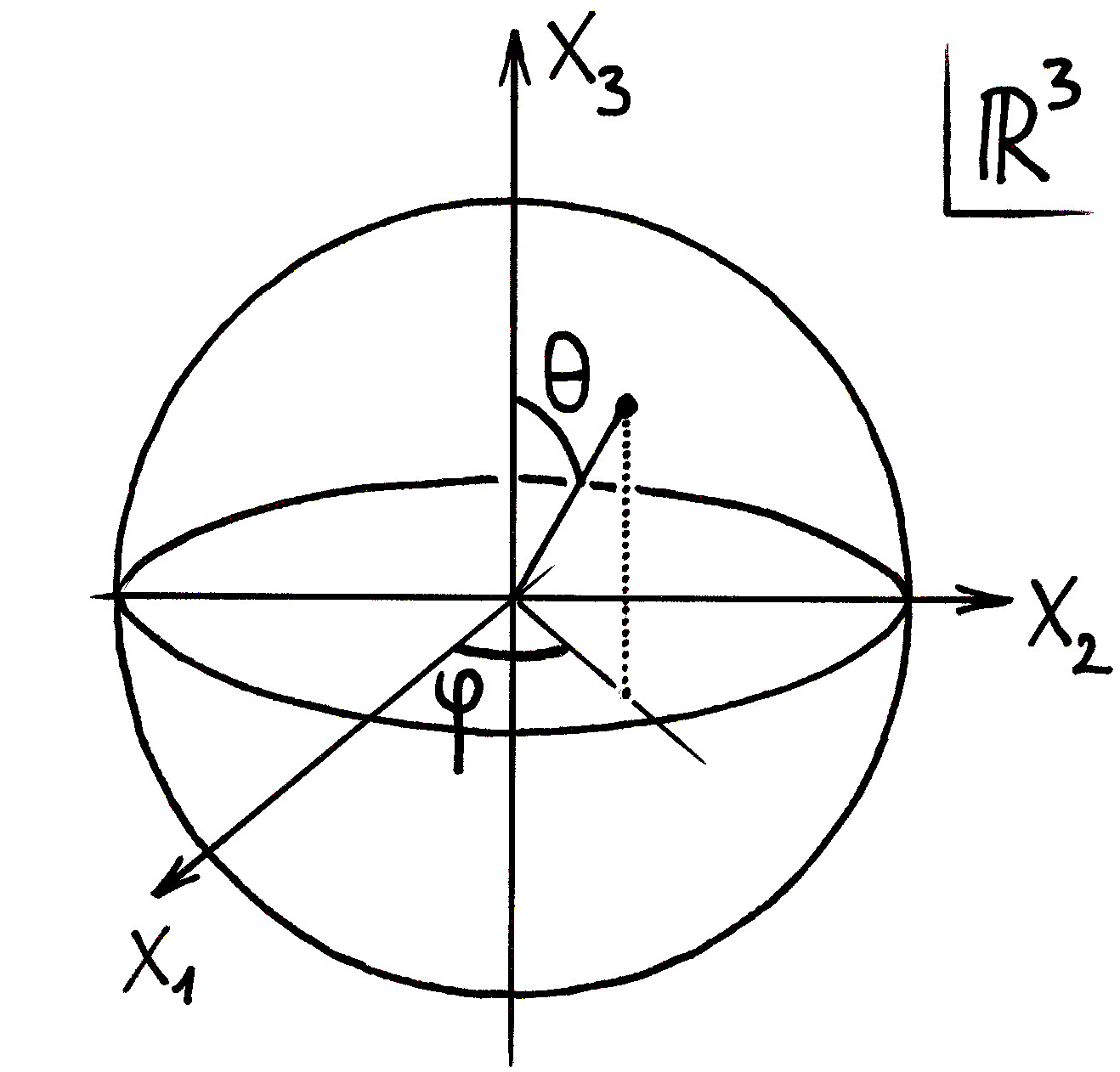}
\caption{A sphere embedded in $\RR^3$, and the polar coordinates $\theta$, $\phii$ used to 
label its points.}
\end{center}
\end{figure}

In the present case, however, it will be more convenient to use so-called {\it stereographic coordinates}, 
which will simplify the treatment of conformal transformations. These coordinates are defined as 
follows.
Consider a point $(x_1,x_2,x_3)$ on the sphere, different from the south pole $(0,0,-r)$. Then, 
there exists a unique straight line in $\RR^3$ passing through that point and the south pole. Explicitly, all 
points belonging to this line have coordinates $(y_1,y_2,y_3)$ of the form
\be
\left\{
\begin{array}{rcl}
y_1 & = & tx_1,\\
y_2 & = & tx_2,\\
y_3 & = & t(x_3+r)-r
\end{array}
\right.
\label{straightLine}
\ee
where $t$ is a parameter running over all real values. (The point corresponding to $t=0$ is the south pole, 
while $t=1$ corresponds to $(x_1,x_2,x_3)$.) The straight line so obtained crosses the equatorial plane 
$\left\{(x_1,x_2,0)|x_1,x_2\in\RR\right\}$ at exactly one point, called the {\it stereographic projection} of 
$(x_1,x_2,x_3)$ through the south pole. The coordinates $(x'_1,x'_2,0)$ of this projection are obtained by 
setting $y_3=0$ in eq.~(\ref{straightLine}), that is, by taking $t=r/(r+x_3)$, which gives
\be
x'_1=\frac{r\,x_1}{r+x_3},\quad x'_2=\frac{r\,x_2}{r+x_3}.
\label{steProj}
\ee
We will refer to $x'_1$ and $x'_2$ as the stereographic coordinates on the sphere. They can be combined into 
a single complex coordinate
\be
z\equiv \frac{x'_1+ix'_2}{r}=\frac{x_1+ix_2}{r+x_3},
\label{steProjBis}
\ee
which is related to polar coordinates through
\be
z=e^{i\phii}\tan(\theta/2).
\label{stePol}
\ee
For future reference, note that the inverse of relation (\ref{steProjBis}) gives $(x_1,x_2,x_3)$ in terms of 
$z$ and $\bar z$ as
\be
x_1=r\frac{z+\bar z}{1+z\bar z},\quad x_2=\frac{r}{i}\frac{z-\bar z}{1+z\bar z},
\quad x_3=r\frac{1-z\bar z}{1+z\bar z},
\label{invSte}
\ee
where we used the fact that $x_1^2+x_2^2+x_3^2=r^2$. Of course, we could have carried out a parallel 
construction by projecting points of the sphere on the 
equatorial plane through the north pole; this would have given formulas analogous to (\ref{steProj}) and 
(\ref{steProjBis}), but with $r+x_3$ replaced by $r-x_3$.

\begin{figure}[H]
\begin{center}
\includegraphics[width=0.50\textwidth]{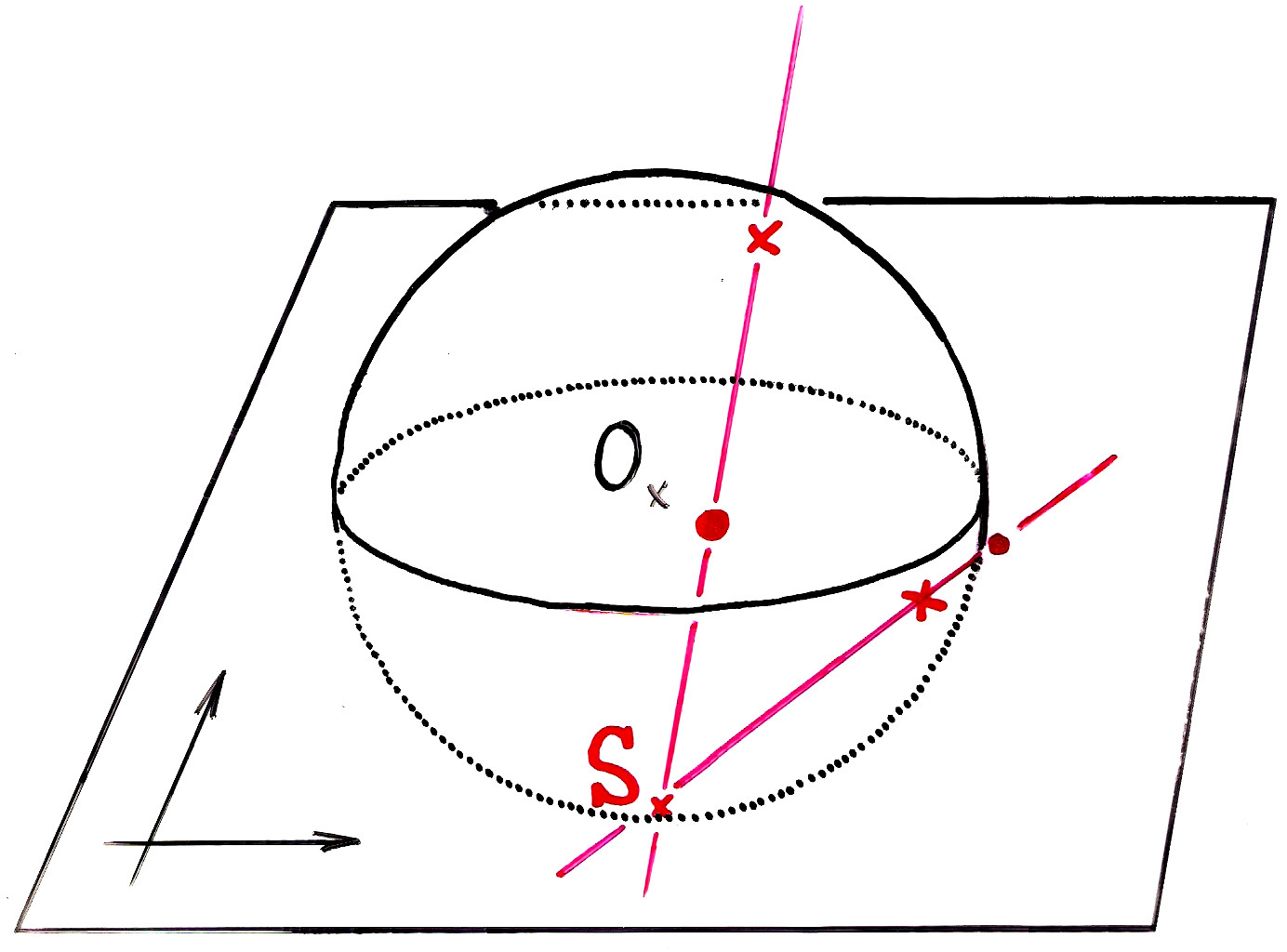}
\caption{The stereographic projection of a sphere centered at $O$ through the south pole $S$. The two red 
crosses are points belonging to the sphere; the projection maps those points on the two red dots on the 
equatorial plane, following straight lines parametrized by eq.~(\ref{straightLine}). The coordinates 
$(x'_1,x'_2,0)$ of the projection of a point $(x_1,x_2,x_3)$ are given by (\ref{steProj}).}
\end{center}
\end{figure}

The stereographic projection is a concrete illustration of the fact that a sphere is locally the same as a 
plane: any point on the sphere, other than the south pole, can be projected to the equatorial plane through 
the south pole. Points that are close to the north pole get projected near the origin $z=0$; the 
whole northern hemisphere is projected in the unit disc $|z|<1$, and the equator is left fixed by the 
projection, corresponding to the unit circle $|z|=1$. Points belonging to the southern hemisphere, on the 
other hand, are projected outside of the unit disc. In particular, points located near the south 
pole are 
projected far from the origin, at large values of $|z|$: as points get closer to the south pole, 
they get projected further and further away. In fact, one may view the infinitely 
remote point on the plane, the ``point at infinity'' $z=\infty$, as the projection of the south pole itself. 
(Of course, the actual projection of the south pole is ill-defined, so the point at infinity does not have a 
well-defined argument.) We conclude that the sphere is diffeomorphic to a plane, up to a point. More 
precisely,
\be
S^2\cong\CC\cup\{\text{point at infinity}\}=\CC\cup\{z=\infty\}.
\label{RiemannSph}
\ee
The representation of the sphere as a plane to which one adds the point at infinity is called the {\it 
Riemann sphere} \cite{Brown}. This relation hints that some of the 
results derived above for conformal transformations of the plane should be 
applicable to the sphere as well. In order to see concretely if this is the case, we first need to express 
the metric of a sphere in terms of the coordinate $z$.

\subsubsection{The metric on a sphere in stereographic coordinates}

The natural metric on a sphere follows from the definition of a sphere as a submanifold of $\RR^3$. 
Namely, endowing $\RR^3$ with the Euclidean metric $dx_1^2+dx_2^2+dx_3^2$, the metric on the sphere is simply
\be
g=\left.\left(dx_1^2+dx_2^2+dx_3^2\right)\right|_{x_1^2+x_2^2+x_3^2=r^2}.
\label{metricSh}
\ee
To express this metric in terms of stereographic coordinates, we use formula (\ref{steProjBis}), from which 
it follows that the differential of $z$ is
\be
dz=\frac{(dx_1+idx_2)(r+x_3)-(x_1+ix_2)dx_3}{(r+x_3)^2}.
\nn
\ee
On the sphere defined by $x_1^2+x_2^2+x_3^2=r^2$, the differentials of $x_1$, $x_2$ and $x_3$ satisfy the 
relation $x_1dx_1+x_2dx_2+x_3dx_3=0$, which can then be used to show that
\be
dzd\bar z
=
\frac{1}{(r+x_3)^2}\left.\left(dx_1^2+dx_2^2+dx_3^2\right)\right|_{x_1^2+x_2^2+x_3^2=r^2}.
\nn
\ee
In the last term of this expression we recognize the metric (\ref{metricSh}) on the sphere, whose expression 
in terms of $z$ thus becomes
\be
g_z=(r+x_3)^2dzd\bar z
=
\frac{4r^2}{(1+z\bar z)^2}dzd\bar z,
\label{Smiley}
\ee
where we used the third relation of (\ref{invSte}) to write $x_3$ as a function of $z$ and $\bar z$. This 
metric is position-dependent, since it explicitly depends on $z$. In fact, up to the factor $4r^2$, it is 
precisely the metric (\ref{gxyy}) that we took as an example earlier on, written in terms of $z=x+iy$. The 
only subtlety is that, in contrast to (\ref{gxyy}) where $x$ and $y$ only take finite values, 
expression (\ref{Smiley}) must be understood as a metric on the Riemann sphere, where $|z|$ may be infinite.\\

Crucially, the metric (\ref{Smiley}) is proportional to the Euclidean metric $dzd\bar z$, which implies that, 
as far as conformal transformations are concerned, we can simply 
repeat the derivation carried out in subsection \ref{confP} for the plane. More precisely, if we demand that 
a diffeomorphism $\phi:\CC\cup\{z=\infty\}\rightarrow\CC\cup\{z=\infty\}:z\mapsto Z(z,\bar z)$ be a conformal 
transformation, the arguments that led to (\ref{CauchyR}) remain true and the function $Z$ must depend either 
only on $z$, or only on $\bar z$. The latter choice 
corresponds to transformations that do not preserve the orientation of the sphere, so we will ignore them. 
Thus, any orientation-preserving conformal transformation of the sphere is a meromorphic function of the form 
$z\mapsto Z(z)$, and locally on the sphere this is all we can say.\\

Globally, of course, this is not yet the end of the story, since we must further require that the function 
$Z(z)$ 
be a diffeomorphism of the sphere $-$ that is, a diffeomorphism of the plane $\CC$ with the point at infinity 
added as in 
(\ref{RiemannSph}). This point will play a key role. Indeed, requiring that $Z(z)$ be regular on 
$\CC\cup\{z=\infty\}$ no longer means that $Z(z)$ is analytic as in (\ref{Zz}); rather, $Z(z)$ now may (and 
should) have at least 
one pole, at $z^*$ say, corresponding to the point that is mapped to the south pole $Z(z^*)=\infty$. 
Thus, $Z(z)$ 
should now be a rational function of the general form
\be
Z(z)=\frac{A+Bz+Cz^2+Dz^3+\cdots}{A'+B'z+C'z^2+D'z^3+\cdots},
\nn
\ee
where the roots of the numerator (resp.~denominator) correspond to the points that are mapped on the origin 
$Z=0$ (resp.~the point at infinity $Z=\infty$), {\it i.e.}~on the north pole (resp.~the south pole). Since 
$Z(z)$ must be an injective map, there must be one, and only one, point that is mapped to the north pole, and 
also exactly one other point that is mapped to the south pole. As in subsection \ref{confP}, this 
requires that both the numerator and the denominator be linear functions of $z$. We can thus write any 
orientation-preserving conformal transformation of the Riemann sphere as
\be
\boxed{Z(z)=\frac{az+b}{cz+d},}
\label{Zconf}
\ee
where $a$, $b$, $c$ and $d$ are complex numbers. Requiring this map to be surjective finally imposes that
\be
\text{det}\begin{pmatrix} a & b \\ c & d \end{pmatrix}\neq0.
\label{abcd}
\ee

This is the classification of conformal transformations of the sphere that we were looking for. Such 
transformations are also called {\it M\"obius transformations}. They obviously contain the set of conformal 
mappings (\ref{confPlane}) of the plane, so 
that translations of $z$, rotations and dilations also represent conformal transformations of the sphere. 
However, there is now an additional two-parameter family of transformations of the form
\be
z\mapsto -b^2/z,\quad b\in\CC^*,
\nn
\ee
corresponding to so-called special conformal transformations \cite{DiFran,Blum}. Such transformations map 
the north pole on the south pole, and vice-versa. Any conformal transformation of the sphere can be obtained 
as the composition of a special conformal transformation, a translation, a rotation and a dilation (possibly 
in a different order).\\

By construction, conformal transformations span a group, so it is worthwile to investigate the 
group structure of the set of M\"obius transformations. Clearly, formula (\ref{Zconf}) is blind to the 
overall 
normalization of the 
matrix in (\ref{abcd}), since multiplying all entries of the matrix by the same non-zero complex number leads 
to 
the same transformation (\ref{Zconf}). We can thus assume, without loss of generality, that the non-zero 
determinant (\ref{abcd}) is actually one, {\it i.e.}~that the matrix $\begin{pmatrix} a 
& b \\ c & d \end{pmatrix}$ belongs to $\SL(2,\CC)$. 
Furthermore, two matrices in $\SL(2,\CC)$ that differ only by their sign define the same conformal 
transformation, so the group of all non-degenerate transformations of the form (\ref{Zconf}) is actually 
isomorphic to the quotient
\be
\SL(2,\CC)/\ZZ_2.
\label{confSphereGroup}
\ee
In other words, according to (\ref{isomL}), the set of orientation-preserving conformal transformations of 
the 
sphere forms a group 
isomorphic to the connected Lorentz group in four dimensions! At this stage, this relation appears just as a 
coincidence of group theory: there seems to be no relation whatsoever between the M\"obius transformations 
(\ref{Zconf}) and the original definition of the Lorentz group as a matrix group acting on 
$\RR^4$. The purpose of the next section will be to show that this apparent coincidence
actually has a geometric origin, rooted in the structure of light-like straight lines in Minkowski 
space-time.

\subsection{An aside: conformal field theories in two dimensions}
\label{CFT}

In the two previous subsections we have seen that any (orientation-preserving) conformal transformation of a 
two-dimensional manifold with a conformally flat metric $\propto dzd\bar z$ can be written as a 
meromorphic function $z\mapsto 
Z(z)$. Demanding that $Z(z)$ be a bijection of the manifold imposes certain restrictions on the function 
$Z$, leading to (\ref{confPlane}) in the case of the plane, and (\ref{Zconf}) in the case of the sphere. 
However, in physical applications, it is often the case that ``global'' requirements such as bijectivity play 
a minor role. This is particularly true in the case of local quantum field theories\footnote{We will not 
explain the meaning of ``quantum field theory'' here. For an introduction, we refer for instance to the 
textbooks \cite{PeS,Weinberg}.}, whose 
properties are mostly determined by {\it local} (as opposed to global) considerations.\\

This feature is of central importance in the context of {\it conformal field theories} in two dimensions 
\cite{DiFran,Blum}. 
By definition, a conformal field theory in $d$ dimensions is a quantum 
field theory, defined on a $d$-dimensional manifold $\calM$ endowed with some metric $g$, that is invariant 
under conformal transformations of $\calM$. In 
the case $\calM=S^2$, with a metric proportional to $dzd\bar z$ in terms of stereographic coordinates, this 
leads to theories that are invariant under all M\"obius transformations (\ref{Zconf}). However, 
the actual 
set of infinitesimal symmetries of such theories ({\it i.e.}~symmetries found without taking global 
issues into account) turns out to be much, much larger than the finite-dimensional 
group (\ref{confSphereGroup}). Indeed, since global requirements such as bijectivity play a secondary role, 
conformal field theories in two dimensions turn out to be invariant under all transformations that can 
be written locally as $z\mapsto Z(z)$, where $Z(z)$ is {\it any} meromorphic function\footnote{At this 
point we should mention that proving conformal invariance of a quantum theory may be a subtle issue when 
the curvature of the underlying manifold does not vanish, due to the Weyl anomaly 
\cite{DiFran,Duff}. We will not discuss these subtleties here.}. This leads to an 
infinite-dimensional symmetry algebra that constrains such theories in a extremely powerful way \cite{BPZ}. 
For instance, when combined with an additional symmetry property called ``modular invariance'', conformal 
invariance of a two-dimensional field theory implies a universal formula for the entropy of that theory, 
known as the Cardy formula \cite{Cardy}. We will briefly return to conformal field theories in the conclusion 
of these notes.

\section{Lorentz group and celestial spheres}
\label{secLorentzSphere}

So far we have seen that the connected Lorentz group in four dimensions, 
$L_+^{\uparrow}=\SO(3,1)^{\uparrow}$, is isomorphic to the quotient $\SL(2,\CC)/\ZZ_2$. We have also 
shown that the latter arises as the group of orientation-preserving conformal transformations of the sphere. 
However, at this stage, the relation between the Lorentz group and the sphere appears as a mere 
coincidence. In particular, since the original Lorentz group is defined by its linear action on a 
four-dimensional space, there is no reason for it to have anything to do with 
certain non-linear transformations of a two-dimensional manifold such as the sphere. The purpose 
of this 
section is to establish this missing link. This will require first defining a notion of 
``celestial spheres'' in Minkowski space-time (subsection \ref{celestSph}), and then computing the action of 
Lorentz transformations on such spheres (subsection \ref{LorentzSphere}). Subsection \ref{boostSphere} is 
devoted to the analysis of the somewhat counterintuitive action of Lorentz boosts in terms of celestial 
spheres. Our approach is motivated by the notion of ``asymptotic symmetries'' in gravity 
\cite{Sachs}, and will rely on a specific choice of coordinates that simplifies the description of null 
infinity in Minkowski space-time. The results as such are well known, and 
coordinate-independent $-$ 
see for instance \cite{Penrose,HeldNewman}. It should be noted that similar 
relations exists also in other space-time dimensions. For instance, in $d=3$ dimensions, the connected 
Lorentz 
group $\SO(2,1)^{\uparrow}$ acts on the celestial {\it circles} at null infinity through projective 
transformations spanning a group $\SL(2,\RR)/\ZZ_2$, in accordance with the isomorphism (\ref{SOSL}). In this 
section, however, we will restrict our attention to the four-dimensional case.

\subsection{Notion of celestial spheres}
\label{celestSph}

As explained in section \ref{secSpec}, inertial observers in special relativity live in Minkowski space-time, 
which may be seen as the vector space $\RR^4$. Inertial coordinates consist of one time coordinate $t$ or 
$x^0=ct$, and three Cartesian space coordinates $(x^1,x^2,x^3)=(x^i)$. (Latin 
indices run over the values $1$, $2$, $3$.) Given 
such coordinates, there is a natural way to define a corresponding family of spheres. Namely, one may 
describe the 
spatial location of an event in terms of spherical, rather than Cartesian, coordinates, defined as
\be
r\equiv\sqrt{(x^1)^2+(x^2)^2+(x^3)^2}=\sqrt{x^ix^i}
\quad\text{and}\quad
z\equiv\frac{x^1+ix^2}{r+x^3},
\label{sphCoo}
\ee
where points on the sphere of radius $r$ are labelled by the stereographic coordinates (\ref{steProjBis}). 
In particular, the spatial coordinates $x^i=x_i$ take the form (\ref{invSte}) when expressed in terms of $z$ 
and $\bar z$. Note that the parity transformation defined by the matrix (\ref{P}) acts on the coordinate $z$ 
according to $z\mapsto-1/\bar z$.\\

For each non-zero $r$, we thus have a spatial sphere naturally associated with the inertial coordinates 
$(x^0,x^1,x^2,x^3)$. Since Lorentz transformations relate different sets of inertial coordinates through 
linear transformations $x\mapsto x'=\Lambda\cdot x$, one might hope that the rewriting of these 
transformations in terms of spherical coordinates (\ref{sphCoo}) could give rise to a ``nice'' action of the 
Lorentz group on spatial spheres, one that would make the relation to conformal transformations more 
apparent. This is not quite the case, however; roughly speaking, the ``celestial sphere'' that we actually 
wish to 
define should be the sphere that 
an inertial observer looks at. This is not achieved by the sphere of radius $r$ defined by (\ref{sphCoo}), 
because radial, ingoing light-rays emitted by the sphere need a non-zero time $r/c$ to get from 
the sphere to the origin at $r=0$ (which we take to be the position of the observer). Thus, we need to work a 
little more: we must somehow combine space and time 
coordinates so as to take into account the finite velocity of light, and define the celestial sphere seen by 
an observer at some moment of time as an object living in the past.\\

The argument just outlined hints at the right definition of what we would like to call a 
``celestial sphere''. Indeed, consider a radial light ray whose trajectory in space-time is described 
by\footnote{Here $z$ is of course the coordinate (\ref{sphCoo}) locating points on the sphere, and {\it not} 
the $z$ coordinate 
of a Cartesian coordinate system.}
\be
r=r_0\pm ct=r_0\pm x^0,\quad z=\text{constant}.
\nn
\ee
In particular, an ingoing radial light ray satisfies $r=r_0-ct$ 
where $r_0$ is some strictly positive initial radius. Along the trajectory of this light ray, the quantity
\be
u\equiv ct+r=x^0+r
\label{defRettime}
\ee
is constant; it represents the time at which the observer located at $r=0$ sees the light ray. One can 
thus parametrize the time of emission of ingoing radial light rays by the value of $u$. Instead of 
using 
coordinates $(x^0,r,z)$ to locate events in space-time, one may then use the {\it Bondi coordinates} 
$(u,r,z)$, in which case $u$ plays the role of a time coordinate and is called {\it advanced time}. The 
situation can be depicted as follows:

\begin{figure}[H]
\begin{center}
\includegraphics[width=0.50\textwidth]{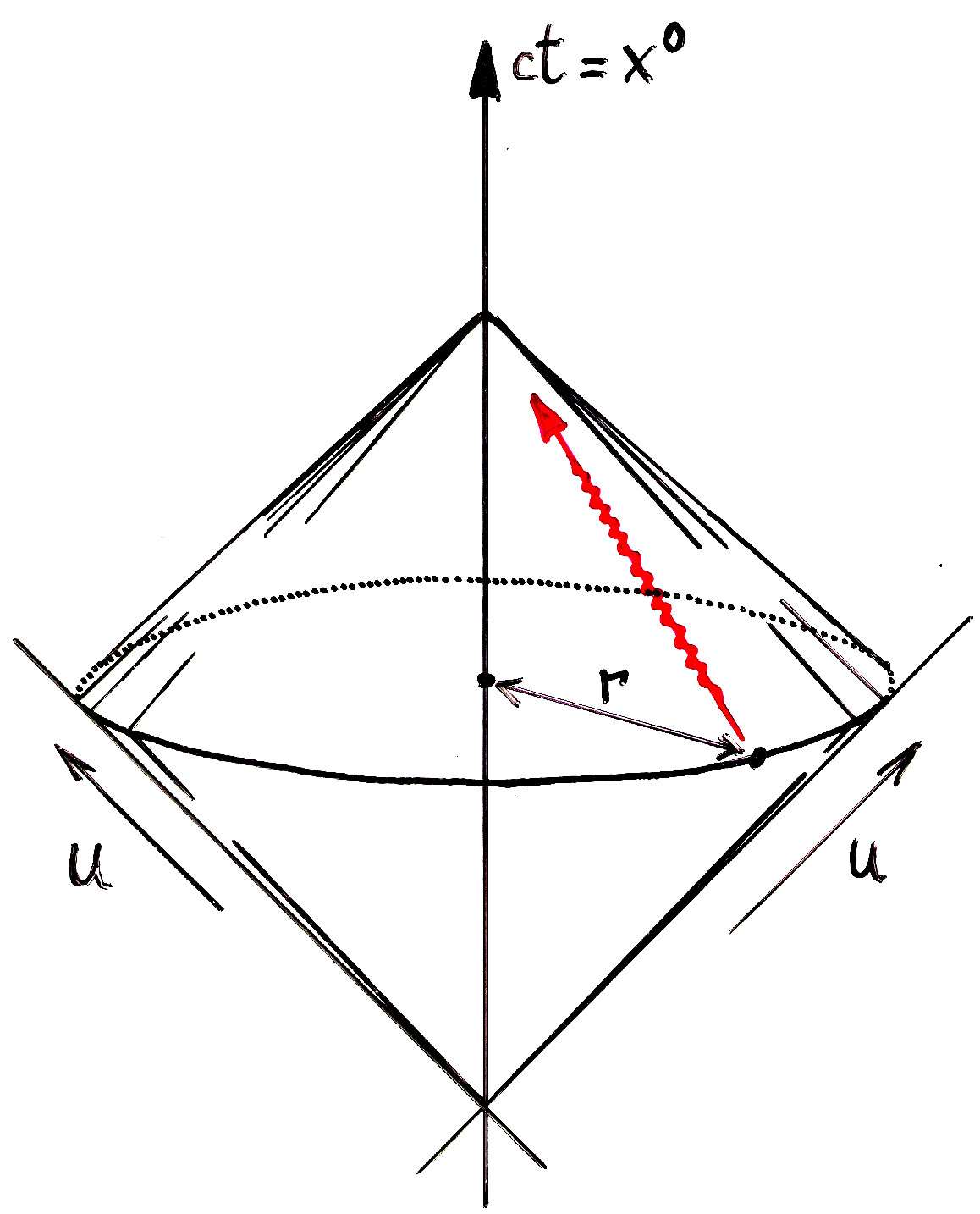}
\caption{The Bondi coordinates $u$ and $r$ in Minkowski space-time. The time coordinate $x^0=ct$ points 
upwards. The wavy red line represents an incoming radial light ray, emitted from some non-zero distance 
$r$ towards the observer located at $r=0$. The light ray moves along one of the generators of the light cone 
defined by $u=\text{cst}$. The figure represents three-dimensional space-time, so the circle of radius $r$ in 
this drawing would in fact be a sphere in our actual, four-dimensional, space-time. That sphere 
is spanned by the stereographic coordinate $z$ in (\ref{sphCoo}).\label{BondiC}}
\end{center}
\end{figure}

In terms of Bondi coordinates, a sphere at constant $r>0$ and constant $u\in\RR$ coincides with the sphere 
seen by an observer sitting at the origin $r=0$ at time $u$. We can then define the {\it celestial sphere} at 
time 
$u$ as the sphere located at an infinite distance, $r\rightarrow+\infty$, and at a fixed value of $u$. It is 
the sphere of all directions 
towards which an observer at $r=0$ can look \cite{Penrose}, the reason for the name ``celestial'' being 
obvious in that 
context. (Celestial spheres are also sometimes called ``heavenly spheres'' \cite{Baez}.) Less obvious 
is the fact that this definition is the one needed to match Lorentz transformations and M\"obius 
transformations, which will be the purpose of the next subsection. The region $\RR\times 
S^2$ spanned by the coordinates $u$ and $z$ at $r\rightarrow+\infty$ is called {\it past null infinity} 
\cite{Sachs}: it consists of events located at an infinite distance from the observer, and it can be reached 
from the line $r=0$ by following a past-directed null vector, that is, a vector whose norm squared vanishes 
with respect to the Minkowski metric (\ref{Triangle}).

\begin{figure}[H]
\begin{center}
\includegraphics[width=0.50\textwidth]{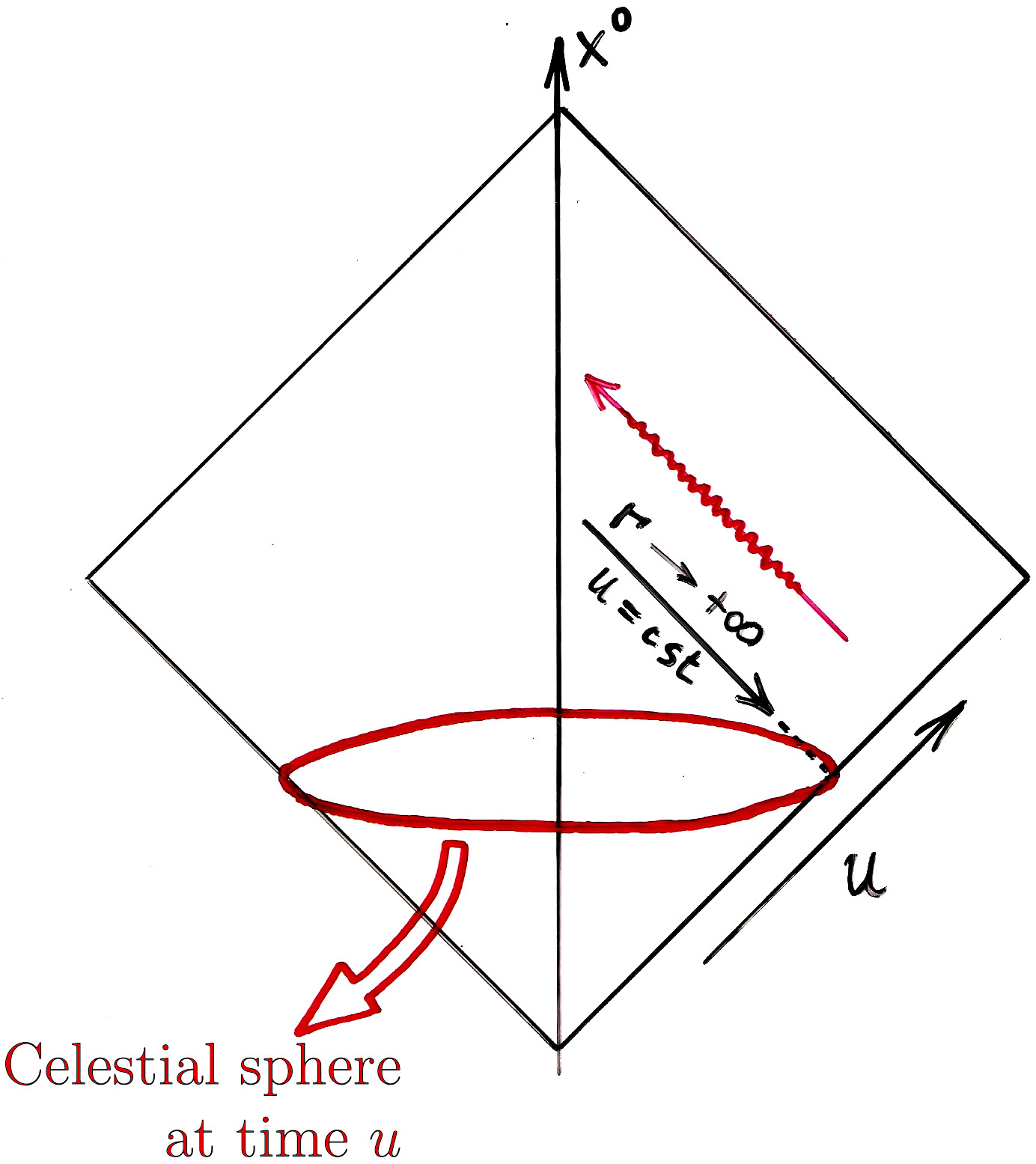}
\caption{A schematic representation of celestial spheres. As in Fig.~\ref{BondiC}, the time coordinate $x^0$ 
points upwards and the wavy red line represents an incoming radial light ray. The picture represents 
three-dimensional space-time. Accordingly, the red circle at the bottom of the image would really be a sphere 
$-$ a celestial sphere $-$ in our four-dimensional space-time. Past null infinity is the cone on the lower 
half of the image; it is a manifold $\RR\times S^2$ spanned by the advanced time $u$ and the stereographic 
coordinate $z$.}
\end{center}
\end{figure}

\paragraph{Remark.} In these notes we define celestial spheres by using a specific set of coordinates 
$(u,r,z)$ in Minkowski space-time. There also exists a different definition, according to which the celestial 
sphere associated with a point in space-time is the projective space of its (past) light-cone, that is, the 
set of past-directed null directions passing through that point \cite{Penrose}. In such terms, the celestial 
sphere at time $u$ that we defined above is the set of past-directed null directions through the point with 
Bondi coordinates $(r=0,u)$. (This could be any point in space-time since the Minkowski metric is invariant 
under all space-time translations.) This sphere can be thought of as the 
complex projective line $\mathbb{CP}^1\cong S^2$, and Lorentz transformations span the group 
$\SL(2,\CC)/\ZZ_2$ of its projective transformations $-$ which are nothing but M\"obius transformations when 
seeing $\mathbb{CP}^1$ as a sphere $S^2$ \cite{Penrose}. The advantage of this projective 
viewpoint is that it is manifestly coordinate-independent, but we will not adopt 
this approach here. (See, however, the end of subsection \ref{boostSphere}.)

\subsection{Lorentz transformations acting on celestial spheres}
\label{LorentzSphere}

The Lorentz group is defined as the set of linear transformations $x\mapsto x'=\Lambda\cdot x$ between 
coordinates of inertial observers in Minkowski space-time. In the previous subsection we have introduced new, 
non-inertial, Bondi coordinates $(u,r,z)$ associated with each choice of inertial coordinates $(x^{\mu})$. In 
order to find the action of the Lorentz group on Bondi coordinates, we must express both $x'$ and $x$ in 
terms of the associated Bondi coordinates, then rewrite the relation $x'=\Lambda\cdot x$ in Bondi 
coordinates and read off the Lorentz transformation properties of $(u,r,z)$. Since the relation between 
inertial coordinates and Bondi coordinates is non-linear, this procedure leads in general to cumbersome 
expressions for $(u',r',z')$ in terms of $(u,r,z)$ and of the matrix elements ${\Lambda^{\mu}}_{\nu}$ 
of a Lorentz transformation. Fortunately, we are not actually interested in the general relation between 
$(u',r',z')$ 
and $(u,r,z)$, but only in its limit $r\rightarrow+\infty$ with finite $u$. Provided Lorentz 
transformations preserve that limit (which is to be expected since they are linear 
in inertial coordinates), keeping $u'$ finite, they correspond to well-defined 
transformations of past null infinity.

\subsubsection{Transformation of the radial coordinate}
% \addcontentsline{toc}{subsubsection}{Transformation of the radial coordinate}

Let us begin by computing the transformation law of the radial coordinate $r$ under Lorentz transformations. 
By definition, the (square of the) radial coordinate $r'$ associated with the inertial coordinates 
$(x'^{\mu})$ is $r'^2=(x'^1)^2+(x'^2)^2+(x'^3)^2$. If now we assume that the coordinates $x'^{\mu}$ are 
obtained by acting on certain coordinates $x^{\mu}$ with 
a Lorentz transformation $\Lambda$, we have $x'^{\mu}={{\Lambda}^{\mu}}_{\nu}x^{\nu}$ and
\bea
r'^2
& = &
\left({\Lambda^{1}}_{0}x^0+{\Lambda^{1}}_{1}x^1+{\Lambda^{1}}_{2}x^2+{\Lambda^{1}}_{3}x^3\right)^2\nn\\
&   &
+
\left({\Lambda^{2}}_{0}x^0+{\Lambda^{2}}_{1}x^1+{\Lambda^{2}}_{2}x^2+{\Lambda^{2}}_{3}x^3\right)^2\nn\\
\label{45}
&   &
+
\left({\Lambda^{3}}_{0}x^0+{\Lambda^{3}}_{1}x^1+{\Lambda^{3}}_{2}x^2+{\Lambda^{3}}_{3}x^3\right)^2.
\eea
The next step consists in expressing the coordinates $x^{\mu}$ in terms of Bondi coordinates $(u,r,z)$ 
through relations (\ref{invSte}) and (\ref{defRettime}). Taking the limit $r\rightarrow+\infty$ while keeping 
$u$ and $z$ fixed, the only terms that survive in the 
parentheses are those proportional to $r$, which gives
\bea
r'^2
& = &
\left(
-{\Lambda^{1}}_{0}r
+
{\Lambda^{1}}_{1}r\frac{z+\bar z}{1+z\bar z}
+
{\Lambda^{1}}_{2}\frac{r}{i}\frac{z-\bar z}{1+z\bar z}
+
{\Lambda^{1}}_{3}r\frac{1-z\bar z}{1+z\bar z}\right)^2\nn\\
&   &
+
\left(-{\Lambda^{2}}_{0}r
+
{\Lambda^{2}}_{1}r\frac{z+\bar z}{1+z\bar z}
+
{\Lambda^{2}}_{2}\frac{r}{i}\frac{z-\bar z}{1+z\bar z}+{\Lambda^{2}}_{3}r\frac{1-z\bar z}{1+z\bar 
z}\right)^2\nn\\
&   &
+
\left(-{\Lambda^{3}}_{0}r
+
{\Lambda^{3}}_{1}r\frac{z+\bar z}{1+z\bar z}
+
{\Lambda^{3}}_{2}\frac{r}{i}\frac{z-\bar z}{1+z\bar z}+{\Lambda^{3}}_{3}r\frac{1-z\bar z}{1+z\bar z}\right)^2
+\calO(r).
\nn
\eea
(In particular, in that limit, we may replace $x^0$ by $-r$.) Here the terms of order $r$ outside the 
parentheses are subdominant with respect to the terms of order 
$r^2$ coming from the parentheses. As the final touch, we take $\Lambda$ to be a proper, orthochronous 
Lorentz transformation, {\it i.e.}~an element of the connected Lorentz group $L_+^{\uparrow}$. We can 
then express all entries ${\Lambda^{\mu}}_{\nu}$ of the Lorentz matrix in terms of complex numbers $a$, 
$b$, $c$, $d$ forming a matrix in $\SL(2,\CC)$, as in eq.~(\ref{THEhomomorphism}). Taking the square root to 
express $r'$ in terms of $r$, this gives the lenghty relation
\bea
r'
& = &
\frac{1}{2}\frac{r}{1+z\bar z}
\bigg[
\Big((\bar ac+\bar bd+a\bar c+b\bar d)(1+z\bar z)+(\bar ad+\bar bc+a\bar d+b\bar c)(z+\bar z)\nn\\
&   &
+(a\bar d-b\bar c-\bar ad+\bar bc)(z-\bar z)-(\bar ac-\bar bd+\bar ac-b\bar d)(1-z\bar z)\Big)^2\nn\\
&   &
-\Big((a\bar c+b\bar d-\bar ac-\bar bd)(1+z\bar z)-(\bar ad+\bar bc-a\bar d-b\bar c)(z+\bar z)\nn\\
&   &
+(a\bar d-b\bar c+\bar ad-\bar bc)(z-\bar z)+(\bar ac-\bar bd-a\bar c+b\bar d)(1-z\bar z)\Big)^2\nn\\
&   &
+\Big((|a|^2+|b|^2-|c|^2-|d|^2)(1+z\bar z)+(a\bar b-c\bar d+\bar ab-\bar cd)(z+\bar z)\nn\\
\label{BigStarr}
&   &
+(a\bar b-c\bar d-\bar ab+\bar cd)(z-\bar z)
-(|a|^2-|b|^2-|c|^2+|d|^2)(1-z\bar z)\Big)^2\bigg]^{1/2}+\calO(1).\quad\quad\quad\quad
\eea
At first sight, this expression looks terrible: if we were to expand all the products 
and parentheses in such 
a way that the argument of the square root be a sum of monomials in $z$, $\bar z$ and the numbers $a$, $b$, 
$c$, $d$ (and their complex conjugates), then the sum would contain about $3\times32!$ terms. Fortunately, as 
one can check by a straighforward but tedious computation, the terms of the sum conspire to give a very 
simple final answer:
\be
r'
=
r\cdot\frac{|az+b|^2+|cz+d|^2}{1+z\bar z}+\calO(1)
\equiv
r\cdot F(z,\bar z)+\calO(1).
\label{r'r}
\ee
This result shows that, as expected, Lorentz transformations do not spoil the limit $r\rightarrow+\infty$: 
the leading effect of Lorentz transformations on $r$ is just an angle-dependent rescaling by some function 
$F(z,\bar z)$. Furthermore, the 
occurrence of combinations such as $az+b$ and $cz+d$ is reminiscent of conformal transformations of the 
sphere, eq.~(\ref{Zconf}). The $\calO(1)$ terms in (\ref{r'r}) are subleading corrections that we will not 
write down, though they will play a role in the transformation law of advanced time.

\subsubsection{Transformation of advanced time}
% \addcontentsline{toc}{subsubsection}{Transformation of advanced time}

Having derived the transformation law of the radial coordinate $r$ (in the large $r$ limit), we now turn to 
the transformation of the remaining coordinates $u$ and $z$. We begin with the former; using the 
definition (\ref{defRettime}), we write
\be
u'
=
r'+ct'
=
r'+x'^{\,0}.
\label{u'}
\ee
We are now supposed to express $r'$ and $x'^{\,0}$ in terms of unprimed coordinates using their Lorentz 
transformation laws, then write everything in terms of 
$r$, $u$ and $z$, and read off the transformation law of $u$ at $r\rightarrow+\infty$. But there 
is a subtlety in carrying out this procedure. Namely, we have just seen that the transformation law 
of $r$ is $r'=F(z,\bar z)\cdot r+\calO(1)$; when plugged into (\ref{u'}), this implies that the 
transformation 
law of $u$ should 
read
\be
u'=F(z,\bar z)\cdot r+\calO(1)+x'^{\,0}.
\nn
\ee
Here the leading $\calO(r)$ term is dangerous: if there is nothing to cancel it, the limit 
$r\rightarrow+\infty$ of the transformation law of $u$ will be ill-defined. The only way to get rid of this 
term 
is to cancel it against the leading $\calO(r)$ term in the transformation law of $x^0$, which is given by
\bea
x'^{\,0}
& = &
{\Lambda^0}_0x^0+{\Lambda^0}_ix^i\nn\\
& \stackrel{\text{(\ref{THEhomomorphism})}}{=} &
\demi(|a|^2+|b|^2+|c|^2+|d|^2)(u-r)
+
\frac{r}{2}
\bigg[
-(a\bar b+c\bar d+\bar a b+\bar c d)\frac{z+\bar z}{1+z\bar z}\nn\\
&   &
-(a\bar b+c\bar d-\bar a b-\bar c d)\frac{z-\bar z}{1+z\bar z}
+(|a|^2-|b|^2+|c|^2-|d|^2)\frac{1-z\bar z}{1+z\bar z}
\bigg]
\nn
\eea
Plugging this in expression (\ref{u'}) and using (\ref{r'r}), one sees that the dangerous terms, proportional 
to $r$, cancel out! This means that the limit $r\rightarrow+\infty$ of the transformation law of $u$ is 
well-defined; in terms of observers, it means that if Alice and Bob are boosted with respect to each other 
and 
if Alice assigns a finite value $u$ of advanced time to some event, then Bob will assign to it a 
Lorentz-transformed value $u'$ which is also finite, though in general different, even if the event is 
located at an infinite distance from both Alice and Bob. This cancellation of potentially divergent terms in 
the transformation law of $u$ is actually the very reason why celestial spheres are defined at {\it null} 
infinity rather than spatial infinity. (The latter would correspond to the limit $r\rightarrow+\infty$ with 
the coordinate $x^0$ being kept finite instead of $u$, and in that case the large $r$ limit of the 
transformation law of $x^0$ would be ill-defined.)\\

Using the fact that the transformation law of $u$ is well-defined at $r\rightarrow+\infty$ (no $\calO(r)$ 
term), we know on dimensional grounds that
\be
u'=G(z,\bar z)\cdot u+\calO(1/r),
\label{u''}
\ee
where $G(z,\bar z)$ is some unknown function. (Indeed, $u$ and $r$ are the 
only Bondi coordinates with dimension of length, so, upon expanding the transformation law of $u$ in 
powers of $r$, the term of order zero in $r$ must be proportional to $u$.) In particular, the effect of 
Lorentz transformations on advanced time at infinity is just an angle-dependent rescaling, just as the 
transformation (\ref{r'r}) of the radial coordinate. The question, then, is to compute the rescaling 
$G(z,\bar z)$.\\

Just as for the radial coordinate, this calculation is straightforward but cumbersome: it requires evaluating the subleading term in the transformation law of $r$, 
which we did not derive in (\ref{r'r}) as it was included in the $\calO(1)$ terms. This subleading term can be found by Taylor-expanding the transformation law (\ref{45}) in powers of $1/r$, keeping $u$ fixed. In practice though, the entire computation can be circumvented thanks to the following argument: the Lorentz-invariant interval $c^2t^2-r^2=u^2-2ur$ reduces to $-2ur$ in the limit $r\rightarrow+\infty$, implying that the product $u\cdot r$ is Lorentz-invariant up to $\calO(1)$ corrections.\footnote{Incidentally, this is an alternative proof that the dangerous $\calO(r)$ term mentioned below (\ref{u'}) vanishes.} This, in turn, implies that the leading coefficient in the transformation law of $u$ is the inverse of the coefficient $F(z,\bar z)$ multiplying the radial coordinate in eq.\ (\ref{r'r}). Accordingly, the Lorentz transformation law of advanced time is \cite{Sachs}
\be
u'=u\cdot\frac{1+z\bar z}{|az+b|^2+|cz+d|^2}+\calO(1/r)
=
\frac{u}{F(z,\bar z)}+\calO(1/r).
\nn
\ee
This is of course of the announced form (\ref{u''}), with $G=F^{-1}$.

\subsubsection{Transformation of stereographic coordinates}
% \addcontentsline{toc}{subsubsection}{Transformation of stereographic coordinates}

The last case to be considered $-$ and the most interesting one for our purposes $-$ is the transformation 
law of the $z$ coordinate under Lorentz transformations in the large $r$ limit. The computation is more or 
less straightforward, as it only involves the dominant piece of the transformation law of $r$, displayed in 
(\ref{r'r}). To begin, one uses (\ref{sphCoo}) to write the transformed $z$ coordinate as
\be
z'=\frac{x'^1+ix'^2}{r'+x'^3},
\nn
\ee
where the primed coordinates on the right-hand side are obtained by acting with a Lorentz 
transformation $\Lambda$ on unprimed coordinates:
\be
z'
=
\frac{{\Lambda^1}_{\mu}x^{\mu}+i{\Lambda^2}_{\mu}x^{\mu}}{F(z,\bar z)\cdot r+{\Lambda^3}_{\mu}x^{\mu}}
+\calO(1/r).
\nn
\ee
(We used eq.~(\ref{r'r}) in writing this.) Expressing the $x^{\mu}$'s in terms of Bondi coordinates 
through relations (\ref{invSte}) and (\ref{defRettime}) and keeping $u$ finite, one finds
\be
\begin{array}{l}
z'=\\
\displaystyle{\frac{
({\Lambda^{1}}_{0}+i{\Lambda^2}_0)(1+z\bar z)
-
({\Lambda^{1}}_{1}+i{\Lambda^2}_1)(z+\bar z)
+
i({\Lambda^{1}}_{2}+i{\Lambda^2}_2)(z-\bar z)
-
({\Lambda^{1}}_{3}+i{\Lambda^2}_3)(1-z\bar z)}
{-|az+b|^2-|cz+d|^2
+
{\Lambda^{3}}_{0}(1+z\bar z)
-
{\Lambda^{3}}_{1}(z+\bar z)
+
i{\Lambda^{3}}_{2}(z-\bar z)
-
{\Lambda^{3}}_{3}(1-z\bar z)}}
\end{array}
\nn
\ee
up to $1/r$ corrections. Finally, 
writing the 
entries ${\Lambda^{\mu}}_{\nu}$ of the Lorentz matrix as in eq.~(\ref{THEhomomorphism}), both the numerator 
and the denominator of the last expression become certain complicated polynomials in $a$, $b$, $c$, $d$, $z$, 
and their complex conjugates. Fortunately,  many terms in these polynomials cancel against each other, 
leading to a simple expression of $z'$ in terms of $z$:
\be
\boxed{z'=\frac{az+b}{cz+d}+\calO(1/r).}
\label{ZconfBis}
\ee
This is precisely the standard expression of M\"obius transformations, eq.~(\ref{Zconf}): Lorentz 
transformations coincide with conformal transformations of the celestial sphere. This is the result we wanted 
to prove.

\paragraph{Remark.} In deriving (\ref{ZconfBis}), our choices of conventions played an important role. 
Indeed, we could have defined the homomorphism $f:\SL(2,\CC)\rightarrow L_+^{\uparrow}$ of 
subsection \ref{LSL} by acting on Hermitian matrices of the form (\ref{XXX}), but with different signs in 
front of the components $x^{\mu}$. (The standard choice \cite{HenneauxGroupe} would correspond to changing 
the sign in front of $x^1$.) This would have led to a different expression of the homomorphism 
(\ref{THEhomomorphism}), which in turn would have given a different formula for the action of Lorentz 
transformations on celestial spheres. For instance, the terms $az+b$ and $cz+d$ would then be replaced by 
combinations such as $az-b$ and $-cz+d$, or $\bar a z+\bar b$ and $\bar c z+\bar d$, or other variations on 
the same theme. But 
the statement that Lorentz transformations act as conformal transformations 
on celestial spheres remains true regardless of one's choices of conventions. Furthermore, the physical 
effect of such conformal transformations is also 
convention-independent; we will see an illustration of such a physical (actually, optical) effect in the next 
subsection.

\subsection{Boosts and optics}
\label{boostSphere}

It is worthwile to analyse the transformation law (\ref{ZconfBis}) for certain 
specific examples of Lorentz transformations. Namely, recall that the homomorphism (\ref{THEhomomorphism}) 
allowed us to represent rotations and boosts by the $\SL(2,\CC)$ matrices (\ref{ExRot}) and 
(\ref{ExBoost}), respectively. We can then plug these matrices in eq.~(\ref{ZconfBis}) and interpret the 
resulting formula as the conformal transformation of the celestial sphere that corresponds to a change 
of frames between two inertial observers, say Alice and Bob. For instance, a rotation by $\theta$ along 
Alice's $x^3$ axis 
corresponds to a rotation of the sphere represented by $z\mapsto e^{-i\theta} z$. This is not surprising: if 
the frames of Alice and Bob are rotated with respect to each other, it is 
obvious that their respective celestial spheres will be identical, up to a rotation.

\subsubsection{Boosts, optics and the {\textit{Millenium Falcon}}}
% \addcontentsline{toc}{subsubsection}{Boosts, optics and the {\textit{Millenium Falcon}}}

A more interesting phenomenon occurs when Bob is boosted with respect to Alice, with rapidity $\chi$ say. 
The stereographic coordinate $z'$ of 
the celestial sphere seen by Bob is then related to the coordinate $z$ of the sphere seen by Alice 
according to
\be
z'=e^{-\chi} z.
\label{surprise}
\ee
Let us take $\chi>0$ for definiteness, {\it i.e.}~let us assume that Bob moves in the direction of positive 
$x^3$, towards the north pole of the sphere, located at $z=0$. Then eq.~(\ref{surprise}) tells us that, 
although Alice and Bob are looking at the same celestial sphere, the points of Bob's sphere are all 
pulled closer to the north pole than the points of Alice's sphere. If we imagine that shining stars are 
glued to the celestial sphere, then the stars seen by Bob are grouped closer to the north pole 
({\it i.e.}~closer to the direction of Bob's motion) than those seen by Alice.

\begin{figure}[H]
\begin{center}
\includegraphics[width=0.30\textwidth]{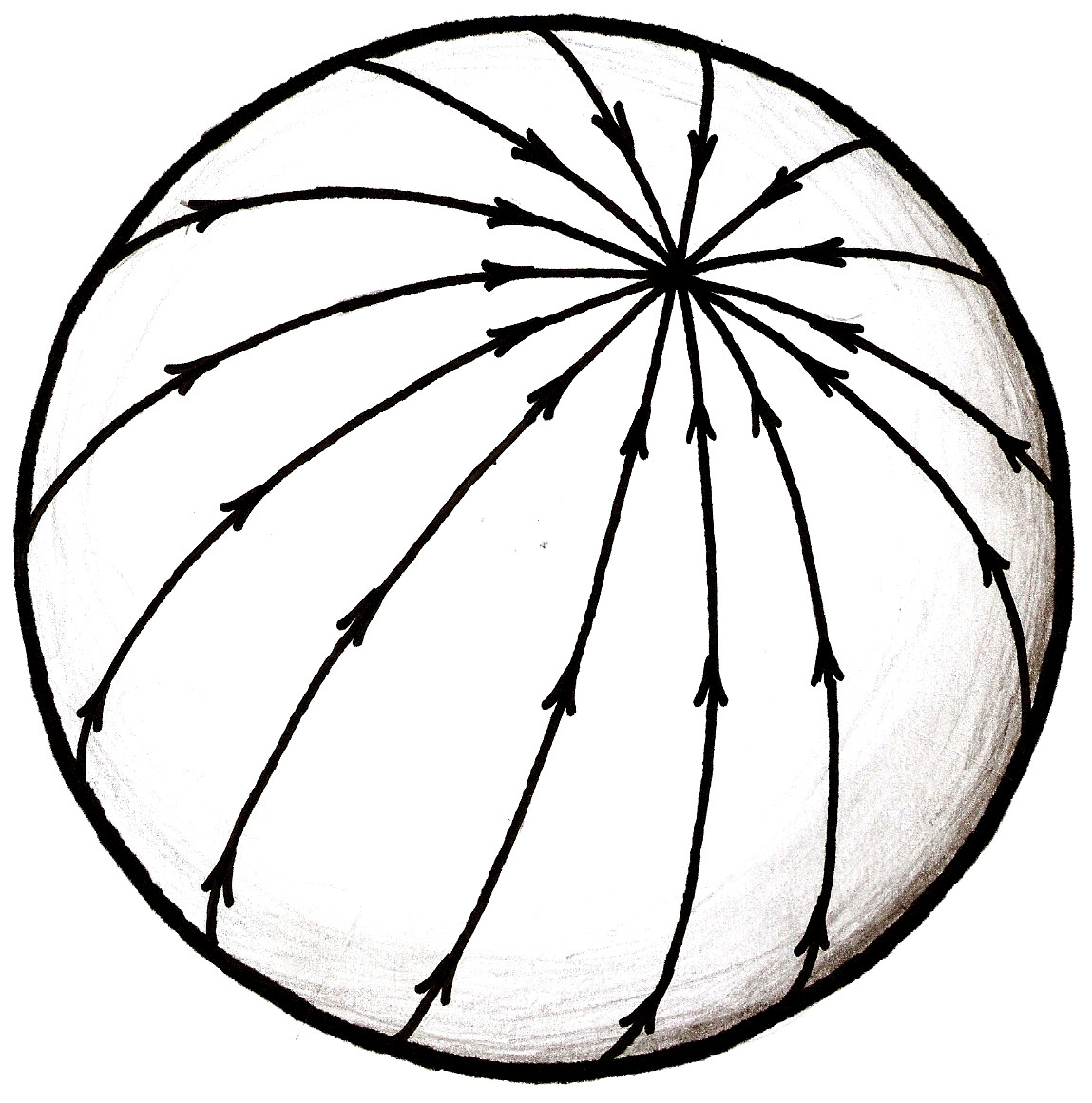}
\caption{The conformal transformation of the celestial sphere corresponding to a boost towards the north 
pole: all points of the sphere are dragged along the arrows, towards the north pole. Equivalently, all 
points are dragged away from the south pole (which is not visible in this picture). In terms of stereographic 
coordinates obtained by projection through the south pole, this transformation corresponds to a contraction 
of 
the Riemann sphere, $z\mapsto e^{-\chi}z$ with $\chi>0$.}
\end{center}
\end{figure}

This result is somewhat counterintuitive, if we base our intuition on our habit of objects flowing 
past us when driving on the highway. Roughly speaking, one would expect that boosting in a given direction 
should make objects spread away from that direction. This intuition is well illustrated in the movie {\it 
Star Wars Episode IV: A New Hope} \cite{StarWars}. In the screenshot reproduced below, Han Solo and Chewbacca 
are sitting in the cockpit of the {\it {\it Millenium Falcon}} spaceship and have just switched on the 
``hyperspace'' 
mode $-$ they are accelerating straight ahead. This acceleration corresponds to a continuous family 
of boosts in the direction of the acceleration. In the picture, these boosts are represented by stars flowing 
{\it away} from the direction of the motion, exactly as dictated by the naive, intuitive expectation just 
described:
\begin{figure}[H]
\begin{center}
\includegraphics[width=0.80\textwidth]{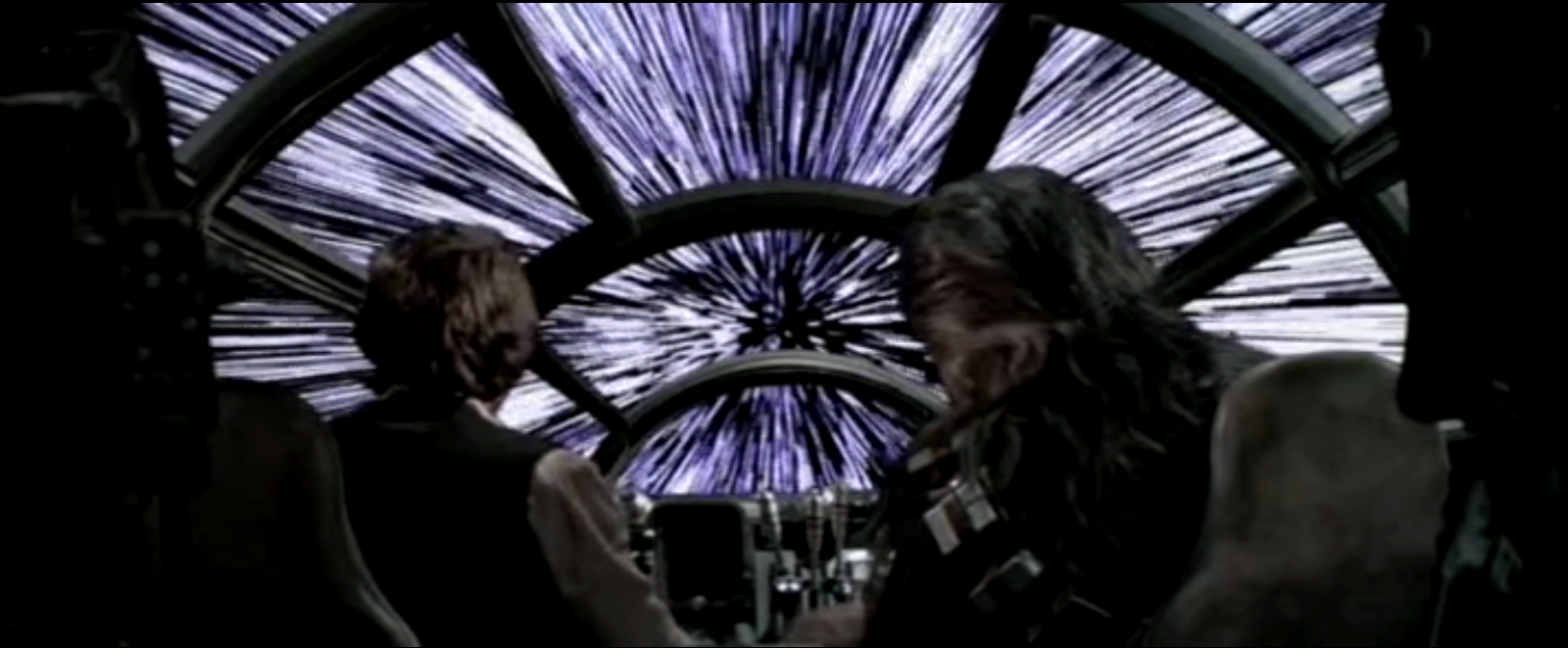}
\end{center}
\end{figure}
Formula (\ref{surprise}) tells us that this representation of the ``jump to lightspeed'' is {\it 
wrong}: what Han Solo and Chewbacca should really see is a {\it contraction} of the sphere at which they are 
looking, towards the direction of their acceleration. In other words, as long as the stars are far enough 
from the 
observer undergoing a boost, they should cluster close to the direction of the boost rather than flow away 
from it.\\

There is of course a subtlety in this argument: our intuition of objects flowing past us when we drive on the 
highway is obviously correct, so how come it contradicts the result (\ref{surprise})? The answer 
is that formula (\ref{surprise}) holds {\it only} in the limit $r\rightarrow+\infty$, when 
the sphere we are talking about is located at an infinite distance from the observer. In that limit, the 
observer's motion does not affect its distance to a point on the celestial sphere; in particular, all points 
on the 
sphere remain at an infinite distance from the observer, and there is no way they could flow past him. 
In real-world applications, however, all objects are necessarily located at some {\it finite} distance, 
in which case the corrections of order $1/r$ neglected in (\ref{ZconfBis}) become relevant. In 
particular, when Bob is moving with respect to Alice, the relation between his Bondi coordinates and 
those of Alice involves some time-dependent factors in the $1/r$ corrections. These corrections 
imply that the objects seen by Bob (be it stars, or cows on the side of the highway) do indeed flow past him 
when he gets close enough to them. In this sense the picture of the {\it Millenium Falcon} cockpit shown 
above is 
not completely wrong. Still, for stars located sufficiently far from Han Solo and Chewbacca, 
the $1/r$ corrections in (\ref{ZconfBis}) are negligible and the optical effect described by the contraction 
(\ref{surprise}) is valid.
% 
% \footnote{To be a little more precise, the large $r$ approximation is valid 
% provided the boosted observer looks at the celestial sphere during a proper time interval $\Delta\tau$ that 
% is much smaller 
% than $r/\gamma(v)c$, where $v$ is the velocity of the boost. When this condition holds, the corrections of 
% order $c\gamma(v)\Delta\tau/r=c\Delta t/r$ to the above formulas are 
% indeed negligible. For example, if Han Solo and Chewbacca are looking at stars located about 10 light-years 
% away and if we assume that their acceleration boosts them up to a velocity $v=0.99 c$ with respect to their 
% original velocity, then $\gamma(v)\simeq 10$ and the $1/r$ corrections are negligible as long as the 
% observation time $\Delta\tau$ measured from the cockpit of the {\it Millenium Falcon} is much smaller than, 
% say, a year.}

\subsubsection{Subtleties}
% \addcontentsline{toc}{subsubsection}{Subtleties}

While $1/r$ corrections are the most obvious source of modifications to the result (\ref{surprise}), there 
are several other caveats in trying to apply Lorentz transformations to realistic situations 
such as the jump to lightspeed in the {\it Millenium Falcon}. The first is the fact that the motion of the 
spaceship during the jump is actually accelerated, so that Han Solo and Chewbacca are 
definitely {\it not} inertial observers! This does not prevent us from guessing what should happen: roughly 
speaking, accelerated motion may be seen as an infinite sequence of infinitesimal boosts, so 
if (\ref{surprise}) remains valid for each infinitesimal boost, one expects the celestial 
sphere seen by an accelerated observer to undergo a time-dependent contraction (in the direction of 
acceleration), with a scaling factor 
that gets smaller and smaller as time goes by. More precisely, since rapidity is the integral (\ref{BigStar}) 
of proper acceleration, the naive application of (\ref{surprise}) to accelerated motion predicts that an 
accelerated observer, looking at the celestial sphere in the direction of his acceleration, should see a 
proper-time-dependent contraction with a scaling factor $\exp[-I(s)/c]$, where $I(s)$ is the integral 
(\ref{Nabla}) of proper acceleration over proper time.\\

The potential problem with this expectation is that special relativity was established for inertial observers 
from the very beginning, so one might fear that 
acceleration invalidates the application of special-relativistic techniques to the {\it Millenium 
Falcon}. Fortunately, there is in principle no obstacle in describing accelerated observers in special 
relativity 
\cite{MTW}. For example, Thomas precession is a well known special-relativistic effect that applies to 
such observers \cite{MTW,Rindler}, and it is precisely derived by thinking of accelerated motion as a 
sequence of 
infinitesimal boosts. The only issue is that the reference frames associated with 
accelerated observers\footnote{Such reference frames are usually defined by attaching a Fermi-Walker 
transported 
tetrad to the world-line of the observer, then using this tetrad to define a local coordinate system; see 
\cite{MTW}, chap.~6.} are not global coordinate systems $-$ they do not cover the whole of space-time. This 
is related to the existence of horizons: for instance, a uniformly accelerated observer in Minkowski 
space-time $-$ a Rindler observer $-$ cannot receive light rays coming from behind his future horizon 
\cite{MTW}. Thus, since 
our definition of celestial spheres relied on the limit $r\rightarrow+\infty$ in Bondi coordinates, one 
may wonder 
whether acceleration invalidates our approach, as the limit may be ill-defined. We will not attempt to 
address 
this issue here, but we will rederive formula (\ref{surprise}) in a local way at the end of this section. 
This will confirm that the optical effects of boosts on the celestial sphere do not actually rely on a large 
$r$ limit, as already mentioned at the end of subsection \ref{celestSph}. In particular, the local nature of 
the derivation implies that it remains valid for an accelerated observer in the sense that acceleration 
deforms the shape of light-cones centered on the observer $-$ as is of course well-known in general 
relativity. Whether this deformation can be seen by an ``asymptotic'' computation analogous to the one 
explained above is another matter, which we will not discuss.\\

A second subtlety to be considered is the fact that the light seen by Han Solo and Chewbacca during the jump 
to lightspeed is blue-shifted due to the Doppler effect. As the velocity of the {\it Millenium Falcon} 
increases, 
the frequency of the light rays hitting the observers inside the cockpit increases as well. Eventually, the 
increase in frequency should become so 
high that the stars actually become invisible $-$ the starlight seen by the pilots of the spaceship has 
reached the ultraviolet region. Thus, the stars seen by Han Solo and Chewbacca not only move in the 
direction of acceleration, but they also change colour, becoming blue, then purple, then 
invisible\footnote{Strictly 
speaking, they become invisible to a human eye $-$ to the best of our knowledge, it is not known whether 
Wookies are able to see a broader spectrum of electromagnetic radiation than human beings: while the stars 
definitely become invisible to Han Solo, they might still be visible to Chewbacca.}.\\

This blue shift applies of course to any electromagnetic radiation reaching the observers 
inside the cockpit. In particular, it applies to the cosmic microwave background radiation\footnote{Here we 
are assuming that the {\it Star Wars} took place in a universe that started off with a Big Bang.}. Thus, at 
sufficiently high velocities, the background radiation should reach the visible spectrum and the actual 
picture seen 
from the cockpit of the {\it Millenium Falcon} should include a fuzzy disc of light centered around the 
direction of the motion \cite{Argyle,Connors}. Upon taking this 
effect into account and recalling that most stars become invisible because of the blue shift, one concludes 
that the actual landscape seen by Han Solo and his hairy companion is indeed far, far away from the image 
shown in the movie.

\subsubsection{A local derivation}
% \addcontentsline{toc}{subsubsection}{A local derivation}

We have just seen that boosting an observer in a given direction affects his celestial sphere by 
contracting all points of the sphere towards that direction. Given the counterintuitive nature of 
this optical phenomenon, it is worthwile to rederive it using a different technique. Namely, consider two 
inertial observers, Alice and Bob, using inertial 
coordinates $(x^{\mu})$ and $(x'^{\mu})$ respectively. We take Bob's coordinates to be boosted, with rapidity 
$\chi$, with respect to those of Alice. For definiteness, we will assume that the boost takes place along the 
$x^1$ direction, so that the relation between Bob's coordinates and Alice's coordinates is 
$x'^{\mu}={\Lambda^{\mu}}_{\nu}x^{\nu}$, with $\Lambda$ the matrix (\ref{boostBis}). 
Now suppose Alice and Bob both see one incoming photon, whose energy-momentum vector is 
$p=(E,-E\cos\theta,-E\sin\theta,0)$ in Alice's coordinates, and $p'=(E',-E'\cos\theta',-E'\sin\theta',0)$ in 
Bob's coordinates. (Here $E$ and $E'$ are the photon's energy in Alice's and Bob's frames, 
respectively.) The angle $\theta$ (resp.~$\theta'$) is the angle between the photon's direction and the axis 
$x^3=x'^3$ in Alice's (resp.~Bob's) frame. The question is: what is the relation between $\theta'$ 
and $\theta$?\\

Since 4-momentum transforms under boosts just as standard inertial coordinates do (the 
energy-momentum vector is a ``four-vector''), we know that 
the photon's 4-momentum in Bob's and Alice's coordinate systems are related by $p'=\Lambda\cdot p$. 
Explicitly, this means that
\be
\begin{pmatrix}
E'\\
-E'\cos\theta'\\
-E'\sin\theta'\\
0
\end{pmatrix}
=
\begin{pmatrix}
\cosh\chi & -\sinh\chi & 0 & 0 \\ -\sinh\chi & \cosh\chi & 0 & 0 \\ 0 & 0 & 1 & 0 \\ 0 & 0 & 0 & 1
\end{pmatrix}
\cdot
\begin{pmatrix}
E\\
-E\cos\theta\\
-E\sin\theta\\
0
\end{pmatrix}.
\nn
\ee
From this we read off $\tan\theta'=\sin\theta/(\sinh\chi+\cosh\chi\,\cos\theta)$, which can be rewritten in 
terms of half angles as
\be
\tan(\theta'/2)-\frac{1}{\tan(\theta'/2)}
=
e^{-\chi}\tan(\theta/2)-\frac{1}{e^{-\chi}\tan(\theta/2)}.
\nn
\ee
This is a quadratic equation for $\tan(\theta'/2)$ as a function of $\tan(\theta/2)$. The solution that 
ensures $\theta'=\theta$ when $\chi=0$ is the simplest one,
\be
\tan(\theta'/2)
=
e^{-\chi}\tan(\theta/2).
\label{theta'}
\ee
Since here $\theta$ and $\theta'$ should be thought of as standard azimuthal coordinates on the sphere in 
unprimed and primed coordinate systems, we can relate them to the stereographic coordinate $z$ 
through relation (\ref{stePol}). The result (\ref{theta'}) thus coincides with the 
contraction (\ref{surprise}), as it should. In particular, provided $\theta$ is in the first quadrant 
(between $0$ and $\pi/2$), $\theta'$ is smaller than $\theta$ when the rapidity $\chi$ is 
positive. (Conversely, when $\theta$ is larger than $\pi/2$, then $\theta'$ is larger than $\theta$, 
corresponding to the fact that points of the celestial sphere located in the direction opposite to the boost 
undergo a dilation.)\\

The important difference between this computation and the one based on Bondi coordinates is the fact that 
here 
we never needed to take a ``large $r$'' limit. The result (\ref{theta'}) is 
valid locally, for any boosted observer detecting a light ray. This implies in particular that an 
accelerated observer looking in the direction of his/her acceleration should see a time-dependent contraction 
of the celestial sphere, just as mentioned above for the case of Han Solo and Chewbacca.

\section{Conclusion}
\label{secCon}

Let us take a look back at what we have done. We have seen how the Lorentz group arises as the set of 
homogeneous coordinate transformations between inertial observers in Minkowski space-time. Since it consists 
of linear transformations, it can be represented in terms of matrices. We have then shown that (the 
maximal connected subgroup of) this matrix group is isomorphic to $\SL(2,\CC)/\ZZ_2$ $-$ a type of relation 
that also 
occurs in other space-time dimensions. As observed in section \ref{secConf}, the group $\SL(2,\CC)/\ZZ_2$ 
also arises, somewhat coincidentally, as the group of conformal transformations of the sphere. The question, 
then, was whether there exists a relation between the action of Lorentz transformations on space-time and that 
of M\"obius transformations on a sphere. We answered this question positively, by showing that the 
difference between the celestial spheres seen by two inertial observers whose coordinates are related by a 
Lorentz transformation is precisely a conformal transformation. Finally, we used this relation to 
discuss the slightly unexpected optical phenomenon associated with boosts: we saw that an observer boosted in 
a given direction sees the stars of his/her celestial sphere being dragged towards that direction. This 
result is applicable, in particular, to the jump to lightspeed as seen from the cockpit of the 
{\it Millenium Falcon}.\\

As emphasized in the introduction, the surprising aspect of the relation between Lorentz and conformal 
transformations is the fact that it links the action of a group on a 
four-dimensional space to its action on a two-dimensional manifold. Of course, from a mathematical viewpoint 
there is nothing wrong with that, but from an intuitive viewpoint it is not {\it a priori} obvious that such 
a connection has any physical meaning $-$ {\it i.e.}~that this relation can actually be seen in a concrete 
experiment, such as accelerating in a spaceship. The purpose of these notes was to unveil that meaning, 
which is well known in the literature but perhaps less well known to undergraduate students following 
a course in special relativity, group theory, or even general relativity.\\

In fact, part of the motivation for these lectures was that the idea of relating some space to a 
lower-dimensional subspace is closely connected to certain recent developments in the study of (quantum) 
gravity, all encompassed under the general name of {\it holography}. Recall that a hologram is a 
two-dimensional surface that produces a three-dimensional image $-$ such an optical device is 
typically found on credit cards or banknotes. The idea of holography in 
quantum gravity \cite{tHooft,Susskind,Maldacena,Witten,Gubser} roughly states that there exists a 
correspondence (and in certain regimes an actual equivalence) between a gravitational system in $d$ 
space-time 
dimensions and some quantum theory living on a lower-dimensional subspace of the gravitational 
system $-$ one says that the two theories are ``dual'' to each other. In particular, according to this idea, 
the four-dimensional world that we see around us might be a ``hologram'' of some lower-dimensional 
theory. This correspondence is motivated by countless computations matching quantities evaluated on the 
high-dimensional, gravitational side, to some other quantities evaluated on the low-dimensional side; the 
interested reader may consult the abundant literature on the subject.\\

The modest result derived in these notes may be seen as a remnant of the holographic principle: we have shown 
that Lorentz invariance in four dimensions becomes conformal invariance in two dimensions upon focusing on 
celestial spheres. In fact, this feature is only part of a much larger construction, that is still under 
study today. Indeed, it was shown in the sixties by Bondi, van der Burg, Metzner and Sachs \cite{BvM,Sachs} 
that the natural symmetry group of four-dimensional ``asymptotically Minkowskian'' 
space-times is an infinite-dimensional extension of the Poincar\'e group, known nowadays as 
the Bondi-Metzner-Sachs (BMS) group. The transformations of space-time generated by this group precisely act 
on ``null infinity'', the region $r\rightarrow+\infty$ that we used in section \ref{secLorentzSphere} to 
define celestial spheres, and extend the natural action of Poincar\'e transformations on that region. In the 
holographic context, the BMS group is to be interpreted as the symmetry 
group of the would-be (as yet conjectural) dual field theory; the latter, if it exists, is expected to be 
some version of a conformal field theory (recall the brief discussion of subsection \ref{CFT}), since 
Lorentz transformations are part of the symmetry group and act as conformal transformations on celestial 
spheres. BMS symmetry has recently been the focus of 
renewed interest, as it was shown that it can be extended to include arbitrary, local conformal 
transformations of the celestial spheres \cite{BarnichTroessaert01,BarnichTroessaert02,Banks}, and 
also that it is related to standard quantum field theory in Minkowski space through certain ``soft 
theorems'' that were known in a completely 
different language ever since the sixties \cite{WeinbergSoft} (see {\it e.g.}~\cite{Strominger01,He,Kapec}, 
references therein, and their follow-ups). Many more open problems remain to be settled, 
both regarding holography in general, and BMS symmetry in particular; the hope of the author is that 
addressing such questions may open the door to a deeper understanding of quantum gravity.

\section*{Acknowledgements}

I am grateful to the organizers of the Brussels Summer School of Mathematics, and in particular to 
T.~Connor, J.~Distexhe, J.~Meyer and P.~Weber, for giving me the opportunity to share some of the beauties of physics 
with the audience of the school. I am also indebted to my teachers M.~Bertelson, J.~Federinov and M.~Henneaux 
for lectures on physics and mathematics of exceptional quality, which were most helpful in preparing these 
notes. Finally, I wish to thank L.~Donnay for pointing out a typo in earlier versions of the formula I wrote for the transformation law of advanced time. This work was supported by Fonds de la Recherche Scientifique-FNRS under grant number FC-95570.

\newpage
% \nocite{*}
% %\bibliographystyle{unsrt}
% \addcontentsline{toc}{section}{References}
% \bibliography{LorentzBib}

\section*{References}
\addcontentsline{toc}{section}{References}

\renewcommand{\section}[2]{}%
 
\def\cprime{$'$}

\providecommand{\href}[2]{#2}\begingroup\raggedright\endgroup

\end{document}